\documentclass{article}%
\usepackage{amssymb}
\usepackage{amsfonts}
\usepackage{amsmath}
\usepackage{graphicx}
\usepackage{caption}
\usepackage{subcaption}
\usepackage{lipsum}
\usepackage{chngcntr}
\counterwithin{figure}{section}
\providecommand{\U}[1]{\protect\rule{.1in}{.1in}}
\newtheorem{theorem}{Theorem}[section]

\newtheorem{corollary}[theorem]{Corollary}
\newtheorem{definition}{Definition}[section]

\newtheorem{lemma}[theorem]{Lemma}
\newtheorem{proposition}[theorem]{Proposition}
\newtheorem{remark}[theorem]{Remark}

\newcommand{\lambdaav}{\lambda_{\text{av}}}
\newcommand\blfootnote[1]{%
	\begingroup
	\renewcommand\thefootnote{}\footnote{#1}%
	\addtocounter{footnote}{-1}%
	\endgroup
}
\begin{document}
	
\title{Optimal dividend strategies for a catastrophe insurer}
\author{Hansj\"{o}rg Albrecher\thanks{Corresponding Author. Department of Actuarial Science, Faculty of
		Business and Economics, University of Lausanne, CH-1015 Lausanne and Swiss
		Finance Institute. Email: hansjoerg.albrecher@unil.ch}, Pablo Azcue\thanks{Departamento de Matematicas, Universidad
		Torcuato Di Tella. Av. Figueroa Alcorta 7350 (C1428BIJ) Ciudad de Buenos
		Aires, Argentina. Email: pazcue@utdt.edu and nmuler@utdt.edu} and Nora Muler$^{\dag}$}
\date{}
\maketitle

\bigskip\abstract{\begin{quote}
		\noindent 
		In this paper we study the problem of optimally paying out dividends from an insurance portfolio, when the criterion is to maximize the expected discounted dividends over the lifetime of the company and the portfolio contains claims due to natural catastrophes, modelled by a shot-noise Cox claim number process. The optimal value function of the resulting two-dimensional stochastic control problem is shown to be the smallest viscosity supersolution of a corresponding Hamilton-Jacobi-Bellman equation, and we prove that it can be uniformly approximated through a discretization of the space of the free surplus of the portfolio and the current claim intensity level. We implement the resulting numerical scheme to identify optimal dividend strategies for such a natural catastrophe insurer, and it is shown that the nature of the barrier and band strategies known from the classical models with constant Poisson claim intensity carry over in a certain way to this more general situation, leading to action and non-action regions for the dividend payments as a function of the current surplus and intensity level. We also discuss some interpretations in terms of upward potential for shareholders when including a catastrophe sector in the portfolio. 	
\end{quote}}	

\section{Introduction\label{Section Introduction}}
Optimal strategies for dividend payout from a surplus process of an insurance portfolio is a classical object of study in risk theory, starting with de Finetti \cite{defin} and Gerber \cite{Ger69}. Many of the existing results in the literature study variations of this problem (in terms of objective functions and constraints) under the assumption that the underlying risk process is a Brownian motion or a classical Cram\'er-Lundberg process, see e.g.\ \cite{AT,avanzi} for an overview. Over the years, it has been noted that the compound Poisson assumption of the Cram\'er-Lundberg process is too restrictive, and that claim number processes of doubly stochastic Poisson type (Poisson processes with stochastic intensity) are a more natural choice for certain application areas. The resulting Cox process can lead to tractable models, particularly under the assumption of a Poisson shot-noise intensity, see e.g.\ Dassios \&  Jang \cite{Dassios Jang} and Albrecher \& Asmussen \cite{AA06}. Such a shot-noise dynamic is for instance a natural model for claim arrivals in the presence of catastrophes, where at Poissonian times a sudden jump of random size increases the intensity, leading to more claims for a certain period, and that additional intensity level then decreases over time as claims due to that catastrophe get reported and settled. In recent years, such models have been studied for various purposes, see e.g.\ Dassios \&  Zhao \cite{DassiosZhao}, Macci \& Torrisi \cite{maTo11}, Jang \& Oh \cite{jangoh} and Pojer \& Thonhauser \cite{PoTh23,PoTh23b} in an insurance context, Boxma \&  Mandjes \cite{BoxMa} for a related model in queueing and Schmidt \cite{Schmidt17} for applications in finance.\\ 
Stochastic control problems for such compound shot-noise Cox processes have, however, to the best of our knowledge not been addressed in the respective literature, with the notable recent exception of Liu \& Cadenillas \cite{LiuCad23}, who study a problem of optimal premiums, retention and prevention strategies which maximize expected utility of the insurer and policyholders in a certain way.\\

In this paper we will study the problem of maximizing expected aggregate discounted dividend payments up to ruin for an insurance risk process of compound shot-noise Cox type. In addition to the mathematical interest in studying such a problem, this allows to some extent to assess and establish profitable strategies for investors of an insurance company which cover catastrophic risks, such as floods and storms, where one exogenous event leads to a sudden and time-transient increase of the claim arrival intensity. The latter is likely to become a more prominent topic in the future, not the least because of climate change and an increased frequency of catastrophes. While the setup of this paper is stationary, it may serve as a benchmark for future studies where also an anticipated increase of the frequency of catastrophes during the period of consideration and its effects on optimal strategies can be considered. Methodologically, the problem studied in this paper is a stochastic control problem in two dimensions. Due to the Markovian structure of the shot-noise process, the risk process is Markovian as a function of the current surplus and the current intensity level. Two-dimensional control problems have recently received quite some attention in risk theory, see e.g.\ \cite{AAM20,AAM22}, \cite{gran} and \cite{gu}. However, the actual techniques needed in the present paper are somewhat different from the ones in the aforementioned papers, since the second dimension affects the dynamics of the process differently, and a substantial amount of technicalities will be needed to tackle the posed optimization problem in a rigorous way. \\

We will prove that the optimal value function of this stochastic control problem is the smallest viscosity supersolution of a Hamilton-Jacobi-Bellman equation and we will also show that it can be approximated uniformly by a sufficiently fine discretization of the surplus process and the intensity process. This will then allow us to determine numerically the value function of this optimal dividend problem together with the optimal dividend payment strategies. It will turn out that the additional variability of the claim intensity process leads to an upward potential for the shareholders of the insurance company. Under the assumption that the intensity process can be observed (which we tacitly assume here, but it is not unrealistic, as the jumps in the shot-noise process of the intensity are the documented catastrophes and the used decay function may be assumed known as a consequence of a modelling approach on past claim settlement experiences), the company can in fact steer the dividend streams according to the present situation of claim intensity and surplus level, and benefit from the already received premiums from the underlying policyholders. The numerical results in this paper allow to quantify this effect, under the albeit somewhat simplistic model assumptions. \\

The structure of the remaining paper is as follows. Section \ref{sec2} introduces the underlying insurance model and the concrete formulation of the considered stochastic control problem. Section \ref{Model and basic results} derives some basic properties of the optimal value function. Section \ref{Section HJB} formulates the corresponding Hamilton-Jacobi-Bellman equation and shows that the optimal value function can be identified as its smallest viscosity supersolution, and Section \ref{Section Asymptotic} briefly discusses some asymptotic properties of the latter. Subsequently, in Section \ref{Section Approximation Disc Surplus} we show that one can uniformly approximate the optimal value function through admissible strategies defined on a discretization grid of the surplus values. In Section \ref{Section Disc Intensity}  we then pave the way for actual numerical solutions of the control problem by discretizing also the intensity space and showing that the optimal value function can be approximated uniformly that way. In Section \ref{Numerical Results} we first describe how to set up the numerical scheme concretely, and afterwards we apply the procedure to a number of parameter settings from classical compound Poisson examples, so that we can study the deviations of the optimal strategies when introducing the shot-noise process for the claim number intensity. Section \ref{seccon} concludes and gives some practical interpretations of the obtained results as well as possible directions for future research. All proofs are delegated to an extensive appendix. 

\section{The model}\label{sec2}
\begin{figure}[tbh]
	\begin{center}
		\includegraphics[width=6cm]{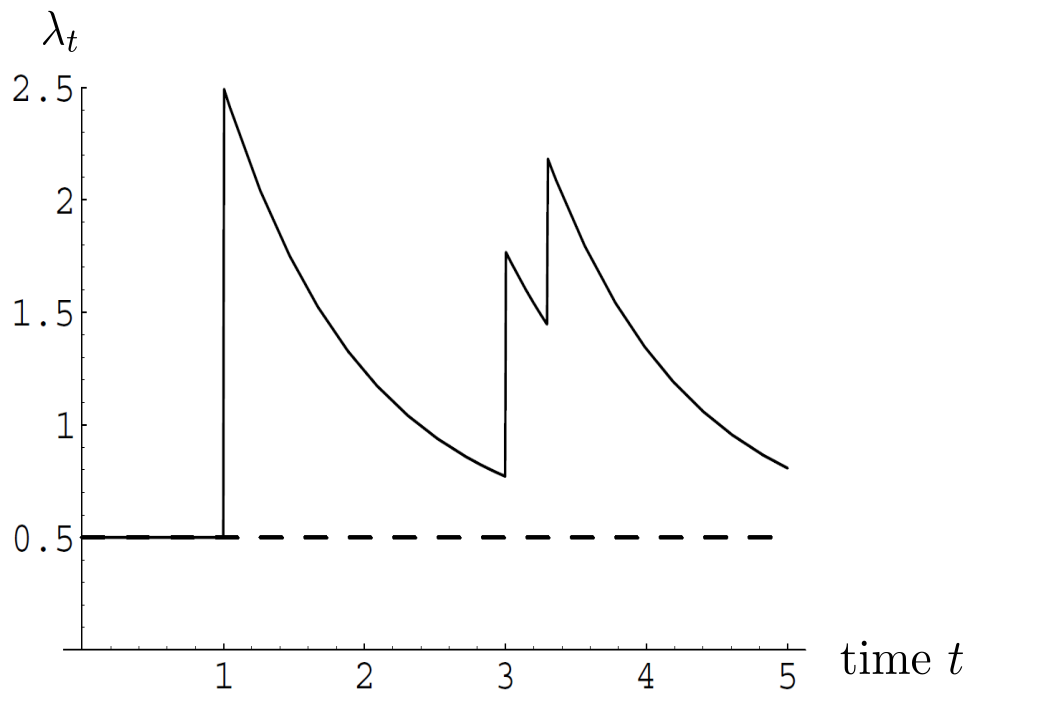}
	\end{center}
	\caption{{\protect\small Sample path of $\lambda_t$}}%
	\label{samplep}%
\end{figure}

Consider a free surplus process of an insurance portfolio given
by%
\begin{equation}
X_{t}=x+pt-\sum_{j=1}^{N_{t}}U_{j}, \label{model}%
\end{equation}
where $x$ is the initial surplus, $p$ is the premium rate and $U_{i}\ $is the
size of the $i$-th claim (arriving at time $\tau_i$). All claims are assumed to be i.i.d.\ positive random
variables with distribution function $F_{U}$ and finite expectation. Let
the process
\[
N_{t}=\#\{j:\tau_{j}\leq t\}
\]
be an inhomogeneous Poisson process with intensity $\lambda_{t}$, corresponding to$\ $the number of claims up to time $t$. The process $N_{t}$\ and the random
variables $U_{i}$ are independent of each other. We assume that the intensity
$\lambda_{t}$ is a shot-noise process, that is%
\begin{equation}
\lambda_{t}=\lambda_{t}^{c}+\sum_{k=1}^{\widetilde{N}_{t}}Y_{k}~e^{-d(t-T_{k}%
)} \label{Lambda t corrida},
\end{equation}
where
\begin{equation}
\lambda_{t}^{c}=\underline{\lambda}+e^{-dt}\left(  \lambda-\underline{\lambda
}\right). \label{Lambda continua}%
\end{equation}
Here $\lambda\geq\underline{\lambda}$ is the initial intensity and
$\widetilde{N}_{t}$ $=\#\{k:T_{k}\leq t\}$ is a Poisson process of constant
intensity $\beta$. Note that $\lambda_{t}\geq\underline{\lambda}$ for all
$t\geq0$ and if the initial intensity is $\lambda>\underline{\lambda}$, then
$\lambda_{t}>\underline{\lambda}$ for all $t\geq0$. See Figure \ref{samplep} for an illustration of a sample path of $\lambda_t$ for $\lambda=\underline{\lambda}$.
$T_{k}$ corresponds to the
arrival times of catastrophes that produce jumps in the intensity $\lambda
_{t}$ and the upward jumps $Y_{k}$ are assumed to be i.i.d.\ positive random
variables with distribution function $F_{Y}$ and finite expectation. In this context, $\underline{\lambda}$ can also be interpreted as a base intensity for incoming claim occurrences, for instance also a 'regular' non-catastrophic policies part in the portfolio. \\

Following Dassios and Jang \cite{Dassios Jang}, we can describe this
model in a rigorous way by defining first the compound Cox filtered
space\textit{. }Given any intensity-filtered probability space\textit{\ }%
$(\Omega^{0},\mathcal{F}^{0},\left(  \mathcal{F}_{t}^{0}\right)  _{t\geq
0},\mathbb{P}^{0})$ and any intensity random process $\mathbf{\lambda}=\left(
\lambda_{t}\right)  _{t\geq0}$ adapted to $\left(  \mathcal{F}_{t}^{0}\right)
_{t\geq0}$, let us define the compound Cox filtered space\textit{\ }%
$(\Omega^{\mathbf{\lambda}},\mathcal{F}^{\mathbf{\lambda}},\left(
\mathcal{F}_{t}^{\mathbf{\lambda}}\right)  _{t\geq0},\mathbb{P}%
^{\mathbf{\lambda}})$ conditional on the process $\mathbf{\lambda}$ in the
following way. The sample set is%
\[
\Omega^{\mathbf{\lambda}}=\Omega^{0}\times\overline{\Omega},
\]
where%
\begin{equation}
\overline{\Omega}=\{(\tau_{n},U_{n})_{n\geq1}\in\lbrack0,\infty)\times
(0,\infty):\tau_{n}<\tau_{n+1}~\text{and }\lim_{n\rightarrow\infty}\tau
_{n}=\infty\}; \label{OmegaBarra}%
\end{equation}
$\mathcal{F}^{\mathbf{\lambda}}$ is the complete $\sigma$-field generated by
the $\sigma$-field $\mathcal{F}^{0}$ and the random variables $\tau_{n}%
:\Omega^{\mathbf{\lambda}}\rightarrow\lbrack0,\infty)$ and $U_{n}%
:\Omega^{\lambda}\rightarrow(0,\infty)$; the filtration $\left(
\mathcal{F}_{t}^{\mathbf{\lambda}}\right)  _{t\geq0},$ where $\mathcal{F}%
_{t}^{\lambda}$ is the complete $\sigma$-field generated by the $\sigma$-field
$\mathcal{F}_{t}^{0}$ and the random variables $\tau_{n}:\Omega^{\lambda
}\rightarrow\lbrack0,\infty)$ and $U_{n}:\Omega^{\lambda}\rightarrow
(0,\infty)$ for $\tau_{n}\leq t$; and $\mathbb{P}^{\mathbf{\lambda}}~$the
probability measure defined in $\mathcal{F}^{\mathbf{\lambda}}$ which satisfies

\begin{enumerate}
\item $(U_{n})_{n\geq1}$ is a sequence of i.i.d.\ random variables with
$\mathbb{P}^{\mathbf{\lambda}}(U_{n}\leq x)=F_{U}(x)$;

\item the counting process\textit{\ }$N_{t}^{\mathbf{\lambda}}:\Omega
^{\mathbf{\lambda}}\rightarrow\mathbb{N}_{0}$ defined by$~N_{t}%
^{\mathbf{\lambda}}=\#\{n:\tau_{n}\leq t\}~$satisfies
\[
\mathbb{P}^{\mathbf{\lambda}}(\left.  N_{t_{2}}^{\mathbf{\lambda}}-N_{t_{1}%
}^{\mathbf{\lambda}}=k\right\vert \lambda_{s},t_{1}\leq s\leq t_{2}%
)=e^{-\int_{t_{1}}^{t_{2}}\lambda_{s}ds}\frac{1}{k!}(\int_{t_{1}}^{t_{2}%
}\lambda_{s}ds)^{k}%
\]
for $t_{1}<\ t_{2}$;

\item the random variables $U_{n}$ are independent of the counting process
$N_{t}^{\mathbf{\lambda}}$.
\end{enumerate}

The shot-noise intensity filtered probability space is then defined as 
\[
(\Omega^{SN},\mathcal{F}^{SN},\left(  \mathcal{F}_{t}^{SN}\right)  _{t\geq
0},\mathbb{P}^{SN}),
\]
where
\[
\Omega^{SN}=\left\{  (T_{k},Y_{k})_{k\geq1}\in\lbrack0,\infty)\times
(0,\infty):T_{k}<T_{k+1}~\text{and }\lim_{k\rightarrow\infty}T_{k}%
=\infty\right\}
\]
and $\mathcal{F}_{t}^{SN}$ is the $\sigma$-field generated by the set
$\{(T_{k},Y_{k})_{k\geq1}:T_{k}\leq t\}$. The Poisson process
\[
\widetilde{N}_{t}=\#\{k:T_{k}\leq t\}
\]
of constant intensity $\beta$ is independent of the random variables $Y_{k}$.
The intensity shot-noise process $\lambda_{t}$ with initial intensity
$\lambda$ is given by (\ref{Lambda t corrida}).\\

Given any $\lambda\geq\underline{\lambda},$ we consider in this paper the
compound Cox filtered space\textit{\ }$(\Omega,\mathcal{F},\left(
\mathcal{F}_{t}\right)  _{t\geq0},\mathbb{P})$ conditional on the intensity
shot-noise process $\left(  \lambda_{t}\right)  _{t\geq0}$ with initial
intensity $\lambda$ given by (\ref{Lambda t corrida}). Here the intensity
filtered probability space\textit{\ }$(\Omega^{0},\mathcal{F}^{0},\left(
\mathcal{F}_{t}^{0}\right)  _{t\geq0},\mathbb{P}^{0})=(\Omega^{SN}%
,\mathcal{F}^{SN},\left(  \mathcal{F}_{t}^{SN}\right)  _{t\geq0}%
,\mathbb{P}^{SN})$, the counting process $N_{t}:\Omega\rightarrow
\mathbb{N}_{0}$ is called the Cox process conditional on the intensity
shot-noise process $\lambda_{t}$ and the surplus given in (\ref{model}) is
called the compound Cox process with drift $p$\ generated by the shot-noise
process $\lambda_{t}$\ with initial intensity $\lambda$\ and initial surplus
$x.$

\bigskip Let us define
\begin{equation}
\Lambda_{t}=\int_{0}^{t}\lambda_{s}ds. \label{Integral de Lambda}%
\end{equation}
Then,
\[
\mathbb{P}(\left.  \tau_{j+1}-\tau_{j}\leq t\right\vert \mathcal{F}_{\tau_{j}%
})=\mathbb{E}\left[  \left.  1-e^{-\int_{0}^{t}\lambda_{s+\tau_{j}}%
ds}\right\vert \mathcal{F}_{\tau_{j}}\right]  =\mathbb{E}\left[  \left.
1-e^{-\left(  \Lambda(t+\tau_{j})-\Lambda(\tau_{j})\right)  }\right\vert
\mathcal{F}_{\tau_{j}}\right]
\]
and in particular%
\[
\mathbb{P}(\tau_{1}\leq t)=\mathbb{E}\left[  1-e^{-\Lambda_{t}}\right]  .
\]

The insurance company uses part of the surplus to pay dividends to the shareholders. Let
us define the dividend strategy $L=\left(  L_{t}\right)  _{t\geq0}$ where
$L_{t}$ denote the cumulative dividends paid up to time $t$. The strategy $\left(  L_{t}\right)  _{t\geq0}$ is admissible if
it is non-decreasing, c\`{a}dl\`{a}g (right continuous with left limits),
adapted with respect to the $\left(  \mathcal{F}_{t}\right)  _{t\geq0}$, and if it 
satisfies $L_{0}\geq0$ and $L_{t}\leq X_{t}$ for $t<\tau^{L},$ where the ruin
time $\tau^{L}$ is defined as
\begin{equation}
\tau^{L}=\inf\left\{  t\geq0:X_{t}-L_{t^{-}}<0\right\}  \text{.}
\label{DefinicionTauRuina}%
\end{equation}
This last condition means that the ruin time can only occur at the arrival of
a claim and that no lump dividend payment can be made at the ruin time. We
define the controlled surplus process as%
\begin{equation}
X_{t}^{L}=X_{t}-L_{t}\text{.} \label{Controled Process}%
\end{equation}
Denote by $\Pi_{x,\lambda}$ the set of admissible dividend strategies
starting with initial surplus level $x\geq0$ and initial intensity
$\lambda\geq\underline{\lambda}.$ For any initial surplus level $x\geq0$
and initial intensity $\lambda\geq\underline{\lambda}$, we can write the
optimal value function as%
\begin{equation}
V(x,\lambda)=\sup_{L\in\Pi_{x,\lambda}}J(L;x,\lambda), \label{DefinitionV}%
\end{equation}
where%

\begin{equation}
J(L;x,\lambda)=\mathbb{E}(\int_{0^{-}}^{\tau^{L}}e^{-qt}dL_{t})\text{.}
\label{Definicion VL}%
\end{equation}
Here $q>0$ is a constant discount factor. We assume that the premium rate $p$
is determined using an expected value principle w.r.t. the asymptotic distribution
of $\lambda_{t}$. That is,
\begin{equation}
p=(1+\eta)\mathbb{E(}U_{1}\mathbb{)}\text{ }\lim_{t\rightarrow\infty
}\mathbb{E}(\frac{\Lambda_{t}}{t}):=(1+\eta)\mathbb{E(}U_{1}\mathbb{)}\lambdaav, \label{Definicion p}%
\end{equation}
where $\eta>0$ is the relative safety loading.

Adapting results of \cite[Cor.2.4]{Dassios Jang}, one can derive the following explicit expressions (we delegate the proof to Appendix \ref{Appendix Section Introduction}).

\begin{proposition}
\label{Proposition E(lambda)}We have that%
\[
\mathbb{E}(\lambda_{t})=\underline{\lambda}(1-e^{-dt})+\lambda e^{-dt}%
+\dfrac{(1-e^{-dt})}{d}\beta~\mathbb{E}(Y_{1}),
\]%
\[
\mathbb{E}(\Lambda_{t})=\underline{\lambda}t-\underline{\lambda}\left(
\frac{1-e^{-dt}}{d}\right)  +\dfrac{(1-e^{-dt})}{d}\lambda+\dfrac
{(e^{-dt}-1+dt)}{d^{2}}\beta~\mathbb{E}(Y_{1})
\]
and%
\[
\lambdaav=\underline{\lambda}+\dfrac{\beta
~\mathbb{E}(Y_{1})}{d}.
\]

\end{proposition}

\section{Basic results \label{Model and basic results}}

The proofs of the propositions in this section are in Appendix \ref{Appendix Section Model and Basic Results}. First, we
show that $V(x,\cdot)$ is non-increasing on $\lambda$ and uniformly continuous.
\begin{proposition}
\label{Continuidad Uniforme en Lambda}(1) If $\lambda_{1}<\lambda_{2}$, then
$V(x,\lambda_{1})\geq V(x,\lambda_{2}).$(2) Given $\varepsilon>0,$ there
exists $\delta>0$ (independent of $x,$ $\lambda_{1}$ and $\lambda_{2}$) such
that if $\lambda_{2}-\lambda_{1}<\delta$, then $V(x,\lambda_{1})-V(x,\lambda
_{2})\leq$ $\varepsilon$. So $V$ is uniformly continuous in $\lambda.$
\end{proposition}
Next, we prove that $V(\cdot,\lambda):\left[  0,\infty
\right)  \rightarrow(0,\infty)$ is locally Lipschitz.

\begin{proposition}
\label{Locally Lipschitz x}Take $x_{2}>x_{1}$, then%
\[
0\leq V(x_{2},\lambda)-V(x_{1},\lambda)\leq V(x_{2},\underline{\lambda}%
)\frac{\beta\lambda+q}{p}(x_{2}-x_{1}).
\]

\end{proposition}

In the next proposition, we show a locally Lipschitz result for $V(x,\cdot)$
in the open set $(\underline{\lambda},\infty)$. At the lower boundary
$\underline{\lambda}$ we only have the uniformly continuity result given in
Proposition \ref{Continuidad Uniforme en Lambda}, because the Lipschitz
constant obtained in the next proposition blows up at $\lambda=\underline
{\lambda}$.

\begin{proposition}
\label{Condicion de Lipschitz en Lambda}Consider $\underline{\lambda}%
<\lambda_{1}<\lambda_{2}$, then%
\[
0\leq V(x,\lambda_{1})-V(x,\lambda_{2})\leq V(x,\lambda_{1})\allowbreak
\frac{\beta\lambda_{2}+q}{d(\lambda_{1}-\underline{\lambda})}\left(
\lambda_{2}-\lambda_{1}\right)  .
\]

\end{proposition}

\begin{remark}
\normalfont\label{Optima Unidimensional} The dividend optimization problem
with constant intensity was studied intensively in the literature (see
e.g.\ Schmidli \cite[Sec.2.4]{Schmidli book 2008} and Azcue and Muler \cite{AM
2005}). Unlike the shot-noise optimization problem, this constant-intensity
problem is one-dimensional. Let us denote by $v^{\lambda}(x)$ the optimal
value function with constant intensity $\lambda$ and premium rate $p$; it is
known that $v^{\lambda}$ is non-decreasing with $v^{\lambda}(x)-x$ positive,
non-decreasing and bounded. Moreover, $\lim_{\lambda\rightarrow\infty
}v^{\lambda}(x)=x$.
\end{remark}

\begin{remark}\normalfont
\label{Crecimiento lineal V} From Proposition
\ref{Continuidad Uniforme en Lambda}, we obtain that $V(x,\lambda)\leq
v^{\underline{\lambda}}(x)$ for all $x\geq0$ and $\lambda\in\lbrack
\underline{\lambda},\infty).$ So, the optimal value function $V$ satisfies
$V(x,\lambda)\leq v^{\underline{\lambda}}(x)$ $\leq x+K$ for some $K>0$.
\end{remark}

We now state the so-called Dynamic Programming Principle (DPP). The proof is
similar to the one given in Lemma 1.2 of {\cite{AM 2014}}.
\begin{lemma}
\label{DPP}For any initial surplus $x\geq0$ and any stopping time $\tau$, we
can write
\[
V(x,\lambda)=\sup\nolimits_{L\in\Pi_{x,\lambda}}\left(  \mathbb{E}(%
{\textstyle\int\nolimits_{0^{-}}^{\tau\wedge\tau^{L}}}
e^{-qs}dL_{s}+e^{-q(\tau\wedge\tau^{L})}I_{\{\tau\wedge\tau^{L}<\tau^{L}%
\}}V(X_{\tau\wedge\tau^{L}}^{L},\lambda_{\tau\wedge\tau^{L}})\right)  .
\]
\end{lemma}
Finally, the asymptotic behavior of $V$ as $\lambda$
goes to infinity can be determined.

\begin{proposition}
\label{Convergencia puntual con lambda a infinito} It holds that for all
$x\geq0$, $\lim_{\lambda\rightarrow\infty}V(x,\lambda)=x$.
\end{proposition}

\section{Hamilton-Jacobi-Bellman equation\label{Section HJB}}

In this section we obtain the Hamilton-Jacobi-Bellman (HJB) equation
associated to the optimization problem (\ref{DefinitionV}). These results are
a generalization to the two-dimensional case of the ones given in 
\cite[Sec.3]{AM 2014} for the one-dimensional case. The proofs of the results of this
section and the auxiliary results are deferred to Appendix \ref{Appendix Section HJB}.

The HJB equation of this optimization problem is given by%

\begin{equation}
\max\{\mathcal{L}(V)(x,\lambda),1-V_{x}(x,\lambda)\}=0, \label{HJB}%
\end{equation}
where%

\begin{equation}%
\begin{array}
[c]{ccc}%
\mathcal{L}(V)(x,\lambda) & = & pV_{x}(x,\lambda)-d\left(  \lambda
-\underline{\lambda}\right)  V_{\lambda}(x,\lambda)-(q+\lambda+\beta
)V(x,\lambda)\\[0.1cm]
&  & +\lambda%
{\textstyle\int\nolimits_{0}^{x}}
V(x-\alpha,\lambda)dF_{U}(\alpha)+\beta%
{\textstyle\int\nolimits_{0}^{\infty}}
V(x,\lambda+\gamma)dF_{Y}(\gamma).
\end{array}
\label{DefinicionL}%
\end{equation}

Since the optimal value function $V$ is locally Lipschitz but could be not
differentiable at some points, we cannot say that $V$ is a solution of the HJB
equation, we prove instead that $V$ is a viscosity solution of the
corresponding HJB equation. Let us define this notion.

\begin{definition}
\label{Viscosity}A uniformly continuous function $\overline{u}:\left[
0,\infty\right)  \times\left[  \underline{\lambda},\infty\right)
\rightarrow\mathbf{R}$\textbf{,} that is locally Lipschitz\ in $(0,\infty
)\times\left[  \underline{\lambda},\infty\right)  $, is a \textit{viscosity
supersolution} of (\ref{HJB})\ at $(x,\lambda)\in(0,\infty)\times
(\underline{\lambda},\infty)$\ if any continuously differentiable function
$\varphi:(0,\infty)\times\left[  \underline{\lambda},\infty\right)
\rightarrow\mathbf{R}\ $such that $\varphi\left(  y,\cdot\right)  $ is bounded
for any $y\geq0,~\varphi(x,\lambda)=\overline{u}(x,\lambda)$ and such that
$\overline{u}-\varphi$\ reaches the minimum at $\left(  x,\lambda\right)
$\ satisfies
\[
\max\left\{  \mathcal{L}(\varphi)(x,\lambda),1-\varphi_{x}(x,\lambda)\right\}
\leq0.\
\]
A uniformly continuous function $\underline{u}:\left[  0,\infty\right)
\times\left[  \underline{\lambda},\infty\right)  \rightarrow\mathbf{R}$, that
is locally Lipschitz\ in $(0,\infty)\times\left[  \underline{\lambda}%
,\infty\right)  $, is a \textit{viscosity subsolution} of (\ref{HJB})\ at
$(x,\lambda)\in(0,\infty)\times(\underline{\lambda},\infty)$\ if any
continuously differentiable function $\ \psi:(0,\infty)\times(\underline
{\lambda},\infty)\rightarrow\mathbf{R}\ $such that $\psi\left(  y,\cdot
\right)  $ is bounded for any $y\geq0$, $\psi(x,\lambda)=\underline
{u}(x,\lambda)$ and such that $\underline{u}-\psi$\ reaches the maximum at
$\left(  x,\lambda\right)  $ satisfies
\[
\max\left\{  \mathcal{L}(\psi)(x,\lambda),1-\psi_{x}(x,\lambda)\right\}
\geq0\text{.}%
\]
A function $u:(0,\infty)\times(\underline{\lambda},\infty)\rightarrow
\mathbf{R}$ which is both supersolution and subsolution at $(x,\lambda
)\in(0,\infty)\times(\underline{\lambda},\infty)$ is called a
\textit{viscosity solution} of (\ref{HJB})\ at $\left(  x,\lambda\right)  $.
\end{definition}

\begin{proposition}
\label{Prop V is a viscosity solution} $V$ is a viscosity solution of the HJB
equation (\ref{HJB}) at any $\left(  x,\lambda\right)  \in(0,\infty
)\times(\underline{\lambda},\infty)$.
\end{proposition}

From Remark \ref{Crecimiento lineal V}, the function $V$ satisfies the 
growth condition%

\begin{equation}
u(x,\lambda)\leq K+x\text{ for all }(x,\lambda)\in(0,\infty)\times
(\underline{\lambda},\infty)\text{.} \label{Growth Condition}%
\end{equation}

We have the following characterization of the optimal value function.

\begin{proposition}
\smallskip\label{MenorSuper} The optimal value function $V$ is the smallest
viscosity supersolution of (\ref{HJB}) among those that are non-increasing in
$\lambda$ and satisfy the growth condition (\ref{Growth Condition}).
\end{proposition}

From the previous proposition we deduce the usual viscosity verification result:

\begin{corollary}
\label{VerificacionDividendos} Consider a family of admissible strategies
$\{L^{x,\lambda}\in\Pi_{x,\lambda}:\left(  x,\lambda\right)  \in
(0,\infty)\times(\underline{\lambda},\infty)\}$. If the function
$W(x,\lambda):=J(L^{x,\lambda};x,\lambda)$ is a viscosity supersolution of
(\ref{HJB}), then $W(x,\lambda)$ is the optimal value function. Also, if for
each $k\geq1$ there exists a family of strategies $\{L_{k}^{x,\lambda}\in
\Pi_{x,\lambda}:\left(  x,\lambda\right)  \in(0,\infty)\times(\underline
{\lambda},\infty)\}$ such that $W(x,\lambda):=\lim_{k\rightarrow\infty}%
J(L_{k}^{x,\lambda};x,\lambda)$ is a viscosity supersolution of the HJB
equation (\ref{HJB}) in $(0,\infty)\times(\underline{\lambda},\infty)$, then
$W$ is the optimal value function $V$.
\end{corollary}

As it is usual for these type of problems, the way in which the optimal value
function $V(x,\lambda)$ solves the HJB equation suggests that the state space
$[0,\infty)\times\lbrack\underline{\lambda},\infty)$ is partitioned into two
regions: a no-action region $\mathcal{NA}$ in which no dividends are paid and
an action region $\mathcal{A}$ in which dividends are paid. Roughly speaking,
the points in the $\mathcal{NA}$ region satisfy $\mathcal{L}(V)=0$ and
$1-V_{x}$ $<0$ and the points in $\mathcal{A}$ satisfy $\mathcal{L}(V)\leq0$
and $1-V_{x}$ $=0$.

\section{Asymptotic properties of the optimal value function
\label{Section Asymptotic}}

In the following proposition, we use Corollary \ref{VerificacionDividendos} in
order to prove that there exists an explicit threshold $p/q$, such that if the
surplus is above this explicit threshold, then one should pay at least the
exceeding surplus as dividends, regardless of the current intensity of the
shot-noise process. Hence $[p/q,\infty)\times\lbrack\underline{\lambda}%
,\infty)\subset$ $\mathcal{A}$ but this inclusion is strict.

The proofs of the results of this section and the auxiliary results are
deferred to Appendix \ref{Appendix Asymptotic}.

\begin{proposition}
\label{Proposicion acotacion de retirar} $V(x,\lambda)=x-{p}/{q}%
+V({p}/{q},\lambda)$ for $x\geq{p}/{q}$.
\end{proposition}

We have the following uniform convergence when the current intensity of the
shot-noise process goes to infinity.

\begin{proposition}
\label{Convergencia Uniforme con Lambda} $\lim_{\lambda\rightarrow\infty
}\left(  \sup_{x\geq0}V(x,\lambda)-x\right)  =0.$
\end{proposition}

\begin{remark}\normalfont
\label{Remark Estrategias Extendidas} For technical reasons, we will also
define an extension of admissible strategies in which the insurance company
can pay all the surplus immediately as dividends and finish the insurance
business at any time. More precisely, let us consider (as before) $L_{t}$ as
the cumulative dividend payment strategy up to time $t$ and $\tau^{F}$ is a
stopping (finite or infinite) time at which the company pays all the surplus
immediately as dividends and finishes the insurance business. We say that the
strategy $\widetilde{\pi}=\left(  L,\tau^{F}\right)  $ is admissible if $L$ is
non-decreasing, c\`{a}dl\`{a}g, adapted with respect to the filtration
generated by the process $\left(  X_{t},\lambda_{t}\right)  $, satisfies
$L_{t}\leq X_{t}$ up to ruin time and $\tau^{F}$ is a stopping time with
respect to the filtration generated by the process $\left(  X_{t},\lambda
_{t}\right)  $. We define $\widetilde{\Pi}_{x,\lambda}$ as the set of all
$\widetilde{\pi}$-admissible strategies. Take any $L\in\Pi_{x,\lambda}$ then
$\widetilde{\pi}=(L,\infty)\in\widetilde{\Pi}_{x,\lambda}$ so we can think
that $\Pi_{x,\lambda}$ is contained in $\widetilde{\Pi}_{x,\lambda}$. Let us
define the value function of any strategy $\widetilde{\pi}\in\widetilde{\Pi
}_{x,\lambda}$ as
\[
J(\widetilde{\pi};x,\lambda)=\mathbb{E}(\int_{0^{-}}^{\tau^{L}\wedge\tau^{F}%
}e^{-qt}dL_{t}+I_{\left\{  \tau^{F}<\tau^{L}\right\}  }e^{-q\tau^{F}}%
X_{\tau^{F}}^{L})\text{.}%
\]
It is straightforward to see that the optimal value function defined in
(\ref{DefinitionV}) satisfies
\begin{equation}
V(x,\lambda)=\sup_{\widetilde{\pi}\in\widetilde{\Pi}_{x,\lambda}}%
J(\widetilde{\pi};x,\lambda) \label{Equivalent optimization problem}%
\end{equation}
because it is never optimal to pay all the current surplus and finish the business.
\end{remark}

\section{Approximation of the value function by a discretization of the
surplus \label{Section Approximation Disc Surplus}}

In this section we show that it is possible to approximate uniformly the
optimal value function $V$ with value functions of admissible strategies that
are constructed through a discrete set on the surplus level. These strategies
are stationary in the sense that they only depend on the current surplus and
intensity. Moreover, the only possible options are to either pay a lump sum of
dividends or not paying any dividends. The result of this section follows
closely the ones in \cite{AM 2021}; in that paper, the intensity of the
arrival of claims was constant and so this approach makes it possible to
approximate numerically the optimal value function. However, in the model of the present paper, the situation is more complicated since the
intensity of the arrival of claims changes over time. In contrast to \cite{AM 2021}, in this section we will only obtain a semi-discrete result
that is not sufficient for the numerical results. In the next section, we will
also discretize the intensity process arrivals in order to obtain a
numerical scheme. The proofs of the results of this section can be found in Appendix \ref{Appendix Disc Surplus}.\\

More precisely, we construct in this section a family of admissible strategies
for any point in a subset of $[0,\infty)\times\lbrack\underline{\lambda
},\infty)$ and then extend it to the whole set $[0,\infty)\times
\lbrack\underline{\lambda},\infty)$. We will show that the value functions of
these strategies approximate uniformly the optimal value function $V$.

Given any approximation parameter $\delta>0$, we define the grid domain%
\begin{equation}
\mathcal{G}_{\delta}:=\left\{  x_{n}^{\delta}=np\delta:n\geq0\right\}  
\label{Gdelta}%
\end{equation}
in the surplus state space $[0,\infty)$; we construct first a subfamily of
admissible strategies $\widetilde{\Pi}_{x_{n}^{\delta},\lambda}^{\delta}$
$\subset\widetilde{\Pi}_{x_{n}^{\delta},\lambda}$ for any point $\left(
x_{n}^{\delta},\lambda\right)  \in$ $\mathcal{G}_{\delta}\times\lbrack
\underline{\lambda},\infty)$, and afterwards a subfamily of admissible
strategies $\widetilde{\Pi}_{x,\lambda}^{\delta}$ $\subset\widetilde{\Pi
}_{x,\lambda}$ for any point $\left(  x,\lambda\right)  \in$ $[0,\infty)$.

Let us define the subfamily $\widetilde{\Pi}_{x,\lambda}^{\delta}$
$\subset\widetilde{\Pi}_{x,\lambda}$ for any point $\left(  x,\lambda\right)
\in$ $[0,\infty)\times\lbrack\underline{\lambda},\infty)$ in a precise way.

Consider first the case in which $(x,\lambda)\in\mathcal{G}_{\delta}%
\times\lbrack\underline{\lambda},\infty),$ so $x=x_{n}^{\delta}$ for some
$n\geq0$. The idea of this construction is to find, at each point in
$\mathcal{G}_{\delta}\times\lbrack\underline{\lambda},\infty)$, the best local
strategy among the ones suggested by the operators of the HJB equation
(\ref{HJB}). These possible local strategies are: either the company pays no
dividends or pays immediately a lump sum $p\delta$ as dividends; moreover, the
company can finish the insurance activity at any time. We modify these local
strategies in such a way that the controlled surplus always lies in
$\mathcal{G}_{\delta}$ immediately after the arrival of a claim. Let $\tau$
and $U$ be the arrival time and the size of the next claim, and $T$ and $Y$ be
the arrival time and the size of the next intensity upward jump. We introduce
the auxiliary function
\begin{equation}
\rho^{\delta}(x):=\max\{x_{n}^{\delta}:x_{n}^{\delta}\leq x\}
\label{Definicion de x en la grilla  por abajo}%
\end{equation}
which gives the closest point of the grid $\mathcal{G}_{\delta}$ below
$x.$ We first define the three possible control actions at any point of
$\mathcal{G}_{\delta}\times\lbrack\underline{\lambda},\infty)$ as follows:

\begin{itemize}
\item Control action $\mathbf{E}_{0}$: Pay no dividends up to the time
$\delta\wedge\tau\wedge T$.

\begin{enumerate}
\item In the case that $\delta<\tau\wedge T$, the surplus at time $\delta$ is
$x_{n+1}^{\delta}\in\mathcal{G}_{\delta}$.

\item If $\delta\wedge T\geq\tau$, the uncontrolled surplus at time $\tau$ is
$x_{n}^{\delta}+\tau p-U;~$if this value is positive, the company pays
immediately the minimum amount of dividends in such a way that the controlled
surplus lies in the closest point of the grid below $x_{n}^{\delta}+\tau p-U$;
this end surplus can be written $\rho^{\delta}(x_{n}^{\delta}+\tau p-U)$ and
the amount paid as dividends is equal to $x_{n}^{\delta}+\tau p-U-\rho
^{\delta}(x_{n}^{\delta}+\tau p-U)$; ruin occurs in case the
surplus $x_{n}^{\delta}+\tau p-U<0$ at time $\tau\leq$ $\delta\wedge T$ .

\item If $T<\delta\wedge\tau,$ the end surplus is $\rho^{\delta}(x_{n}%
^{\delta}+Tp)$ and the amount paid as dividends is $x_{n}^{\delta}%
+Tp-\rho^{\delta}(x_{n}^{\delta}+Tp).$
\end{enumerate}

\item Control actions $\mathbf{E}_{1}$: The company pays immediately $p\delta$
as dividends, so the controlled surplus becomes $x_{n-1}^{\delta}%
\in\mathcal{G}_{\delta}$. The control action $\mathbf{E}_{1}$\textbf{\ }can
only be applied for current surplus $x_{n}^{\delta}>0$.

\item Control action $\mathbf{E}_{F}$: The manager opts to pay the current
surplus $x_{n}^{\delta}$ as dividends and to close the company.
\end{itemize}

We denote the space of control actions as
\begin{equation}
\mathcal{E}=\{\mathbf{E}_{F},\mathbf{E}_{1},\mathbf{E}_{0}\}.
\label{Conjunto E de Estrategias.}%
\end{equation}
Consider $\widetilde{\Pi}_{x_{n}^{\delta},\lambda}^{\delta}$ $\subset
\widetilde{\Pi}_{x_{n}^{\delta},\lambda}$ as the set of all the admissible
strategies with initial surplus $x_{n}^{\delta}\in\mathcal{G}_{\delta}$ which
can be obtained by a sequence of control actions in $\mathcal{E}.$ Note that
if $\mathbf{E}_{0}$ is chosen, the next control action depends on the end
surplus and intensity. The length of this sequence could be finite or
infinite, in the case that is finite the last control action is either
$\mathbf{E}_{0}$ in the case that ruin occurs, or $\mathbf{E}_{F}$ because
the control action $\mathbf{E}_{F}$ finishes the sequence.

Finally, we introduce $\widetilde{\Pi}_{x,\lambda}^{\delta}$ $\subset
\widetilde{\Pi}_{x,\lambda}$ for any point $\left(  x,\lambda\right)  \in$
$[0,\infty)$ in the case that $x\notin\mathcal{G}_{\delta}$. Indeed,
$\widetilde{\Pi}_{x,\lambda}^{\delta}$ is the subfamily of admissible
strategies $\widetilde{\mathbb{\pi}}$ which pays $x-\rho^{\delta}(x)$ as
dividends immediately and then follows any strategy $\widetilde{\mathbb{\pi}%
}_{1}$ in $\widetilde{\Pi}_{\rho^{\delta}(x),\lambda}^{\delta}$ with
$\rho^{\delta}(x)\in\mathcal{G}_{\delta},$ so we have that
\[
J(\widetilde{\pi};x,\lambda)=J(\widetilde{\mathbb{\pi}}_{1};\rho^{\delta
}(x),\lambda)+x-\rho^{\delta}(x).
\]
We define
\begin{equation}
V^{\delta}(x,\lambda)=\sup_{\widetilde{\pi}\in\widetilde{\Pi}_{x,\lambda
}^{\delta}}J(\widetilde{\pi};x,\lambda)=V^{\delta}(\rho^{\delta}%
(x),\lambda)+x-\rho^{\delta}(x). \label{vdelta}%
\end{equation}
We will show that, in a certain sense, $\lim_{\delta\rightarrow0}V^{\delta}=V$ uniformly.

It is straightforward that $V^{\delta}(x,\lambda)$ is non-decreasing in
$x\ $and non-increasing in $\lambda$. In the next proposition, we find a bound
on the variations of $V^{\delta}$, which show in particular that $V^{\delta
}(x,\cdot)$ is locally Lipschitz.

\begin{proposition}
\label{Lipschitz Vdelta} The function $V^{\delta}(x,\lambda)$ satisfies
\[
V^{\delta}(x_{2},\lambda)-V^{\delta}(x_{1},\lambda)\leq V^{\delta}%
(x_{2},\underline{\lambda})\frac{e^{\left(  \beta+q\right)  \delta+\int
_{0}^{\delta}\lambda_{u}^{c}du}-1}{p\delta}\left(  \rho^{\delta}(x_{2}%
)-\rho^{\delta}(x_{1})\right)  +\delta p
\]
for $x_{2}\geq x_{1}$ $\geq0.$ Also%
\[
0\leq V^{\delta}(x,\lambda_{1})-V^{\delta}(x,\lambda_{2})\leq V^{\delta
}(x,\lambda_{1})\frac{\beta\lambda_{2}+q}{d(\lambda_{1}-\underline{\lambda}%
)}\left(  \lambda_{2}-\lambda_{1}\right)
\]
for $\lambda_{2}\geq\lambda_{1}>\underline{\lambda}$ and $\lambda_{2}%
-\lambda_{1}\leq d(\lambda_{1}-\underline{\lambda})$.\bigskip
\end{proposition}

As in \cite{AM 2021}, we will show that the function $V^{\delta}$ restricted
to $\mathcal{G}_{\delta}\times\lbrack\underline{\lambda},\infty)$ is a
solution of a discrete version of the HJB equation (\ref{HJB}), given in
(\ref{Delta HJB}). In order to define this discrete HJB equation, let us
introduce the operators related to the control actions in $\mathcal{E}$.
Consider the operators $\mathcal{T}_{0}$, $\mathcal{T}_{1}$ and $\mathcal{T}%
_{F}$ in the set of functions $\mathcal{W}^{\delta}=\left\{  w:\mathcal{G}%
_{\delta}\times\lbrack\underline{\lambda},\infty)\rightarrow\mathbf{[}%
0,\infty)\mathbf{~}\text{which are Lebesgue measurable}\right\}  $ defined
as follows%

\begin{equation}
\mathcal{T}_{0}(w)(x_{n}^{\delta},\lambda):=\mathbb{P}(\delta\wedge
T\wedge\tau=\delta)e^{-q\delta}w(x_{n+1}^{\delta},\lambda_{\delta}%
^{c})+\mathcal{I}^{\delta}(w)(x_{n}^{\delta},\lambda), \label{Definicion TE}%
\end{equation}%
\begin{equation}%
\begin{array}
[c]{ccc}%
\mathcal{T}_{1}(w)(x_{n}^{\delta},\lambda):=w(x_{n-1}^{\delta},\lambda)+\delta
p & \text{and} & \mathcal{T}_{F}(w)(x_{n}^{\delta},\lambda):=x_{n}^{\delta}.
\end{array}
\label{Definicion Ti TS}%
\end{equation}
Here,%
\begin{equation}%
\begin{array}
[c]{l}%
\mathcal{I}^{\delta}(w)(x,\lambda)\\%
\begin{array}
[c]{l}%
:=\mathbb{E}(I_{\delta\wedge T\wedge\tau=\tau}e^{-q\tau}w(\rho^{\delta}%
(x_{n}^{\delta}+p\tau-U),\lambda_{\tau}^{c}))\\[0.1cm]
+\mathbb{E}(I_{\delta\wedge T\wedge\tau=T}e^{-qT}w(x_{n}^{\delta},\lambda
_{T}^{c}+Y))\\[0.1cm]
+\mathbb{E}(I_{\delta\wedge T\wedge\tau=\tau}e^{-q\tau}(x+p\tau-U-\rho
^{\delta}(x_{n}^{\delta}+p\tau-U)))\\[0.1cm]
+\mathbb{E}(I_{\delta\wedge T\wedge\tau=T}e^{-qT}pT),
\end{array}
\end{array}
\label{Definicion Idelta(W)}%
\end{equation}
where $\tau$ and $U$ are the arrival time and the size of the next claim, and
$T$ and $Y$ are the arrival time and the size of the next intensity upward
jump. We also introduce the operator $T$ in $\mathcal{W}^{\delta}$ as%
\begin{equation}
\mathcal{T}:=\max\{\mathcal{T}_{0},\mathcal{T}_{1},\mathcal{T}_{F}\}.
\label{Definicion T}%
\end{equation}

Given any family of admissible strategies%
\[
\widetilde{\mathbf{\pi}}=\left\{  \widetilde{\pi}_{x_{n}^{\delta},\lambda}%
\in\widetilde{\Pi}_{x_{n}^{\delta},\lambda}^{\delta}\text{ for }\left(
x_{n}^{\delta},\lambda\right)  \in\mathcal{G}_{\delta}\times\lbrack
\underline{\lambda},\infty)\right\}  \text{,}%
\]
we define the value function $W:\mathcal{G}_{\delta}\times\lbrack
\underline{\lambda},\infty)\rightarrow\mathbf{R}$ of $\widetilde{\mathbf{\pi}%
}$ as
\[
W(x_{n}^{\delta},\lambda):=J(\widetilde{\pi}_{x_{n}^{\delta},\lambda}%
;x_{n}^{\delta},\lambda)\text{.}%
\]
$W\in\mathcal{W}^{\delta},$ $\mathcal{T}_{0}(W)(x_{n}^{\delta},\lambda)$
and $\mathcal{T}_{1}(W)(x_{n}^{\delta},\lambda)$ are the values of the
strategies with initial surplus $x_{n}^{\delta}\in\mathcal{G}_{\delta}$ and
initial intensity $\lambda$ which consist of applying first the control
actions $\mathbf{E}_{0}$ and $\mathbf{E}_{1}\in$ $\mathcal{E}$
respectively\textbf{,} and afterwards applying the strategy in the family
$\widetilde{\mathbf{\pi}}$ corresponding to the end surplus and intensity.
Also, $\mathcal{T}_{F}(W)(x_{n}^{\delta},\lambda)$ is the value function of
the control action $\mathbf{E}_{F}\in$ $\mathcal{E}$.

We define the \textit{discrete HJB equation in }$\mathcal{G}_{\delta}%
\times\lbrack\underline{\lambda},\infty)$ as%
\begin{equation}
\mathcal{T}(W)-W=0. \label{Delta HJB}%
\end{equation}
Assume that there exists $\widetilde{\pi}_{x_{n}^{\delta},\lambda}%
\in\widetilde{\Pi}_{x_{n}^{\delta},\lambda}^{\delta}$ such that $V^{\delta
}(x_{n}^{\delta},\lambda)=J(\widetilde{\pi}_{x_{n}^{\delta},\lambda}%
;x_{n}^{\delta},\lambda)$ for all $\left(  x_{n}^{\delta},\lambda\right)
\in\mathcal{G}_{\delta}\times\lbrack\underline{\lambda},\infty)$. Then, since
$V^{\delta}\in$ $\mathcal{W}^{\delta}$ by Proposition \ref{Lipschitz Vdelta},
it is straightforward to see that $\mathcal{T}(V^{\delta})(x_{n}^{\delta
},\lambda)=V^{\delta}(x_{n}^{\delta},\lambda)$. Also we can find which is the
optimal control action in $\mathcal{E}$ at each $\left(  x_{n}^{\delta
},\lambda\right)  \in\mathcal{G}_{\delta}\times\lbrack\underline{\lambda
},\infty).$ In Proposition \ref{Limite de Vl}, we will show that $\mathcal{T}%
(V^{\delta})=V^{\delta}$ without the former assumption. In Proposition
\ref{Proposicion Estrategia Optima Discreta}, we will show that indeed,
$V^{\delta}(x_{n}^{\delta},\lambda)$ is the value function of an optimal
strategy within $\widetilde{\Pi}_{x_{n}^{\delta},\lambda}^{\delta}$ and this
strategy is stationary.

By definitions (\ref{Definicion TE}), (\ref{Definicion Ti TS}),
(\ref{Definicion Idelta(W)}) and
(\ref{Definicion T}), we obtain immediately the following result.

\begin{proposition}
\label{Ts crecientes} The operators $\mathcal{T}_{0}$, $\mathcal{T}_{1},$
$\mathcal{T}_{F}$ and $\mathcal{T}$ are non-decreasing and $T$ satisfies
\[%
\sup\nolimits_{\left(  x_{n}^{\delta},\lambda\right)  \in\mathcal{G}_{\delta
}\times\lbrack\underline{\lambda},\infty)}\left\vert \mathcal{T}(W_{1}%
)(x_{n}^{\delta},\lambda)-\mathcal{T}(W_{2})(x_{n}^{\delta},\lambda
)\right\vert 
\leq\sup\nolimits_{\left(  x_{n}^{\delta},\lambda\right)  \in\mathcal{G}%
_{\delta}\times\lbrack\underline{\lambda},\infty)}\left\vert W_{1}%
(x_{n}^{\delta},\lambda)-W_{2}(x_{n}^{\delta},\lambda)\right\vert 
\]
for any $W_{1}$ and $W_{2}$ in $\mathcal{W}^{\delta}.$
\end{proposition}

Given $l\geq1$, let us define $\widetilde{\Pi}_{x_{n}^{\delta},\lambda
}^{\delta,l}$ as the set of all the admissible strategies in $\widetilde{\Pi
}_{x_{n}^{\delta},\lambda}^{\delta}$ with initial surplus $x_{n}^{\delta}%
\in\mathcal{G}_{\delta}$ and initial intensity $\lambda\geq\underline{\lambda
}$ which can be obtained by a sequence of exactly $l$ local control actions in
$\mathcal{E}$.

We define%
\begin{equation}
V_{l}^{\delta}(x,\lambda)=\sup\nolimits_{\widetilde{\pi}\in\widetilde{\Pi
}_{x,\lambda}^{\delta,l}}J(\widetilde{\pi};x,\lambda). \label{Definicionvk}%
\end{equation}

Since $\widetilde{\Pi}_{x_{n}^{\delta},\lambda}^{\delta,1}=\{\mathbf{E}_{F}\}$
then $V_{1}^{\delta}(x,\lambda)=x$. By Proposition \ref{Ts crecientes}, we
have that $V_{1}^{\delta}\in$ $\mathcal{W}^{\delta}$ and $V_{2}^{\delta
}(x,\lambda)=\mathcal{T}(V_{l}^{\delta})(x_{n}^{\delta},\lambda)\geq
\mathcal{T}(V_{l}^{\delta})(x_{n}^{\delta},\lambda)$. So, we can conclude, by a
recursive argument, the following result.

\begin{proposition}
\label{Properties of Vl} It holds that $V^{\delta}\geq V_{l+1}^{\delta}\geq
V_{l}^{\delta}$, $V_{l}^{\delta}\in$ $\mathcal{W}^{\delta}$ and $\mathcal{T}%
(V_{l}^{\delta})(x_{n}^{\delta},\lambda)=$ $V_{l+1}^{\delta}(x_{n}^{\delta
},\lambda)$ for $l\geq1$.
\end{proposition}

In the next proposition, we show that $V^{\delta}:\mathcal{G}_{\delta}%
\times\lbrack\underline{\lambda},\infty)\rightarrow\lbrack0,\infty]$ is a
solution of the discrete HJB equation (\ref{Delta HJB}).

\begin{proposition}
\label{Limite de Vl} It holds that $\lim_{l\rightarrow\infty}V_{l}^{\delta
}=V^{\delta}$ and $\mathcal{T}(V^{\delta})(x_{n}^{\delta},\lambda)=$
$V^{\delta}(x_{n}^{\delta},\lambda)$.
\end{proposition}

\begin{remark}\normalfont 
\label{Delta HJB con dos operadores} Since $\mathcal{T}_{F}(V^{\delta}%
)(x_{n}^{\delta},\lambda)=x_{n}^{\delta}<V^{\delta}(x_{n}^{\delta},\lambda)$,
we can write the discrete HJB equation (\ref{Delta HJB}) as
\[
\max\{\mathcal{T}_{0}(W)-W,\mathcal{T}_{1}(W)-W\}=0\text{,}%
\]
disregarding the operator $\mathcal{T}_{F}$.
\end{remark}

\begin{proposition}
\label{Menor Supersolucion Discreta} Given any $\widetilde{\pi}\in
\widetilde{\Pi}_{x_{n}^{\delta},\lambda}^{\delta}$ and any supersolution
$\overline{W}:$ $\mathcal{G}_{\delta}\times\lbrack\underline{\lambda}%
,\infty)\rightarrow\mathbf{R}$ of (\ref{Delta HJB}), we have that
$J(\widetilde{\pi};x_{n}^{\delta},\lambda)\leq\overline{W}(x_{n}^{\delta
},\lambda).$
\end{proposition}

From the previous proposition, we deduce the next corollary.

\begin{corollary}
\label{Smallest supersolution}The $\mathcal{G}_{\delta}$-optimal value
function $V^{\delta}$ $:\mathcal{G}_{\delta}\times\lbrack\underline{\lambda
},\infty)\rightarrow\mathbf{R}$ can be characterized as the smallest
supersolution of the discrete HJB equation (\ref{Delta HJB}) with growth
condition (\ref{Growth Condition}).
\end{corollary}

\begin{definition}
\label{Partition} Given any partition the set $\mathcal{G}_{\delta}%
\times\lbrack\underline{\lambda},\infty)$ into three measurable subsets
$\mathcal{P}=\left(  \mathcal{A},\mathcal{NA},\mathcal{B}\right)$, we define
for any point $(x_{n}^{\delta},\lambda)\in\mathcal{G}_{\delta}\times
\lbrack\underline{\lambda},\infty)$ the local control action $S(x_{n}^{\delta
},\lambda)\in\mathcal{E}$ in the following way:

\begin{itemize}
\item If $(x_{n}^{\delta},\lambda)\in\mathcal{B}$, take $S_{\mathcal{P}}%
(x_{n}^{\delta},\lambda)=\mathbf{E}_{F}$. $\mathcal{B}$ is called \textit{finish-the-business set}. 

\item If $(x_{n}^{\delta},\lambda)\in\mathcal{NA}$, take $S_{\mathcal{P}%
}(x_{n}^{\delta},\lambda)=\mathbf{E}_{0}$. $\mathcal{NA}$ is called the non-action set. 

\item And if $(x_{n}^{\delta},\lambda)\in\mathcal{A}$, take $S_{\mathcal{P}%
}(x_{n}^{\delta},\lambda)=\mathbf{E}_{1}$\textbf{.} Note that $(x_{0}^{\delta
},\lambda)=(0,\lambda)\notin\mathcal{A}$. $\mathcal{A}$ is called the action set.
\end{itemize}

The $\mathcal{G}_{\delta}$-\textit{\ }strategy $\pi_{x_{n}^{\delta},\lambda
}^{\mathcal{P}}\in\Pi_{x_{n}^{\delta},\lambda}^{\delta}$ associated to
$\mathcal{P}$~in the initial surplus and intensity $(x_{n}^{\delta}%
,\lambda)\in\mathcal{G}_{\delta}\times\lbrack\underline{\lambda},\infty)$ is
defined inductively as follows: Let us call $m_{1}=(x_{n}^{\delta},\lambda)$
and $s_{1}=S_{\mathcal{P}}(x_{n}^{\delta},\lambda)$; assuming that
$m_{1},m_{2},..,m_{k-1}\in\mathcal{G}_{\delta}\times\lbrack\underline{\lambda
},\infty)$ and $s_{1},s_{2},..,s_{k-1}$ $\in\mathcal{E}$ are defined and the
process does not stop at step $k-1$, we define $m_{k}\in$ $\mathcal{G}%
_{\delta}\times\lbrack\underline{\lambda},\infty)$ as the end surplus of
$s_{k-1}$ and $s_{k}=S_{\mathcal{P}}(m_{k})\in\mathcal{E}$.
\end{definition}

Note that the family $\mathbf{\pi}^{\mathcal{P}}=\left(  \pi_{x_{n}^{\delta
},\lambda}^{\mathcal{P}}\right)  _{(x_{n}^{\delta},\lambda)\in\mathcal{G}%
_{\delta}\times\lbrack\underline{\lambda},\infty)}$ is stationary in
$\mathcal{G}_{\delta}\times\lbrack\underline{\lambda},\infty)$ in the sense
that the local control actions depend only on the point of the grid at which
the current surplus lies and also on the current intensity. Moreover, if we
define the associated value function $W_{\mathcal{P}}(x_{n}^{\delta}%
,\lambda)=J(\pi_{x_{n}^{\delta},\lambda}^{\mathcal{P}};x_{n}^{\delta}%
,\lambda),$ then $\mathcal{T}_{F}(W_{\mathcal{P}})=W_{\mathcal{P}}$ in
$\mathcal{B}$, $\mathcal{T}_{1}(W_{\mathcal{P}})=W_{\mathcal{P}}$ in
$\mathcal{A}$ and $\mathcal{T}_{0}(W_{\mathcal{P}})=W_{\mathcal{P}}$ in
$\mathcal{NA}$. We extend the definition of the value function $W_{\mathcal{P}%
}:[0,\infty)\times\lbrack\underline{\lambda},\infty)\rightarrow\lbrack
0,\infty)$ as%
\begin{equation}
W_{\mathcal{P}}(x,\lambda)=W_{\mathcal{P}}(\rho^{\delta}(x),\lambda
)+x-\rho^{\delta}(x), \label{Extension WP}%
\end{equation}
this corresponds to pay the minimum amount of dividends so the surplus lies in
$\mathcal{G}_{\delta}.$

Given the function $V^{\delta}$, since $V^{\delta}=\mathcal{T}(V^{\delta}) $
and $V^{\delta}$ is Lebesgue measurable, we define the $\mathcal{G}_{\delta}%
$-partition $\mathcal{P}_{\delta}^{\ast}=\left(  \mathcal{A}_{\delta}^{\ast
},\left(  \mathcal{NA}\right)  _{\delta}^{\ast},\mathcal{B}_{\delta}^{\ast
}\right)  $ as%

\begin{equation}%
\begin{array}
[c]{lll}%
\mathcal{A}_{\delta}^{\ast} & = & \left\{  \left(  x_{n}^{\delta}%
,\lambda\right)  \in\mathcal{G}_{\delta}\times\lbrack\underline{\lambda
},\infty):\mathcal{T}_{1}(V^{\delta})(x_{n}^{\delta},\lambda)=V^{\delta}%
(x_{n}^{\delta},\lambda)\right\}  ,\\[0.1cm]
\left(  \mathcal{NA}\right)  _{\delta}^{\ast} & = & \left\{  \left(
x_{n}^{\delta},\lambda\right)  \in\left(  \mathcal{G}_{\delta}\times
\lbrack\underline{\lambda},\infty)\right)  -\mathcal{A}_{\delta}^{\ast
}:\mathcal{T}_{0}(V^{\delta})(x_{n}^{\delta},\lambda)=V^{\delta}(x_{n}%
^{\delta},\lambda)\right\}  ,\\[0.1cm]
\mathcal{B}_{\delta}^{\ast} & = & \left\{  \left(  x_{n}^{\delta}%
,\lambda\right)  \in\left(  \mathcal{G}_{\delta}\times\lbrack\underline
{\lambda},\infty)\right)  -\left(  \mathcal{A}_{\delta}^{\ast}\cup\left(
\mathcal{NA}\right)  _{\delta}^{\ast}\right)  :\mathcal{T}_{F}(V^{\delta
})(x_{n}^{\delta},\lambda)=V^{\delta}(x_{n}^{\delta},\lambda)\right\}  .
\end{array}
\label{Optimal Partition delta}%
\end{equation}

Note that, by Remark \ref{Delta HJB con dos operadores}, $\mathcal{B}_{\delta
}^{\ast}$ is empty, so $(x_{0}^{\delta},\lambda)=(0,\lambda)\in\left(
\mathcal{NA}\right)  _{\delta}^{\ast}$. Since the value function
$W_{\mathcal{P}_{\delta}^{\ast}}(x_{n}^{\delta},\lambda)$ is a supersolution
of the discrete HJB equation (\ref{Delta HJB}) with growth condition
(\ref{Growth Condition}), we deduce from Corollary
\ref{Smallest supersolution}, the following result.

\begin{proposition}
\label{Proposicion Estrategia Optima Discreta}We have that $V^{\delta
}=W_{\mathcal{P}_{\delta}^{\ast}}$ in $\mathcal{G}_{\delta}\times
\lbrack\underline{\lambda},\infty)$, and so $\mathbf{\pi}^{\mathcal{P}%
_{\delta}^{\ast}}=\left(  \pi_{x_{n}^{\delta},\lambda}^{\mathcal{P}_{\delta
}^{\ast}}\right)  _{(x_{n}^{\delta},\lambda)\in\mathcal{G}_{\delta}%
\times\lbrack\underline{\lambda},\infty)}$ is the $\mathcal{G}_{\delta}
$\textit{-optimal strategy. }$\mathcal{P}_{\delta}^{\ast}$ is called the
optimal $\mathcal{G}_{\delta}$-partition.
\end{proposition}

Let us prove now that the optimal value function $V$ can be approximated
uniformly by $V^{\delta}$ for some $\delta$ small enough. Since we have a
monotonicity condition on the embedded grids $\mathcal{G}_{\delta/2}%
\subset\mathcal{G}_{\delta}$, so $x_{n}^{\delta}=x_{2n}^{\delta/2}$,
$\widetilde{\Pi}_{x_{n}^{\delta},\lambda}^{\delta}\subset\widetilde{\Pi
}_{x_{2n}^{\delta/2},\lambda}^{\delta/2}$, and this implies that $V^{\delta
}(x_{n}^{\delta},\lambda)\leq V^{\delta/2}(x_{2n}^{\delta/2},\lambda)$. Take
$\delta_{k}:=\delta/2^{k}$ for $k\geq0$. We will see that $V^{\delta_{k}}$
$\nearrow V$ locally uniformly as $k$ goes to infinity. Consider the dense set
in $\mathbf{R}_{+}^{n}$, $\mathcal{G}:=\bigcup\nolimits_{k\geq0}%
\mathcal{G}_{\delta_{k}}$. Note that $\mathcal{G}_{\delta_{k}}\subset
\mathcal{G}_{\delta_{k+1}}$, so
\[
V^{\delta_{k}}\leq V^{\delta_{k+1}}\leq V.
\]
Now we conclude the main result of the paper. The proof is in Appendix \ref{Appendix Disc Surplus}.

\begin{theorem}
\label{Main Theorem} For any $\delta>0$, the functions $V^{\delta_{k}}$
$\nearrow V$ uniformly as $k$ goes to infinity.
\end{theorem}

\section{Discretization on the intensity process
\label{Section Disc Intensity}}

As we pointed out before, the construction of $V^{\delta}$ given in the
previous section only uses a discretization on the surplus space$.$ In this
section, we propose a numerical scheme using a discretization on the intensity
space as well. We will find a partition into an action set, a non-action set and a finish-the-business set as introduced in Definition
\ref{Partition}, depending on both the discretization in the surplus space and
the discretization in the intensity space, whose value function approximates
the value function $V$ uniformly. The proofs of the results of this
section are deferred to Appendix \ref{Appendix Disc Intensity}.

\bigskip For any parameter $\Delta>0$, let us introduce the following
discretization on the intensity space,
\begin{equation}
\mathcal{H}_{\Delta}=\left\{  \lambda_{0}^{\Delta}=\underline{\lambda}%
,\lambda_{1}^{\Delta},\lambda_{2}^{\Delta},\lambda_{3}^{\Delta},...\right\}
\subset\lbrack\underline{\lambda},\infty). \label{Definicion Grilla Lambda}%
\end{equation}
where $\lambda_{m}^{\Delta}=\underline{\lambda}+m\Delta$, and consider the function%

\begin{equation}
\sigma^{\Delta}\left(  \lambda\right)  =\min\left\{  \lambda_{m}^{\Delta}%
\in\mathcal{H}_{\Delta}:\lambda_{m}^{\Delta}\geq\lambda\right\}
\in\mathcal{H}_{\Delta}. \label{Definicion Lambda en grilla}%
\end{equation}

\begin{definition}
\label{delta-delta partition}Given $\delta>0$ and $\Delta>0,$ we say that
$\mathcal{P}=\left(  \mathcal{A},\mathcal{NA},\mathcal{B}\right)  $ is a
$(\delta,\Delta)$-partition if the three subsets $\mathcal{A}$, $\mathcal{NA}$
and $\mathcal{B}$ satisfy the following condition: if $(x_{n}^{\delta}%
,\lambda_{m}^{\Delta})$ is in a subset for $m\geq1$, then $\left\{
x_{n}^{\delta}\right\}  \times(\lambda_{m-1}^{\Delta},\lambda_{m}^{\Delta}]$
is also in this subset. So, the $(\delta,\Delta)$-partitions only depend on
the discrete grid $\mathcal{G}_{\delta}\times\mathcal{H}_{\Delta}$.
\end{definition}

In order to find a $(\delta,\Delta)$-partition whose associated value function
approximates the optimal value function, we modify the intensity process, 
defining a new intensity process in the grid domain $\mathcal{G}%
_{\delta}\times\mathcal{H}_{\Delta}$.

Given the process $\lambda_{t}$ from (\ref{Lambda t corrida}) with
initial value $\lambda_{0}=\lambda$, let us define the auxiliary intensity process
$\widehat{\lambda}_{t}$ for given parameters $\delta$, $\Delta$ and initial
$\widehat{\lambda}_{0}=\sigma^{\Delta}\left(  \lambda\right)  =\lambda
_{m}^{\Delta}\in\mathcal{H}_{\Delta}$ for some $m\geq0$ as%
\begin{equation}
\widehat{\lambda}_{t}:=\sigma^{\Delta}\left(  \lambda_{t}\right)
\in\mathcal{H}_{\Delta}. \label{Proceso Lambda Discreto Por arriba}%
\end{equation}
By definition, $\Delta\geq\widehat{\lambda}_{t}-\lambda_{t}\geq0$. We define
$\widehat{\Pi}_{x_{n}^{\delta},\lambda_{m}^{\Delta}}^{\delta,\Delta}$ as the
set of all the admissible strategies with initial surplus $x_{n}^{\delta}%
\in\mathcal{G}_{\delta}$, initial intensity $\widehat{\lambda}_{0}=\lambda
_{m}^{\Delta}\in\mathcal{H}_{\Delta}$ and discrete intensity process
$\widehat{\lambda}_{t}$ which can be obtained by a sequence of
control actions in $\mathcal{E}$ as in (\ref{Conjunto E de Estrategias.}).

We can replicate the construction of $V^{\delta}$ in the previous section, but
considering the compound Cox filtered space of the auxiliary intensity
process$\ \widehat{\lambda}_{t}\in\mathcal{H}_{\Delta}$ instead of the process
$\lambda_{t}.$ Consider for any $(x_{n}^{\delta},\lambda_{m}^{\Delta}%
)\in\mathcal{G}_{\delta}\times\mathcal{H}_{\Delta},$
\begin{equation}
\widehat{V}^{\delta,\Delta}(x_{n}^{\delta},\lambda_{m}^{\Delta})=\sup_{\pi
\in\widehat{\Pi}_{x_{n}^{\delta},\lambda_{m}}^{\delta,\Delta}}J_{\widehat
{\lambda}}(\pi;x_{n}^{\delta},\lambda_{m}^{\Delta}),
\label{optimal value function auxiliar}%
\end{equation}
where%
\begin{equation}
J_{\widehat{\lambda}}(\pi;x_{n}^{\delta},\lambda_{m}^{\Delta})=\mathbb{E}%
(\int_{0^{-}}^{\tau L}e^{-qt}dL_{t}+I_{\tau<\tau_{F}}e^{-q\tau_{F}}\widehat
{X}_{\tau_{F}}^{L}) \label{Value function auxiliar}%
\end{equation}
and%
\begin{equation}
\widehat{X}_{t}^{L}=\widehat{X}_{t}-L_{t} \label{controlled surplus auxiliar}%
\end{equation}
with
\begin{equation}
\widehat{X}_{t}=x_{n}^{\delta}+pt-\sum_{j=1}^{\widehat{N}_{t}}U_{j},
\label{Uncontrolled Surplus Intensity Discrete}%
\end{equation}
and $\widehat{N}_{t}$ has intensity $\widehat{\lambda}_{t}.$

One can extend the definition of $\widehat{V}^{\delta,\Delta}\ $to
$[0,\infty)\times\lbrack\underline{\lambda},\infty)$ as%

\[
\widehat{V}^{\delta,\Delta}(x,\lambda)=(x-\rho^{\delta}(x))+\widehat
{V}^{\delta,\Delta}(\rho^{\delta}(x),\sigma^{\Delta}\left(  \lambda\right)
)
\]
for all $x\geq0$ and $\lambda\geq\underline{\lambda}$, where $\rho^{\delta}$
is defined in (\ref{Definicion de x en la grilla por abajo}).\\

Let us show now that $\widehat{V}^{\delta,\Delta}$ converges uniformly to
$V^{\delta}$ as $\Delta\rightarrow0$. The proof is in Appendix \ref{Appendix Disc Intensity}.

\begin{proposition}
\label{cercania funciones especiales} It holds that $\widehat{V}%
^{\delta,\Delta}\leq V^{\delta}$ and $\lim_{\Delta\rightarrow0}\left(
\sup_{x\geq0,\lambda\geq\underline{\lambda}}V^{\delta}(x,\lambda)-\widehat
{V}^{\delta,\Delta}(x,\lambda)\right)  =0.$
\end{proposition}

As in Section \ref{Section Approximation Disc Surplus}, but considering
$\widehat{\lambda}_{t}$ instead of $\lambda_{t}$, we introduce the following
operators in $\mathcal{W}^{\delta,\Delta}=\left\{  w:\mathcal{G}_{\delta
}\times\mathcal{H}_{\Delta}\rightarrow\lbrack0,\infty)\right\}  $:
$\mathcal{T}_{1}\ $and $\mathcal{T}_{F}$ as defined in (\ref{Definicion Ti TS}%
); the operator
\[
\widehat{\mathcal{T}}_{0}(w)(x_{n}^{\delta},\lambda_{m}^{\Delta}%
):=\mathbb{P}(\delta\wedge T\wedge\tau=\delta)e^{-q\delta}w(x_{n+1}^{\delta
},\sigma^{\Delta}(\lambda_{\delta}^{c}))+\widehat{\mathcal{I}}(w)(x_{n}%
^{\delta},\lambda_{m}^{\Delta}),
\]
where%

\[%
\begin{array}
[c]{l}%
\widehat{\mathcal{I}}(w)(x_{n}^{\delta},\lambda_{m}^{\Delta})\\%
\begin{array}
[c]{l}%
:=\mathbb{E}(I_{\delta\wedge T\wedge\tau=\tau}I_{x+p\tau-U\geq0}e^{-q\tau
}w(\rho^{\delta}(x_{n}^{\delta}+p\tau-U),\widehat{\lambda}_{\tau}))\\[0.1cm]
+\mathbb{E}(I_{\delta\wedge T\wedge\tau=T}~e^{-qT}w(x_{n}^{\delta}%
,\widehat{\lambda}_{T}))\\[0.1cm]
+\mathbb{E}(I_{\delta\wedge T\wedge\tau=\tau}e^{-q\tau}(x_{n}^{\delta}%
+p\tau-U-\rho^{\delta}(x_{n}^{\delta}+p\tau-U)))\\[0.1cm]
+\mathbb{E}(I_{\delta\wedge T\wedge\tau=T}e^{-qT}pT);
\end{array}
\end{array}
\]
and the operator%
\[
\widehat{\mathcal{T}}:=\max\{\widehat{\mathcal{T}}_{0},\mathcal{T}%
_{1},\mathcal{T}_{F}\}.
\]

In the next two propositions, we obtain the mirror results of Propositions \ref{Limite de Vl} and
\ref{Proposicion Estrategia Optima Discreta}, with similar proofs.

\begin{proposition}
\label{Limite de VL doble discreto} It holds that $\widehat{\mathcal{T}%
}(\widehat{V}^{\delta,\Delta})=$ $\widehat{V}^{\delta,\Delta}\ $in
$\mathcal{G}_{\delta}\times\mathcal{H}_{\Delta}$.
\end{proposition}

Since $\widehat{V}^{\delta,\Delta}=\mathcal{T}(\widehat{V}^{\delta,\Delta})$,
we can define, as in (\ref{Optimal Partition delta}), the following partition
$\widehat{\mathcal{P}}^{\delta,\Delta}$ in the grid $\mathcal{G}_{\delta
}\times\mathcal{H}_{\Delta}$:%

\begin{equation}%
\begin{array}
[c]{lll}%
\widehat{\mathcal{A}}_{\delta,\Delta} & = & \left\{  \left(  x_{n}^{\delta
},\lambda_{m}^{\Delta}\right)  \in\mathcal{G}_{\delta}\times\mathcal{H}%
_{\Delta}:\mathcal{T}_{1}(\widehat{V}^{\delta,\Delta})(x_{n}^{\delta}%
,\lambda_{m}^{\Delta})=\widehat{V}^{\delta,\Delta}(x_{n}^{\delta},\lambda
_{m}^{\Delta})\right\}  ,\\
\widehat{\mathcal{NA}}_{\delta,\Delta} & = & \left\{  \left(  x_{n}^{\delta
},\lambda_{m}^{\Delta}\right)  \in\left(  \mathcal{G}_{\delta}\times
\mathcal{H}_{\Delta}\right)  -\widehat{\mathcal{A}}_{\delta,\Delta}%
:\widehat{\mathcal{T}}_{0}(\widehat{V}^{\delta,\Delta})(x_{n}^{\delta}%
,\lambda_{m}^{\Delta})=\widehat{V}^{\delta,\Delta}(x_{n}^{\delta},\lambda
_{m}^{\Delta})\right\}  ,\\
\widehat{\mathcal{B}}_{\delta,\Delta} & = & \left\{  \left(  x_{n}^{\delta
},\lambda_{m}^{\Delta}\right)  \in\left(  \mathcal{G}_{\delta}\times
\mathcal{H}_{\Delta}\right)  -\left(  \widehat{\mathcal{A}}_{\delta,\Delta
}\cup\widehat{\mathcal{NA}}_{\delta,\Delta}\right)  :\mathcal{T}_{F}%
(\widehat{V}^{\delta,\Delta})(x_{n}^{\delta},\lambda_{m}^{\Delta})=\widehat
{V}^{\delta,\Delta}(x_{n}^{\delta},\lambda_{m}^{\Delta})\right\}  .
\end{array}
\label{Partition doble delta}%
\end{equation}

Moreover, $\widehat{\mathcal{B}}_{\delta,\Delta}=\varnothing$ and $\left(
x_{0}^{\delta},\lambda_{m}^{\Delta}\right)  \in\widehat{\mathcal{NA}}%
_{\delta,\Delta}$. And, as in Definition \ref{Partition}, we can define from
this partition, a family of strategies
\[
\widehat{\mathbf{\pi}}^{\delta,\Delta}=\left\{  \widehat{\pi}_{x_{n}^{\delta
},\lambda_{m}^{\Delta}}^{\delta,\Delta}\in\widehat{\Pi}_{x_{n}^{\delta
},\lambda_{m}}^{\delta,\Delta}\text{ for }(x_{n}^{\delta},\lambda_{m}^{\Delta
})\in\mathcal{G}_{\delta}\times\mathcal{H}_{\Delta}\right\}  .
\]
These strategies are stationary in $\mathcal{G}_{\delta}\times\mathcal{H}%
_{\Delta}$ in the sense that the local control actions depend only on the
point of the grid at which the current surplus and current intensity lie. We
obtain the existence of the optimal strategy in $\widehat{\Pi}_{x_{n}^{\delta
},\lambda_{m}}^{\delta,\Delta}$.

\begin{proposition}
\label{Proposicion Estrategia Optima Doble Discreta} $\widehat{V}%
^{\delta,\Delta}(x_{n}^{\delta},\lambda_{m}^{\Delta})=J_{\widehat{\lambda}%
}(\widehat{\pi}_{x_{n}^{\delta},\lambda_{m}^{\Delta}}^{\delta,\Delta}%
;x_{n}^{\delta},\lambda_{m}^{\Delta})$ for any $(x_{n}^{\delta},\lambda
_{m}^{\Delta})\in\mathcal{G}_{\delta}\times\mathcal{H}_{\Delta}.$
\end{proposition}

Now we use the partition given in (\ref{Optimal Partition delta}), to find a
$(\delta,\Delta)$-partition in $\mathcal{G}_{\delta}\times\lbrack
\underline{\lambda},\infty)$ as defined in Definition
\ref{delta-delta partition} whose value function approximates the value value
function $V$ uniformly. This is the main result of the section.

\begin{definition}
\label{Extension Particion discreta a no}Given the partition
(\ref{Partition doble delta}) in $\mathcal{G}_{\delta}\times\mathcal{H}%
_{\Delta},$ we extend to the partition $\mathcal{P}^{\delta,\Delta
}=(\mathcal{A}_{\delta,\Delta},\left(  \mathcal{NA}\right)  _{\delta,\Delta
},\mathcal{B}_{\delta,\Delta})$ in $\mathcal{G}_{\delta}\times\lbrack
\underline{\lambda},\infty)$ as
\[%
\begin{array}
[c]{lll}%
\mathcal{A}_{\delta,\Delta} & = & \left(
{\textstyle\bigcup\nolimits_{\left\{  (n,m):\left(  x_{n}^{\delta},\lambda
_{m}^{\Delta}\right)  \in\widehat{\mathcal{A}}_{\delta,\Delta},\text{ }%
m\geq1\right\}  }}
\left(  \left\{  x_{n}^{\delta}\right\}  \times(\lambda_{m-1}^{\Delta}%
,\lambda_{m}^{\Delta}]\right)  \right)  \cup\left(
{\textstyle\bigcup\nolimits_{\left\{  n:\text{ }\left(  x_{n}^{\delta
},\underline{\lambda}\right)  \in\widehat{\mathcal{A}}_{\delta,\Delta
}\right\}  }}
\left\{  \left(  x_{n}^{\delta},\underline{\lambda}\right)  \right\}  \right)
\\
\left(  \mathcal{NA}\right)  _{\delta,\Delta} & = & \left(
{\textstyle\bigcup\nolimits_{\left\{  (n,m):\left(  x_{n}^{\delta},\lambda
_{m}^{\Delta}\right)  \in\widehat{\mathcal{NA}}_{\delta,\Delta},\text{ }%
m\geq1\right\}  }}
\left(  \left\{  x_{n}^{\delta}\right\}  \times(\lambda_{m-1}^{\Delta}%
,\lambda_{m}^{\Delta}]\right)  \right)  \cup\left(
{\textstyle\bigcup\nolimits_{\left\{  n:\text{ }\left(  x_{n}^{\delta
},\underline{\lambda}\right)  \in\widehat{\mathcal{NA}}_{\delta,\Delta
}\right\}  }}
\left\{  \left(  x_{n}^{\delta},\underline{\lambda}\right)  \right\}  \right)
\\
\mathcal{B}_{\delta,\Delta} & = & \varnothing.
\end{array}
\]
\end{definition}

Let us consider the $\mathcal{G}_{\delta}$-\textit{\ }strategy $\pi
_{x_{n}^{\delta},\lambda}^{\mathcal{P}^{\delta,\Delta}}\in\widetilde{\Pi
}_{x_{n}^{\delta},\lambda}^{\delta}$ associated to $\mathcal{P}^{\delta
,\Delta}$ for each $(x_{n}^{\delta},\lambda)\in\mathcal{G}_{\delta}%
\times\lbrack\underline{\lambda},\infty)$ and the associated function
$W_{\mathcal{P}^{\delta,\Delta}}:[0,\infty)\times\lbrack\underline{\lambda
},\infty)\rightarrow\lbrack0,\infty)$ defined in (\ref{Extension WP}).\\

The next theorem states that $V$ can be approximated uniformly by
$W_{\mathcal{P}^{\delta,\Delta}}$ for some $\delta$ and $\Delta$ small enough.
In Section \ref{Numerical Results}, we use this result in order to obtain
numerically $(\delta,\Delta)$-partitions whose associated value functions
approximate uniformly the optimal value function $V$.

\begin{theorem}
\label{Teorema Aproximacion Uniforme}For any $\varepsilon>0$ there exists
$\delta$ and $\Delta$ small enough so that $0\leq V-W_{\mathcal{P}^{\delta,\Delta}%
}\leq V-\widehat{V}^{\delta,\Delta}\leq$ $\varepsilon$ in $[0,\infty
)\times\lbrack\underline{\lambda},\infty).$
\end{theorem}

\section{Numerical results \label{Numerical Results}}

In this section, we show some numerical results. Proposition
\ref{Convergencia Uniforme con Lambda} suggests that we can approximate the
optimal value function with a value function of a partition with
$(x_{n}^{\delta},\lambda_{m}^{\Delta})\in\mathcal{B}_{\delta,\Delta}$ for
$m\leq m_{1}$ with $m_1$ large enough and $n\geq0$. From Proposition
\ref{Proposicion acotacion de retirar}, we can assume that the partition
satisfies $(x_{n}^{\delta},\lambda_{m}^{\Delta})\in\left(  \mathcal{NA}%
\right)  _{\delta,\Delta}$ for $x_{n}^{\delta}>p/q\ $. This allows to make a
numerical computation because we stay in a finite grid.

So, given $\delta$, $\Delta~$small enough and $m_{1}$ large enough, we proceed
in the following way: The numerical scheme should choose if the local action
in each of the points in the grid
\[
\overline{\mathcal{G}}_{\delta}\times\overline{\mathcal{H}}_{\Delta}=\left\{
(x_{n}^{\delta},\lambda_{m}^{\Delta}):x_{n}^{\delta}\leq p/q\ \text{and }m\leq
m_{1}\right\}
\]
is either $\mathbf{E}_{0}$ or $\mathbf{E}_{1}$ or $\mathbf{E}_{F}$. In order to do that, we proceed
inductively, as follows:

\begin{enumerate}
\item We define the initial partition $\widehat{\mathcal{P}}_{1}$ as
$\mathcal{B}_{1}=\overline{\mathcal{G}}_{\delta}\times\overline{\mathcal{H}%
}_{\Delta},$ $\widehat{\mathcal{NA}}_{1}=\widehat{\mathcal{A}}_{1}%
=\varnothing$ for the sake of simplicity. So, the value function is $\widehat
{W}_{\widehat{\mathcal{P}}_{1}}(x_{n}^{\delta},\lambda_{m}^{\Delta}%
)=x_{n}^{\delta}.$

\item If the partition $\widehat{\mathcal{P}}_{l}$ and the associated value
function $\widehat{W}_{\widehat{\mathcal{P}}_{l}}$ are given, we define the
next partition $\widehat{\mathcal{P}}_{l+1}$ as
\[%
\begin{array}
[c]{lll}%
\widehat{\mathcal{A}}_{l+1} & = & \{(x_{n}^{\delta},\lambda_{m}^{\Delta}%
)\in\overline{\mathcal{G}}_{\delta}\times\overline{\mathcal{H}}_{\Delta
}:\widehat{\mathcal{T}}(\widehat{W}_{\widehat{\mathcal{P}}_{l}})(x_{n}%
^{\delta},\lambda_{m}^{\Delta})=\widehat{\mathcal{T}}_{1}(\widehat
{W}_{\widehat{\mathcal{P}}_{l}})(x_{n}^{\delta},\lambda_{m}^{\Delta})\},\\[0.1cm]
\widehat{\mathcal{NA}}_{l+1} & = & \{(x_{n}^{\delta},\lambda_{m}^{\Delta}%
)\in\overline{\mathcal{G}}_{\delta}\times\overline{\mathcal{H}}_{\Delta
}-\widehat{\mathcal{A}}_{l+1}:\widehat{\mathcal{T}}(\widehat{W}_{\widehat
{\mathcal{P}}_{l}})(x_{n}^{\delta},\lambda_{m}^{\Delta})=\widehat{\mathcal{T}%
}_{0}(\widehat{W}_{\widehat{\mathcal{P}}_{l}})(x_{n}^{\delta},\lambda
_{m}^{\Delta})\},
\end{array}
\]
and we put $\widehat{\mathcal{B}}_{l+1}=\varnothing$, because $\mathbf{E}_{F} $
is never optimal.
\end{enumerate}

We have, as in Propositions \ref{Properties of Vl} and \ref{Limite de Vl},
that $\widehat{W}_{\widehat{\mathcal{P}}_{l+1}}%
=\widehat{\mathcal{T}}(\widehat{W}_{\widehat{\mathcal{P}}_{l}})\geq\widehat
{W}_{\widehat{\mathcal{P}}_{l}}$ in $\overline{\mathcal{G}}_{\delta}%
\times\overline{\mathcal{H}}_{\Delta}$; that $\widehat{\mathcal{P}}%
_{l+1}=\widehat{\mathcal{P}}_{l}$ $=\widehat{\mathcal{P}}\ $for $l$ large
enough because here the grid $\overline{\mathcal{G}}_{\delta}\times
\overline{\mathcal{H}}_{\Delta}$ is finite; that $\lim_{l\rightarrow\infty
}\widehat{W}_{\widehat{\mathcal{P}}_{l}}$ $=\widehat{W}_{\widehat{\mathcal{P}%
}}$ and that $\widehat{\mathcal{T}}(\widehat{W}_{\widehat{\mathcal{P}}%
})=\widehat{W}_{\widehat{\mathcal{P}}}\ $ in $\overline{\mathcal{G}}_{\delta
}\times\overline{\mathcal{H}}_{\Delta}$. In our numerical scheme, we also
check that the limit action region of $\widehat{\mathcal{P}}$ does not change
when $m_{1}$ is enlarged.

By Definition \ref{Extension Particion discreta a no} and Theorem
\ref{Teorema Aproximacion Uniforme}, we know that when $\delta$ and $\Delta$
are small enough we obtain a near-optimal value function and $\mathcal{G}%
_{\delta}$-\textit{\ }strategy. We check in our numerical scheme that
$\delta~$and $\Delta$ are small enough by comparing the value functions
$\widehat{W}_{\widehat{\mathcal{P}}}$ for $(\delta,\Delta)$ with the one of
$(\delta/2,\Delta/2)$. In the following examples, we show in gray the
non-action region and in black the change region of the partition
$\mathcal{P}$ in $\overline{\mathcal{G}}_{\delta}\times\lbrack\underline
{\lambda},\lambda_{m_{1}}^{\Delta}]$ associated to the partition
$\widehat{\mathcal{P}}$ in $\overline{\mathcal{G}}_{\delta}\times
\overline{\mathcal{H}}_{\Delta}$ as in Definition
\ref{Extension Particion discreta a no}. We also show the approximation of the
optimal value function $V$ (using the value function of the partition
$\widehat{\mathcal{P}}$ in $\overline{\mathcal{G}}_{\delta}\times
\overline{\mathcal{H}}_{\Delta})$ and finally, we compare the approximation of
our optimal value function with one of the classical dividend
optimization problem with constant intensity, i.e.\ the situation without jumps in the claim arrival intensity. \\


\subsection{Example 1: Exponential claim sizes}
Let us first consider $\underline{\lambda}=1/4$, the intensity jump distribution being exponential with cdf $F_{Y}(x)=1-e^{-x/2}$, the intensity of jump intensity arrivals being $\beta=1/2$ (so we expect a catastrophe every two years), $d=7/10$ (so that the additional claim arrival intensity due to a catastrophe is halved after one year), the claim size distribution being exponential with cdf $F_{U}(x)=1-e^{-10x}$, the discount rate being $~q=2/10$ and the insurance safety loading applied in the policies equal to $\eta=2/10$. We then obtain from (\ref{Definicion p}) that $p=141/700.$ For the grid parameters, it turns out that $\delta=28/423$, $\Delta=23/240$ and $m_{1}=60$ is appropriate here. \\

\begin{figure}[ptb]
	\begin{subfigure}{.45\textwidth}
		\centering
		\includegraphics[width=5.5cm]{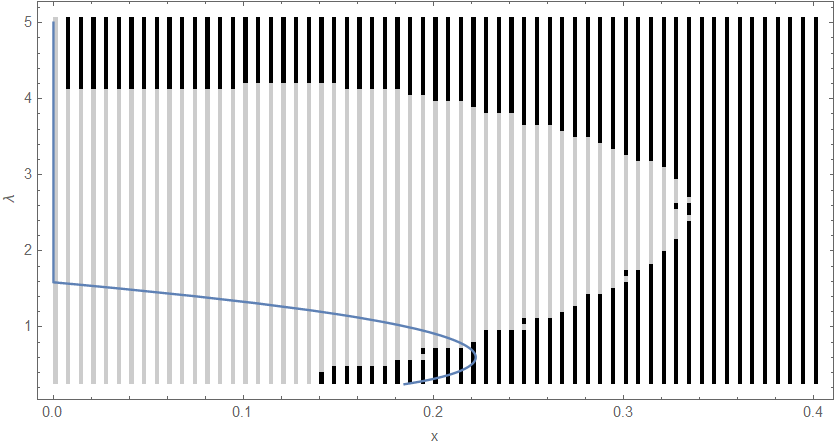}
		\caption{Action regions (dark) and non-action regions (light) of the optimal strategy as a function of $\lambda$ and $x$}
		\label{Ex1_Str}
	\end{subfigure} \begin{subfigure}{.45\textwidth}
		\centering
		\includegraphics[width=5.5cm]{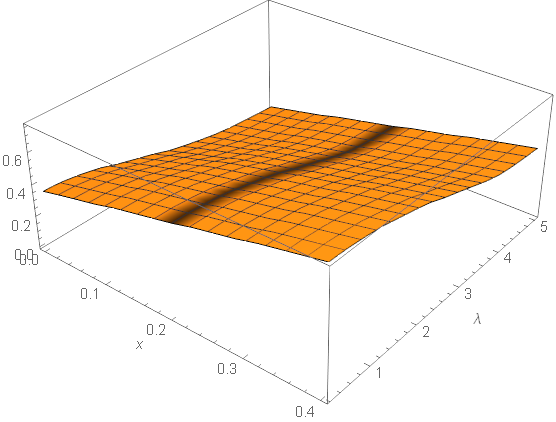}
		\caption{Value function $V(x,\lambda)$}
		\label{Ex1_Val}
	\end{subfigure}\\[0.2cm]
		\begin{subfigure}{.45\textwidth}
		\centering
		\includegraphics[width=5.5cm]{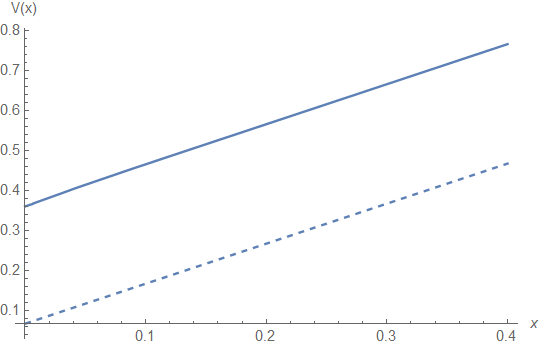}
		\caption{Comparison of the value functions $V(x,\underline{\lambda})$\\  (solid) and $V_{\text{CL}}(x,\underline{\lambda})$ (dashed)}
		\label{Ex1_Comp}
	\end{subfigure}
	\begin{subfigure}{.45\textwidth}
		\centering
		\includegraphics[width=5.5cm]{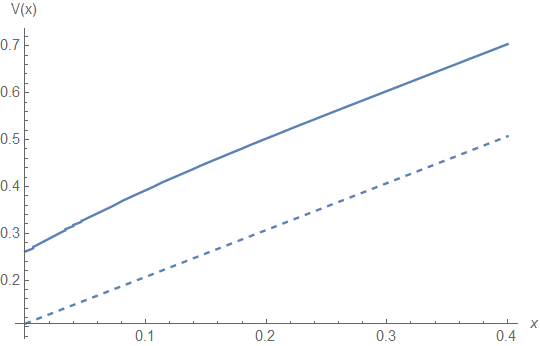}
		\caption{Comparison of the value functions $V(x,\lambda_{\text{av}})$ \\(solid) and $V_{\text{CL}}(x,\lambda_{\text{av}})$ (dashed)}
		\label{Ex1_CompAv}
	\end{subfigure}
	\caption{\protect\small Optimal dividends for a compound shot-noise Cox process with exponential claim sizes}%
	\label{fig1}%
\end{figure}

Figure \ref{Ex1_Str} depicts the action and non-action region and Figure \ref{Ex1_Val} the
approximation of the optimal value function $V$ as a function of initial surplus $x$ and claim initial intensity level $\lambda$. In the absence of catastrophe jumps in the intensity process (that is, for the classical Cram\'er-Lundberg process with a compound Poisson claim process), it is well-known that for exponential claim sizes, a barrier strategy is optimal (cf.\ Gerber \cite{Ger69}). With the additional presence of a shot-noise component in the intensity process, one observes from Figure \ref{Ex1_Str} that a barrier strategy is still optimal, but its value changes dynamically as a function of current claim intensity $\lambda$. Concretely, for small values of $\lambda$, a barrier strategy is optimal that pays all the surplus above the barrier as dividends, and this barrier first increases with $\lambda$ and eventually decreases for larger values of $\lambda$. Finally, there is a critical value of $\lambda$, above which the situation becomes too risky, and the barrier level goes down to zero, i.e. all surplus is paid out as dividends. For comparison, in Figure \ref{Ex1_Str} we also depict the optimal barrier level of the classical risk model for the same premium income $p$, but varying actual (constant in time) level $\lambda$ (blue solid line), and one sees that there the optimal barrier level already goes down to zero for a considerably smaller value of $\lambda$. One can nicely see from this comparison how the dynamic change of the level of $\lambda$ through time allows to be more adaptive in the strategy. \\
In Figure \ref{Ex1_Comp} we depict the resulting value function as a function of initial capital $x$, where we choose $\lambda=\underline{\lambda}=1/4$ as the initial level (solid line, which is in fact just the cross-section $V(x,0.4)$ from Figure \ref{Ex1_Val}) and compare it to the value function $V_{\text{CL}}(x,\underline{\lambda})$ for the classical risk model with homogeneous intensity $\underline{\lambda}$ throughout (dashed line), which corresponds to the case $\beta=0$ of the shot-noise case here. Note that the same relative safety loading $\eta$ is applied in the two cases, so the absolute value of $p$ does not coincide for the two models (i.e., $V_{\text{CL}}(x,\underline{\lambda})$ does not correspond to $v^{\lambda}(x)$, for which that was the case). In other words, Figure \ref{Ex1_Comp} compares the economic value of the insurance company for the shareholders with and without additional catastrophe insurance business, and the plot shows that including that business line is an advantage. This can be interpreted as follows: the additional insurance business linked to catastrophe claims (above the baseline intensity $\underline{\lambda}$) leads to additional premium income, which is upfront (as potential claims will only appear later when the intensity indeed jumped up), so that one has additional degrees of freedom to steer the dividend streams according to the current level of intensity and surplus. In fact, the additional variability in the intensity process stays advantageous even when starting at higher levels of the initial intensity (and the comparison then is  whether a volatile claim intensity can be preferable to a constant one, for the same number of policies). To see this, Figure \ref{Ex1_CompAv} compares the two value functions for the long-term average value $\lambda_{\text{av}}=1/4+2/(2\cdot 0.7)=1.679$ as the initial intensity level (solid line) or as homogeneous intensity level throughout (dashed line), respectively (again with safety loading $\eta=0.2$ for both, which now also leads to the same $p$, so here $V_{\text{CL}}(x,\lambdaav)=v^{\lambdaav}(x)$). Clearly, the additional variability of the claim intensity process (accompanied by the optimal associated dividend strategy) leads still to a significantly larger value function. \\

\begin{figure}[tbh]
	\begin{center}
		\includegraphics[width=6cm]{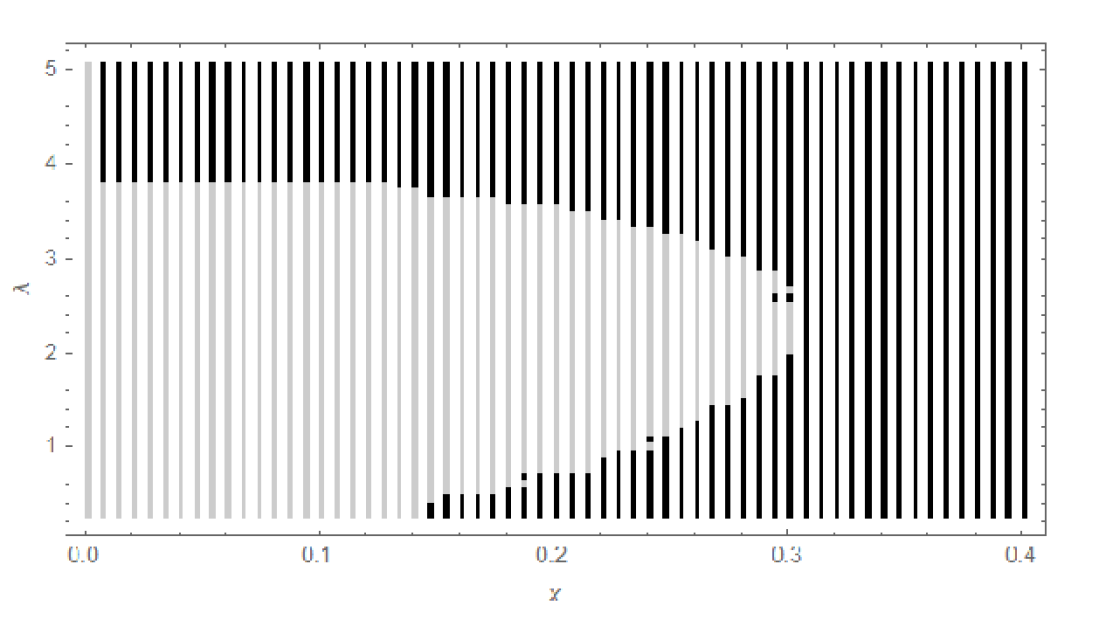}
	\end{center}
	\caption{{\protect\small Counterpart of Figure \ref{Ex1_Str} for deterministic intensity jumps of size $2$}}%
	\label{fig1b}%
\end{figure}

If one replaces the exponential intensity jump distribution by a deterministic jump of the same expected value (i.e.\ $F_{Y}(x)=I_{[2,\infty)}$), the modified  action/non-action regions are shown in Figure \ref{fig1b}, which shows that the results are not really sensitive to the choice of the concrete intensity jump distribution (as long as the expected jump size is maintained). The resulting value function is even visually indistinguishable from Figure \ref{Ex1_Val}, so that we do not include it here. \\

\begin{figure}[ptb]
	\begin{subfigure}{.45\textwidth}
		\centering
		\includegraphics[width=5.5cm]{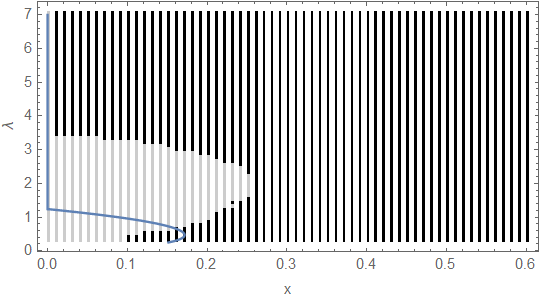}
		\caption{$\beta=0.4$}
		\label{sens_betadown}
	\end{subfigure} \begin{subfigure}{.45\textwidth}
		\centering
		\includegraphics[width=5.5cm]{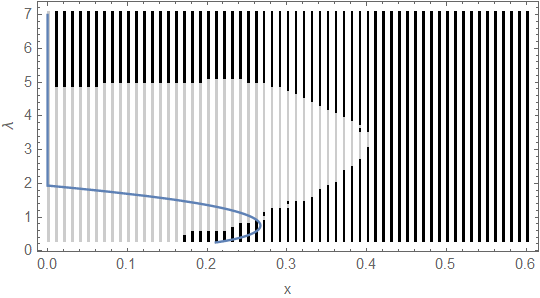}
		\caption{$\beta=0.6$}
		\label{sens_betaup}
	\end{subfigure}\\[0.2cm]
	\begin{subfigure}{.45\textwidth}
		\centering
		\includegraphics[width=5.5cm]{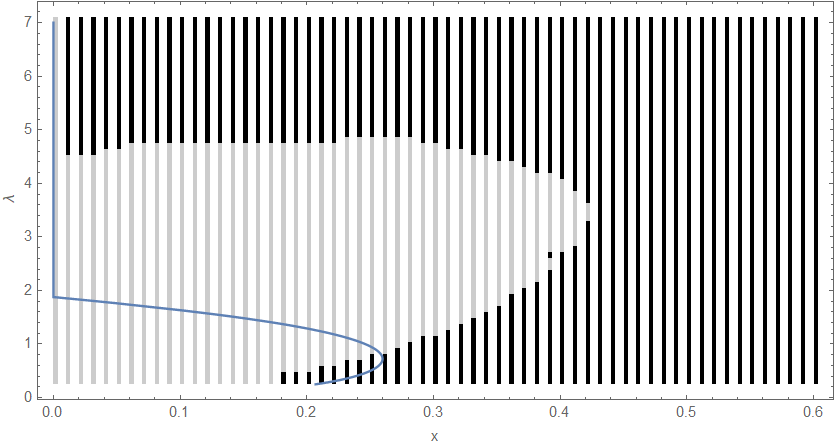}
		\caption{$d=0.6$}
		\label{sens_ddown}
	\end{subfigure}
	\begin{subfigure}{.45\textwidth}
		\centering
		\includegraphics[width=5.5cm]{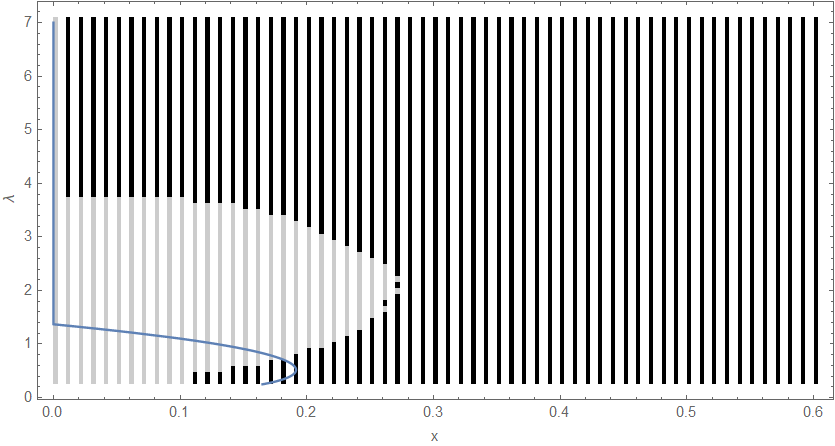}
		\caption{$d=0.8$}
		\label{sens_dup}
	\end{subfigure}\\[0.2cm]
	\begin{subfigure}{.45\textwidth}
		\centering
		\includegraphics[width=5.5cm]{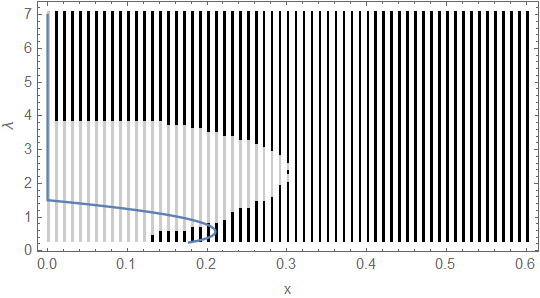}
		\caption{$\eta=0.15$}
		\label{sens_betadown}
	\end{subfigure} \begin{subfigure}{.45\textwidth}
		\centering
		\includegraphics[width=5.5cm]{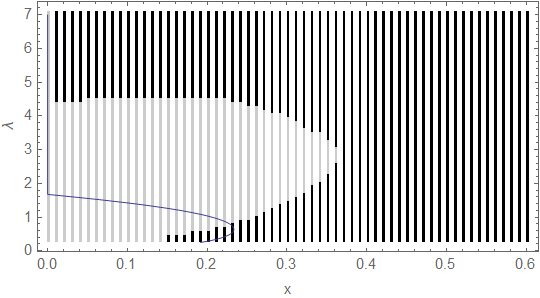}
		\caption{$\eta=0.25$}
		\label{sens_betaup}
	\end{subfigure}\\[0.2cm]
	\begin{subfigure}{.45\textwidth}
		\centering
		\includegraphics[width=5.5cm]{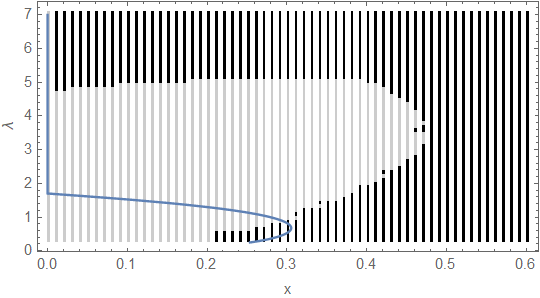}
		\caption{$q=0.15$}
		\label{sens_ddown}
	\end{subfigure}
	\begin{subfigure}{.45\textwidth}
		\centering
		\includegraphics[width=5.5cm]{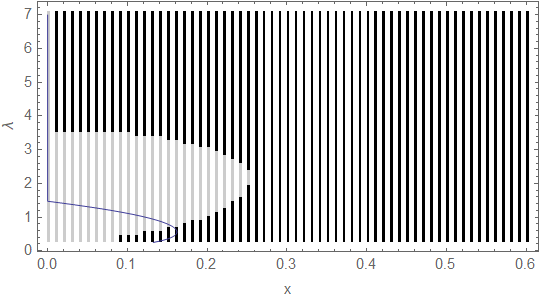}
		\caption{$q=0.25$}
		\label{sens_dup}
	\end{subfigure}
	\caption{\protect\small  Counterpart of Figure \ref{Ex1_Str} for modified $\beta$ (first row), modified $d$ (second row), modified $\eta$ (third row) and modified $q$ (last row)}%
	\label{figsens}%
\end{figure}

In order to assess the sensitivity of the action region with respect to changes of other parameters, we depict in Figure \ref{figsens} the counterpart of Figure \ref{Ex1_Str} for slight changes of the parameters in either direction. We also plot the optimal barrier of the classical Cramér-Lundberg model for the same premium level $p$ for each case. One can see that increasing or decreasing the intensity $\beta$ of catastrophe arrivals has a considerable effect on the optimal barrier as a function of $x$ and $\lambda$. A smaller value of the decay rate $d$ in the intensity function leads to a more dangerous situation, enlarging the barrier levels, where for increasing $d$ the intensity growth due to a catastrophe disappears quicker and we get closer to the classical barrier level (in blue), which would be reached for $d\to\infty$. A higher safety loading $\eta$ leads to larger barrier levels (one may interpret that this is due to the fact that one wants to benefit from the positive drift longer, exceeding the advantage of paying out profits early). Finally, an increased discount rate makes earlier payments more important and attractive, leading to lower barrier levels. 

\subsection{Example 2: Erlang(2) claim sizes}
Next, we consider a situation of a certain Erlang(2) claim size distribution, for which we know from \cite{AM 2014} that a two-band strategy maximizes the expected discounted dividend payments in the absence of shot-noise jumps in the Poisson intensity. Consider therefore the parameters choices of \cite{AM 2014} $\underline{\lambda}=10$, $F_{U}(x)=1-(1+x)e^{-x}$, $~q=1/10$ and $\eta=7/100$, to which we add now a catastrophe shot-noise component with exponential intensity jumps with cdf $F_{Y}(x)=1-e^{-x/2}$, $\beta=2/10$ and $d=2/10$. We obtain from (\ref{Definicion p}) that $p=642/25.$\\

\begin{figure}[ptb]
	\begin{subfigure}{.32\textwidth}
		\centering
		\includegraphics[width=5.5cm]{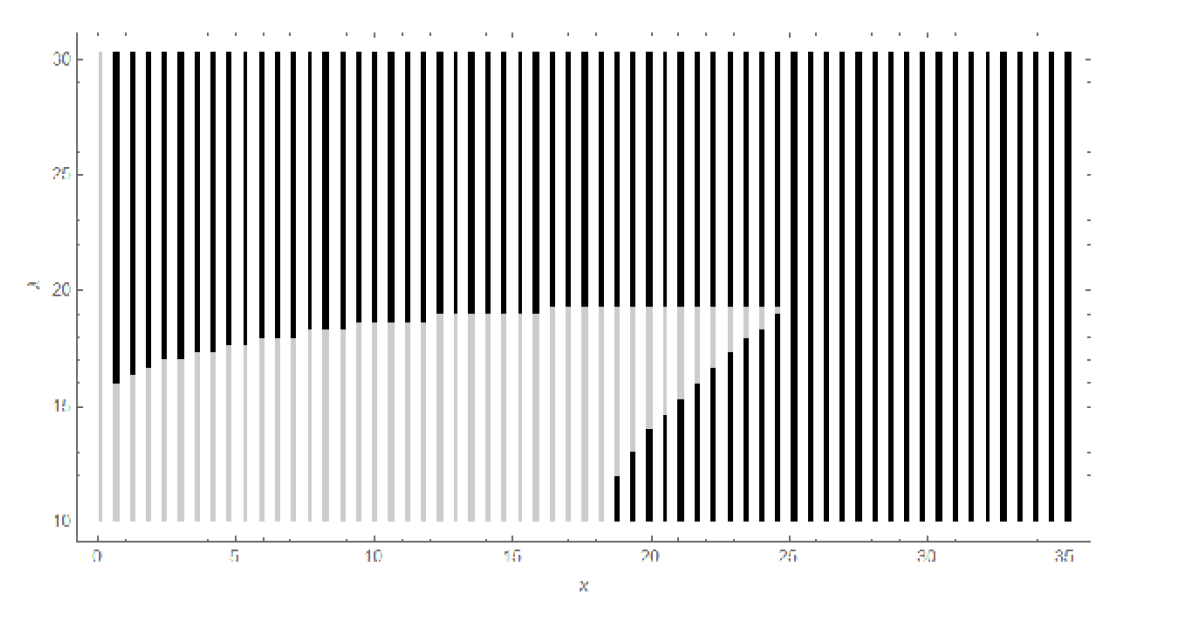}
		\caption{Action regions (dark) and non-action regions (light) of the optimal strategy as a function of $\lambda$ and $x$}
		\label{Ex2_Str}
	\end{subfigure} \begin{subfigure}{.32\textwidth}
		\centering
		\includegraphics[width=5.5cm]{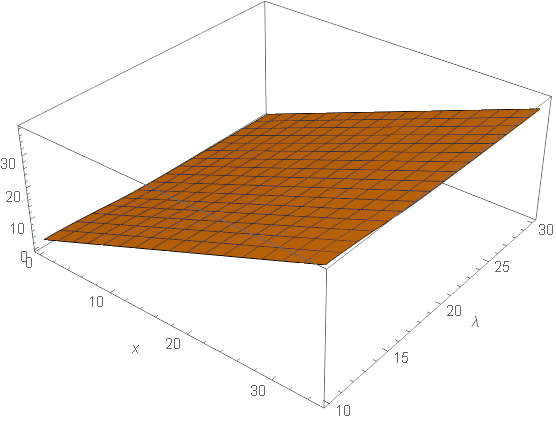}
		\caption{Value function $V(x,\lambda)$}
		\label{Ex2_Val}
	\end{subfigure}
	\begin{subfigure}{.32\textwidth}
		\centering
		\includegraphics[width=5.5cm]{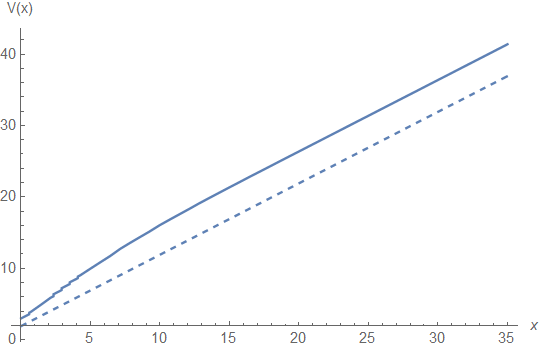}
		\caption{Comparison of the value functions $V(x,\lambda_{\text{av}})$ (solid) and $V_{\text{CL}}(x,\lambda_{\text{av}})$ (dashed)}
		\label{Ex2_Comp}
	\end{subfigure}
	\caption{\protect\small Optimal dividends for a compound shot-noise Cox process with Erlang(2) claim sizes}%
	\label{fig2}%
\end{figure}

For the grid parameters, the values $\delta=25/963$, $\Delta=1/2$ and $m_{1}=60$ turn out to be appropriate here. Figure \ref{Ex2_Str} depicts the action and non-action region. One observes that the optimality of the two-band regime is in fact retained here also, but only for a small strip of $\lambda$-values, and in that region the corresponding band values vary with $\lambda$ also. For values of $\lambda$ below that regime, the two-band strategy collapses to a barrier strategy (the fact that this band strategy is very sensitive to the particular model assumptions is well-known, see e.g.\ also \cite{ABT11} for such an effect under discrete surplus observations). For values of $\lambda$ above that regime, the optimal strategy is again of take-the-money-and-run type, i.e.\ pay all the surplus as dividends immediately (as the risk of facing many claims in the near future diminishing the surplus is too high). Figure \ref{Ex2_Val} depicts the resulting value function as a function of current levels of $x$ and $\lambda$. Finally, Figure \ref{Ex2_Comp} compares $V(x,\lambdaav)$ to the value function of the classical risk model with constant intensity $\lambdaav$ and the same premium income. Again, we observe that the additional variability introduced in the portfolio by having catastrophe insurance business (rather than only non-catastrophic one with constant claim intensity) is in fact an advantage for the expected discounted dividend payments until ruin. 

\subsection{Example 3: Deterministic claim sizes}
\begin{figure}[ptb]
	\begin{subfigure}{.32\textwidth}
		\centering
		\includegraphics[width=5.5cm]{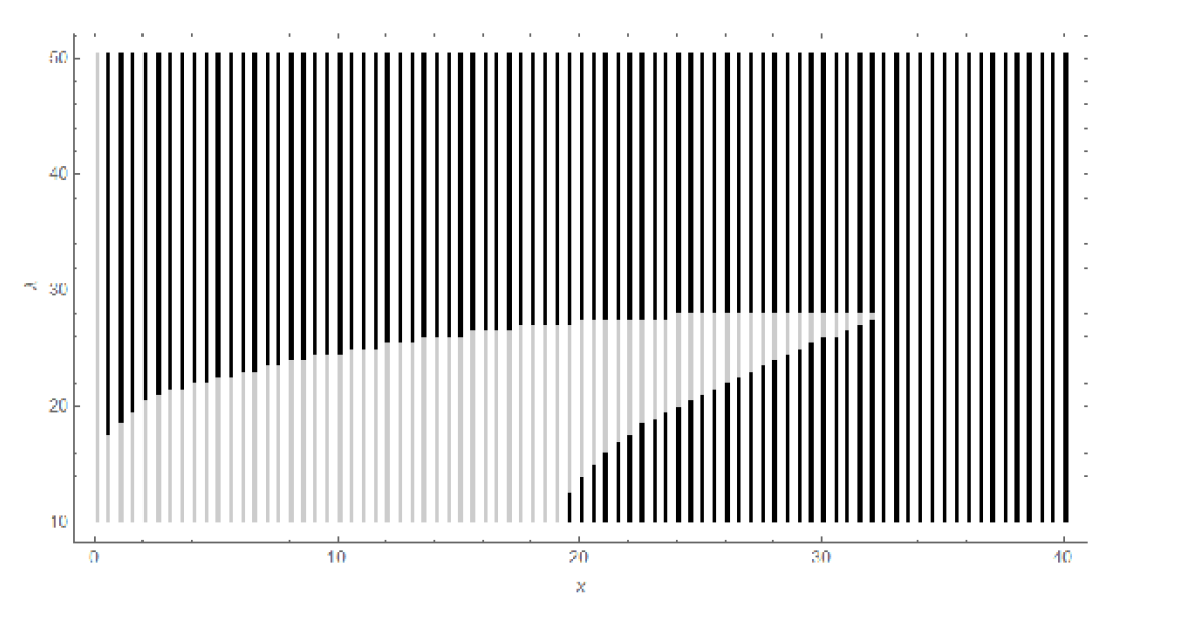}
		\caption{Action regions (dark) and non-action regions (light) of the optimal strategy as a function of $\lambda$ and $x$}
		\label{Ex3_Str}
	\end{subfigure} \begin{subfigure}{.32\textwidth}
		\centering
		\includegraphics[width=5.5cm]{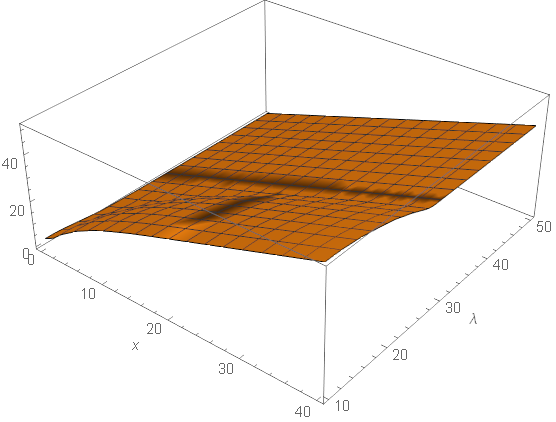}
		\caption{Value function $V(x,\lambda)$}
		\label{Ex3_Val}
	\end{subfigure}
		\begin{subfigure}{.32\textwidth}
		\centering
		\includegraphics[width=5.5cm]{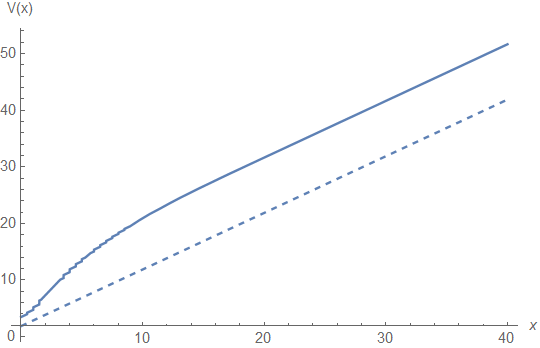}
		\caption{Comparison of the value functions $V(x,\lambda_{\text{av}})$ (solid) and $V_{\text{CL}}(x,\lambda_{\text{av}})$ (dashed)}
		\label{Ex3_Comp}
	\end{subfigure}
	\caption{\protect\small Optimal dividends for a compound shot-noise Cox process with deterministic claim sizes for the parameters of Example 2}%
	\label{fig3}%
	\end{figure}

In addition to a fine interplay of many factors, one intuitive reason for the optimality of band (rather than barrier) strategies can be attributed to the presence of modes in the claim size density (see e.g.\ \cite{berdel,AG23} where in the latter reference even a situation with an optimal strategy consisting of 4 bands was identified). Following that line of thinking, one might expect a 2-band strategy to remain optimal when replacing the Erlang claim size distribution by a deterministic claim size equal to its expected value. We therefore reconsider the situation of Example 2, solely changing the cdf of the claim size distribution to $F_{U}(x)=I_{[2,\infty)}$. We use the same grid parameters as above: $\delta=25/963$, $\Delta=1/2$ and $m_{1}=60.$ Figure \ref{Ex3_Str} depicts the action and non-action region. Indeed, the result is very similar to Figure \ref{Ex2_Str}, so for a certain range of $\lambda$-values, a two-band strategy is optimal. Yet, the lack of variability of the claim size is reflected in different numerical values of the $\lambda$-dependent bands. Figure \ref{Ex3_Val} depicts the resulting value function as a function of current levels of $x$ and $\lambda$, which again has a very similar shape, but different absolute values. As before, Figure \ref{Ex3_Comp} compares $V(x,\lambdaav)$ to the value function of the model without a catastrophe component.  \\

\begin{figure}[ptb]
	\begin{subfigure}{.32\textwidth}
		\centering
		\includegraphics[width=5.5cm]{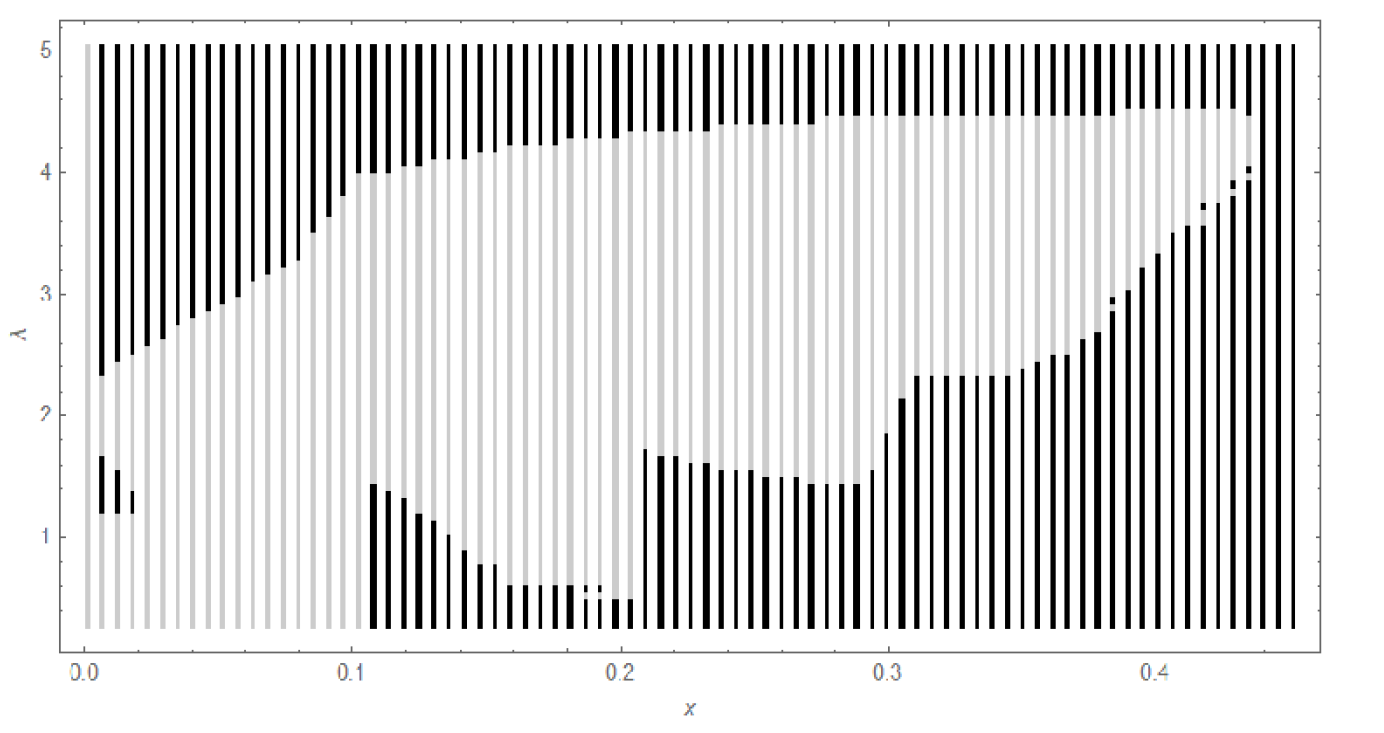}
		\caption{Action regions (dark) and non-action regions (light) of the optimal strategy as a function of $\lambda$ and $x$}
		\label{Ex3b_Str}
	\end{subfigure} \begin{subfigure}{.32\textwidth}
		\centering
		\includegraphics[width=5.5cm]{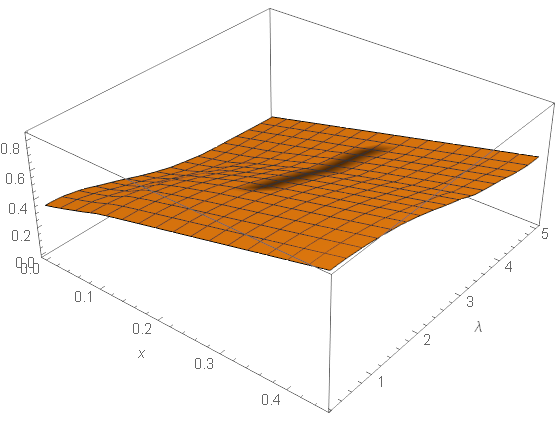}
		\caption{Value function $V(x,\lambda)$}
		\label{Ex3b_Val}
	\end{subfigure}
		\begin{subfigure}{.32\textwidth}
		\centering
		\includegraphics[width=5.5cm]{Figures/Ex3_Comp_Average.png}
		\caption{Comparison of the value functions $V(x,\lambda_{\text{av}})$ (solid) and $V_{\text{CL}}(x,\lambda_{\text{av}})$ (dashed)}
		\label{Ex3b_Comp}
	\end{subfigure}
	\caption{\protect\small Optimal dividends for a compound shot-noise Cox process with deterministic claim sizes for the parameters of Example 1}%
	\label{fig3b}%
\end{figure}

If instead we replace the exponential claim size distribution in Example 1 above by its deterministic counterpart equal to the expected value 1/10 (i.e.\  $F_{U}(x)=I_{[1/10,\infty)}$), but keep all other parameters as in Example 1, we arrive at Figure \ref{Ex3b_Str}. In this example one can nicely see the signature of the deterministic claim size equal to 0.1 in the shape of the action/non-action regions. For fixed $\lambda$, one can also see how multiple bands can appear here, leading to a surprisingly aesthetic butterfly shape. Note, however, that when the optimal strategy is applied dynamically, both the values of $x$ and $\lambda$ change instantaneously, so that one will in fact not literally apply a band strategy in the classical sense. Figure \ref{Ex3b_Val} gives the corresponding value function as a function of initial $x$ and $\lambda$, and Figure \ref{Ex3b_Comp} compares it for $\lambda=\lambdaav$ with the one of the classical risk model with homogeneous intensity. Notice that in all the examples of this section the additional variability due to the shot-noise component is in fact advantageous for the shareholders.

\section{Conclusion and Outlook}\label{seccon}
In this paper we solved the two-dimensional stochastic control problem of optimizing expected discounted dividends until ruin of an insurance portfolio, when the claim number process is a Cox process with shot-noise intensity function. We identified the optimal value function as the smallest viscosity supersolution of a Hamilton-Jacobi-Bellman equation and provided a numerical scheme to uniformly approximate it via a discretization of the surplus space and the intensity space. In the numerical implementations, we then investigated to what extent the optimal dividend strategies deviate from the classical ones where the Poisson intensity is constant. As it turns out, the additional variability of the claim occurrence pattern can be used to the advantage of the insurer as far as a valuation of the company in terms of expected discounted future dividends is concerned. Concerning the concrete dividend strategies, the nature of barrier and band strategies remains in principle valid, although with barrier and band levels that now depend on the current level of the claim occurrence intensity $\lambda$, so that one needs to react adaptively to its change over time. \\

The fact that additional insurance business linked to catastrophe claims can be beneficial is somewhat promising in times when even reinsurers become more reluctant to include natural catastrophe claims in their portfolio and risk structure, and reinsurance premiums increase globally, see e.g.\ \cite{Schneider}. While the model assumptions employed in this paper are clearly somewhat too simple for practical implementation in catastrophe insurance practice, the results still demonstrate that a good understanding of the underlying actuarial risks in the catastrophe lines of business also has upward potential. Clearly, the present paper is only one first step into this direction, and many generalizations are possible and desirable. As already mentioned in the introduction, a prominent one is to go beyond the stationary model world and allow for risk factors that change over time, a feature that will allow to more explicitly model the effects of climate change on the profitability of an insurance portfolio. Also, the profitability criterion of maximizing dividends over the lifetime of the company should be accompanied by some constraints on solvency, for instance along approaches that have been pursued in the classical optimal dividend literature already. Also, with respect to regulation, a more refined model as to when received and unused premiums can be paid out as company profits may be pursued. Moreover, other interpretations of the reason and nature of the shot noise Cox feature of the claim number model are naturally possible, and each one will lead to different or additional features of the model to be studied. Finally, it may be possible to apply and extend some features of the techniques developed in this paper to the problem of optimizing dividends for a cyber risk insurer, where in addition to the shot-noise feature one typically also has a self-exciting component in the intensity process, see e.g.\ Zeller \& Scherer \cite{Scherer}.

\appendix
\section{Appendices}
\subsection{Proofs of Section \ref{Section Introduction}
\label{Appendix Section Introduction}}

\textit{Proof of Proposition \ref{Proposition E(lambda)}.}
Let us define $\lambda_{t}^{0}$ and $\Lambda_{t}^{0}$ as the processes defined
in (\ref{Lambda t corrida}) and (\ref{Integral de Lambda}) in the particular
case of $\underline{\lambda}=0,$ then we have%
\[
\lambda_{t}=\underline{\lambda}(1-e^{-dt})+\lambda_{t}^{0}\text{, }\Lambda
_{t}=\underline{\lambda}\int_{0}^{t}(1-e^{-ds})ds+\mathbb{E}(\Lambda_{t}%
^{0}).
\]
From \cite[Cor.2.4]{Dassios Jang},we get%
\[
\mathbb{E}(e^{-v~\Lambda_{t}^{0}})=e^{-\frac{v~\lambda}{d}(1-e^{-dt}%
)}e^{-\beta\int\nolimits_{0}^{t}(1-g(\frac{v}{d}(1-e^{-d~(t-s)})))ds},
\]
where
\[
g(u)=\mathbb{E}(e^{-uY_{1}}).
\]
So we get

\[%
\begin{array}
[c]{lll}%
\mathbb{E}(e^{-v~\Lambda_{t}}) & = & \mathbb{E}(e^{-v~\Lambda_{t}^{0}})\cdot
e^{-v\underline{\lambda}\int_{0}^{t}(1-e^{-ds})ds}\\
& = & e^{-\frac{v~\lambda}{d}(1-e^{-dt})}\cdot e^{-\beta\int\nolimits_{0}%
^{t}(1-g(\frac{v}{d}(1-e^{-d~(t-s)})))ds}\cdot e^{-v(\underline{\lambda
}t-\underline{\lambda}\left(  \frac{1-e^{-dt}}{d}\right)  )}.
\end{array}
\]
We also have%

\[%
\begin{array}
[c]{lll}%
\mathbb{E}(\Lambda_{t}^{0}) & = & -\left.  \partial_{v}\mathbb{E}%
(e^{-v\Lambda_{t}^{0}})\right\vert _{v=0}\\
& = & \dfrac{1-e^{-dt}}{d}\lambda+\dfrac{e^{-dt}-1+dt}{d^{2}}\beta
~\mathbb{E}(Y_{1}),
\end{array}
\]
so
\[%
\begin{array}
[c]{lll}%
\mathbb{E}(e^{-v~\Lambda_{t}}) & = & \mathbb{E}(e^{-v~\Lambda_{t}^{0}})\cdot
e^{-v\underline{\lambda}\int_{0}^{t}(1-e^{-ds})ds}\\
& = & e^{-\frac{v~\lambda}{d}(1-e^{-dt})}\cdot e^{-\beta\int\nolimits_{0}%
^{t}(1-g(\frac{v}{d}(1-e^{-d~(t-s)})))ds}\cdot e^{-v(\underline{\lambda
}t-\underline{\lambda}\left(  \frac{1-e^{-dt}}{d}\right)  )}.%
\end{array}
\]
In addition, 
\[%
\begin{array}
[c]{lll}%
\mathbb{E}(\lambda_{t}^{0}) & = & \partial_{t}\mathbb{E}(\Lambda_{t})\\
& = & \lambda e^{-dt}+\dfrac{(1-e^{-dt})}{d}\beta~\mathbb{E}(Y_{1}),
\end{array}
\]
and we conclude that%
\[%
\begin{array}
[c]{lll}%
\mathbb{E}(e^{-v~\Lambda_{t}}) & = & \mathbb{E}(e^{-v~\Lambda_{t}^{0}})\cdot
e^{-v\underline{\lambda}\int_{0}^{t}(1-e^{-ds})ds}\\
& = & e^{-\frac{v~\lambda}{d}(1-e^{-dt})}\cdot e^{-\beta\int\nolimits_{0}%
^{t}(1-g(\frac{v}{d}(1-e^{-d~(t-s)})))ds}\cdot e^{-v(\underline{\lambda
}t-\underline{\lambda}\left(  \frac{1-e^{-dt}}{d}\right)  )}%
\end{array}
\]
and%
\[%
\begin{array}
[c]{lll}%
\mathbb{E}(\lambda_{t}) & = & \underline{\lambda}(1-e^{-dt})+\mathbb{E}%
(\lambda_{t}^{0})\\
& = & \underline{\lambda}(1-e^{-dt})+\lambda e^{-dt}+\dfrac{(1-e^{-dt})}%
{d}\beta~\mathbb{E}(Y_{1}),
\end{array}
\]
as well as
\[%
\begin{array}
[c]{lll}%
\mathbb{E}(\Lambda_{t}) & = & \underline{\lambda}\int_{0}^{t}(1-e^{-ds}%
)ds+\mathbb{E}(\Lambda_{t}^{0})\\
& = & \underline{\lambda}\int_{0}^{t}(1-e^{-ds})ds+\dfrac{(1-e^{-dt})}%
{d}\lambda+\dfrac{(e^{-dt}-1+dt)}{d^{2}}\beta~\mathbb{E}(Y_{1})\\
& = & \underline{\lambda}t-\underline{\lambda}\left(  \frac{1-e^{-dt}}%
{d}\right)  +\dfrac{(1-e^{-dt})}{d}\lambda+\dfrac{(e^{-dt}-1+dt)}{d^{2}}%
\beta~\mathbb{E}(Y_{1})%
\end{array}
\]
\hfill$\blacksquare$

\subsection{Proofs of Section \ref{Model and basic results}
\label{Appendix Section Model and Basic Results}}

In order to prove Propositions \ref{Continuidad Uniforme en Lambda},
\ref{Locally Lipschitz x} and \ref{Condicion de Lipschitz en Lambda}, we need
to make a definition and give two lemmas.

\begin{definition}
Given two sequences $(\tau_{i}^{1},U_{i}^{1})_{i\geq1}\in\overline{\Omega}$
and$~(\tau_{j}^{2},U_{j}^{2})_{j\geq1}\in\overline{\Omega}$ introduced in
(\ref{OmegaBarra}), we define the ordered union of the two sequences as
\[
(\tau_{i}^{1},U_{i}^{1})_{i\geq1}\amalg(\tau_{j}^{2},U_{j}^{2})_{j\geq1}%
=(\tau_{n},U_{n})_{n\geq1}\in\overline{\Omega},
\]
where $\left\{  (\tau_{n},U_{n}):n\geq1\right\}  =\left\{  (\tau_{i}^{1}%
,U_{i}^{1}):i\geq1\right\}  \cup\left\{  (\tau_{j}^{2},U_{j}^{2}%
):j\geq1\right\}  $ and $\tau_{n}\leq\tau_{n+1}.$
\end{definition}

\begin{definition}
Given any intensity-filtered probability space\textit{\ }$(\Omega
^{0},\mathcal{F}^{0},\left(  \mathcal{F}_{t}^{0}\right)  _{t\geq0}%
,\mathbb{P}^{0})$ and two intensity random process $\mathbf{\lambda}%
^{1}=\left(  \lambda_{t}^{1}\right)  _{t\geq0}$ and $\mathbf{\lambda}%
^{2}=\left(  \lambda_{t}^{2}\right)  _{t\geq0}$ adapted to $\left(
\mathcal{F}_{t}^{0}\right)  _{t\geq0}$, we define the superposition
\[
(\Omega^{\mathbf{\lambda}^{1}\oplus\mathbf{\lambda}^{2}},\mathcal{F}%
^{\mathbf{\lambda}^{1}\oplus\mathbf{\lambda}^{2}},\left(  \mathcal{F}%
_{t}^{\mathbf{\lambda}^{1}\oplus\mathbf{\lambda}^{2}}\right)  _{t\geq
0},\mathbb{P}^{\mathbf{\lambda}^{1}\oplus\mathbf{\lambda}^{2}})
\]
of the two compound Cox filtered spaces $(\Omega^{\mathbf{\lambda}^{1}%
},\mathcal{F}^{\mathbf{\lambda}^{1}},\left(  \mathcal{F}_{t}^{\mathbf{\lambda
}^{1}}\right)  _{t\geq0},\mathbb{P}^{\mathbf{\lambda}^{1}})$ and
$(\Omega^{\mathbf{\lambda}^{2}},\mathcal{F}^{\mathbf{\lambda}^{2}},\left(
\mathcal{F}_{t}^{\mathbf{\lambda}^{2}}\right)  _{t\geq0},\mathbb{P}%
^{\mathbf{\lambda}^{2}})$ in the following way:

\begin{itemize}
\item $\Omega^{\mathbf{\lambda}^{1}\oplus\mathbf{\lambda}^{2}}=\left\{
(\omega,\tau_{i}^{1},U_{i}^{1})_{i\geq1}\amalg(\tau_{j}^{2},U_{j}^{2}%
)_{j\geq1})~\text{s.t. }(\omega,(\tau_{n}^{i},U_{n}^{i})_{n\geq1})\in
\Omega^{\mathbf{\lambda}^{i}}~\text{for }i=1,2\right\}  $;

\item $\mathcal{F}^{\mathbf{\lambda}^{1}\oplus\mathbf{\lambda}^{2}}$is the
complete $\sigma$-field generated by the $\sigma$-field $\mathcal{F}^{0}$ and
the random variables $\tau_{n}^{1},~U_{n}^{1},\tau_{m}^{2},~U_{m}^{2}$;

\item $\mathcal{F}_{t}^{\mathbf{\lambda}^{1}\oplus\mathbf{\lambda}^{2}}$ is
the complete $\sigma$-field generated by the $\sigma$-field $\mathcal{F}%
_{t}^{0}$ and the random variables $\tau_{n}^{1},~U_{n}^{1}$ for $\tau_{n}%
^{1}\leq t$ and $\tau_{m}^{2}$ and$~U_{m}^{2}$ for $\tau_{m}^{2}\leq t$;

\item The probability measure $\mathbb{P}^{\mathbf{\lambda}^{1}\oplus
\mathbf{\lambda}^{2}}$ is the probability which satisfies that

\begin{enumerate}
\item the processes $N_{t}^{\mathbf{\lambda}_{1}}=\#\{k:\tau_{k}^{1}\leq t\}$,
$N_{t}^{\mathbf{\lambda}_{2}}=\#\{k:\tau_{k}^{2}\leq t\}$ and the random
variables $(U_{n}^{1})_{n\geq1}$ and $(U_{n}^{2})_{n\geq1}$ are independent,

\item the random variables $U_{n}^{1}$ and $U_{m}^{2}$ with $n,m\in\mathbb{N}$
are i.i.d., $\mathbb{P}^{\mathbf{\lambda}^{1}\oplus\mathbf{\lambda}^{2}}%
(U_{n}^{i}\leq x)=F_{U}(x)$ for $i=1,2$ and%
\[
\mathbb{P}^{\mathbf{\lambda}^{1}\oplus\mathbf{\lambda}^{2}}(N_{t}%
^{\mathbf{\lambda}_{1}}+N_{t}^{\mathbf{\lambda}_{2}}=n)=\sum_{s=0}%
^{n}\mathbb{P}^{\mathbf{\lambda}^{1}}(N_{t}^{\mathbf{\lambda}_{1}}%
=s)\cdot\mathbb{P}^{\mathbf{\lambda}^{2}}(N_{t}^{\mathbf{\lambda}_{2}}=n-s).
\]

\end{enumerate}
\end{itemize}
\end{definition}

We have the following lemma (cf. for instance \cite[Sec.3]{Moller}).

\begin{lemma}
The set $\Omega_{0}^{\mathbf{\lambda}^{1}\oplus\mathbf{\lambda}^{2}}=\left\{
\tau_{n}^{1}\neq\tau_{m}^{2}\text{ for any }n,m\in\mathbb{N}\right\}
\in\mathcal{F}^{\mathbf{\lambda}^{1}\oplus\mathbf{\lambda}^{2}} $has full
measure. Also, the restriction of the \textit{superposition probability
space}%
\[
(\Omega^{\mathbf{\lambda}^{1}\oplus\mathbf{\lambda}^{2}},\mathcal{F}%
^{\mathbf{\lambda}^{1}\oplus\mathbf{\lambda}^{2}},\left(  \mathcal{F}%
_{t}^{\mathbf{\lambda}^{1}\oplus\mathbf{\lambda}^{2}}\right)  _{t\geq
0},\mathbb{P}^{\mathbf{\lambda}^{1}\oplus\mathbf{\lambda}^{2}})
\]
to $\Omega_{0}^{\mathbf{\lambda}^{1}\oplus\mathbf{\lambda}^{2}}$coincides with
the {compound Cox filtered space }$(\Omega^{\mathbf{\lambda}%
},\mathcal{F}^{\mathbf{\lambda}},\left(  \mathcal{F}_{t}^{\mathbf{\lambda}%
}\right)  _{t\geq0},\mathbb{P}^{\mathbf{\lambda}})$ \textit{conditional on the
intensity process} $\mathbf{\lambda=\lambda}_{1}+\mathbf{\lambda}_{2}=\left(
\lambda_{t}^{1}+\lambda_{t}^{2}\right)  _{t\geq0}$.

Moreover, if $P_{1}:\Omega_{0}^{\mathbf{\lambda}^{1}\oplus\mathbf{\lambda}%
^{2}}\rightarrow\Omega^{\mathbf{\lambda}^{1}}$ and $P_{2}:\Omega
_{0}^{\mathbf{\lambda}^{1}\oplus\mathbf{\lambda}^{2}}\rightarrow
\Omega^{\mathbf{\lambda}^{2}}$ are the projections defined as
\[
P_{i}\left(  (\omega,(\tau_{n}^{1},U_{n}^{1})_{n\geq1}\amalg(\tau_{n}%
^{2},U_{n}^{2})_{n\geq1})\right)  =(\omega,(\tau_{n}^{i},U_{n}^{i})_{n\geq
1})\in\Omega^{\mathbf{\lambda}^{i}},%
\]
then
\begin{enumerate}
\item For any $A\in\mathcal{F}^{\mathbf{\lambda}_{i}},$ we have that the
preimage $P_{i}^{-1}(A)\in\mathcal{F}^{\mathbf{\lambda}^{1}\oplus
\mathbf{\lambda}^{2}}$ and $\mathbb{P}^{\mathbf{\lambda}^{1}\oplus
\mathbf{\lambda}^{2}}(P_{i}^{-1}(A))=\mathbb{P}^{\mathbf{\lambda}^{i}}(A)$.

\item For any $A\in\mathcal{F}_{t}^{\mathbf{\lambda}_{i}},$we have that the
preimage $P_{i}^{-1}(A)\in\mathcal{F}_{t}^{\mathbf{\lambda}^{1}\oplus
\mathbf{\lambda}^{2}}$.

\item $N_{t}^{\mathbf{\lambda}_{1}}\circ P_{1}+N_{t}^{\mathbf{\lambda}_{2}%
}\circ P_{2}=N_{t}^{\mathbf{\lambda}_{1}+\mathbf{\lambda}_{2}}:\Omega
_{0}^{\mathbf{\lambda}^{1}\oplus\mathbf{\lambda}^{2}}\rightarrow\mathbb{N}%
_{0}$.
\end{enumerate}
\end{lemma}

\begin{lemma}
\label{Descomposicion de lambdas} As a consequence of the previous lemma,
consider any intensity-filtered probability space \\$(\Omega
^{0},\mathcal{F}^{0},\left(  \mathcal{F}_{t}^{0}\right)  _{t\geq0}%
,\mathbb{P}^{0})$ and two intensity random processes $\mathbf{\lambda}%
^{1}=\left(  \lambda_{t}^{1}\right)  _{t\geq0}$ and $\mathbf{\lambda}%
^{2}=\left(  \lambda_{t}^{2}\right)  _{t\geq0}$ adapted to $\left(
\mathcal{F}_{t}^{0}\right)  _{t\geq0}$ with $\mathbf{\lambda}^{1}%
\leq\mathbf{\lambda}^{2}$; writing $\mathbf{\lambda}^{2}=\mathbf{\lambda}%
^{1}+(\mathbf{\lambda}^{2}-\mathbf{\lambda}^{1})$ we can consider the
projections $\overline{P}_{1}:\Omega^{\mathbf{\lambda}^{2}}\rightarrow
\Omega^{\mathbf{\lambda}^{1}}$ and $\overline{P}_{2}:\Omega^{\mathbf{\lambda
}^{2}}\rightarrow\Omega^{\mathbf{\lambda}^{2}-\mathbf{\lambda}^{1}}$. Also,
the processes $N_{t}^{\mathbf{\lambda}_{1}}\circ\overline{P}_{1}$ and
$N_{t}^{\mathbf{\lambda}^{2}-\mathbf{\lambda}^{1}}\circ\overline{P}_{2}$ are
independent and satisfy $N_{t}^{\mathbf{\lambda}_{1}}\circ\overline{P}%
_{1}+N_{t}^{\mathbf{\lambda}^{2}-\mathbf{\lambda}^{1}}\circ\overline{P}%
_{2}=N_{t}^{\mathbf{\lambda}_{2}}$. We also can write%
\[
(\tau_{n}^{2},U_{n}^{2})_{n\geq1}=(\tau_{n}^{1},U_{n}^{1})_{n\geq1}%
\amalg(\widehat{\tau}_{m},\widehat{U}_{m})_{m\geq1}.
\]

\end{lemma}

\noindent\textit{Proof of Proposition \ref{Continuidad Uniforme en Lambda}}. 
\noindent\underline{Proof of (1):} Take $\lambda_{1}<\lambda_{2}$ and consider the shot-noise
processes $\mathbf{\lambda}^{i}=(\lambda_{t}^{i})_{t\geq0}$ with initial
intensity $\lambda_{i}$ and the associated compound Cox process $X_{t}^{i}$
with drift $p$\ generated by the shot-noise process $\lambda_{t}^{i}$\ with
initial intensity $\lambda_{i}$\ and initial surplus $x$ for $i=1,2$ as
defined in (\ref{model}) and (\ref{Lambda t corrida}). Since
\[
\lambda_{t}^{2}=\lambda_{t}^{1}+e^{-dt}\left(  \lambda_{2}-\lambda_{1}\right)
,
\]
we have that $\mathbf{\lambda}^{1}\leq\mathbf{\lambda}^{2}.$ We use the
results of Lemma \ref{Descomposicion de lambdas}, calling $N_{t}^{1}%
=N_{t}^{\mathbf{\lambda}^{1}}\circ\overline{P}_{1},$ $N_{t}^{2}=N_{t}%
^{\mathbf{\lambda}^{2}}$ and $\widehat{N}_{t}=N_{t}^{\mathbf{\lambda}%
^{2}-\mathbf{\lambda}^{1}}\circ\overline{P}_{2}$, we have
\[
X_{t}^{2}=x+pt-\sum_{n=1}^{N_{t}^{2}}U_{n}^{2}~=x+pt-\left(  \sum_{n=1}%
^{N_{t}^{1}}U_{n}^{1}+\sum_{m=1}^{\widehat{N}_{t}}\widehat{U}_{m}\right)
=X_{t}^{1}-\sum_{m=1}^{\widehat{N}_{t}}\widehat{U}_{m}.
\]
Now take $\left(  L_{t}\right)  _{t\geq0}\in\Pi_{x,\lambda_{2}}$ with
\[
V(x,\lambda_{2})\leq J(L;x,\lambda_{2})+\varepsilon.
\]
Let $r_{i}=\min\left\{  t:X_{t}^{i}-L_{t}<0\right\}  ;$ since $X_{t}^{1}\geq
X_{t}^{2}$, we get $r_{1}\geq r_{2}$;%
\[
V(x,\lambda_{1})\geq J(L;x,\lambda_{1})=E(\int_{0^{-}}^{r_{1}}e^{-qs}%
dL_{s})\geq E(\int_{0^{-}}^{r_{2}}e^{-qs}dL_{s})=J(L;x,\lambda_{2})\geq
V(x,\lambda_{2})-\varepsilon,~
\]
so we conclude (1).\\

\noindent\underline{Proof of (2):} Take $\lambda_{1}<\lambda_{2}$ and $L=\left(  L_{t}\right)
_{t\geq0}\in\Pi_{x,\lambda_{1}}$ with
\[
V(x,\lambda_{1})\leq J(L;x,\lambda_{1})+\varepsilon.
\]

Note that%

\[%
\begin{array}
[c]{lll}%
\mathbb{P}(\widehat{N}_{t}=k)=\mathbb{P}(N_{t}^{2}-N_{t}^{1}=k) & = &
e^{-\int_{0}^{t}e^{-dt}\left(  \lambda_{2}-\lambda_{1}\right)  ds}\frac{1}%
{k!}(\int_{0}^{t}e^{-dt}\left(  \lambda_{2}-\lambda_{1}\right)  ds)^{k}%
\text{.}%
\end{array}
\]

Let $r_{i}=\min\left\{  t:X_{t}^{i}-L_{t}<0\right\}  $, we have that
$r_{1}\geq r_{2}$ and since $X_{t}^{1}=x+pt-\sum_{n=1}^{N_{t}^{1}}U_{n}^{1}$
and%
\[%
\begin{array}
[c]{lll}%
X_{t}^{2} & = & x+pt-\sum_{n=1}^{N_{t}^{2}}U_{n}^{2}\\
& = & ~x+pt-\left(  \sum_{n=1}^{N_{t}^{1}}U_{n}^{1}~+\sum_{m=1}^{\widehat
{N}_{t}}\widehat{U}_{m}~\right) \\
& = & X_{t}^{1}-\sum_{m=1}^{\widehat{N}_{t}}\widehat{U}_{m}%
\end{array}
\]
we have that%
\[
X_{r_{2}}^{1}=X_{r_{2}}^{2}+\sum_{m=1}^{\widehat{N}_{r_{2}}}\widehat{U}%
_{m}-L_{r_{2}}<\sum_{m=1}^{\widehat{N}_{r_{2}}}\widehat{U}_{m}.
\]
Take $T$ such that $e^{-qT}\left(  \frac{1}{d}\mathbb{E(}U_{1}\mathbb{)+}%
\frac{p}{q}\right)  \leq\frac{\varepsilon}{2}$ and consider $\lambda_{2}%
\leq\lambda_{1}+1$ and
\[
\lambda_{2}-\lambda_{1}\leq\min\{\frac{-\log(1-\frac{\varepsilon}{2\left(
\frac{1}{d}\mathbb{E(}U_{1}\mathbb{)+}p/q\right)  })}{T\frac{(1-e^{-dT})}{dT}.%
},1\}.
\]
Let us show that 
\[%
\begin{array}
[c]{lll}%
J(L;x,\lambda_{1})-J(L;x,\lambda_{2}) & = & \mathbb{E}(\mathbb{E}\left(
\left.  \int_{r2}^{r_{1}}e^{-qs}dL_{s}\right\vert \mathcal{F}_{r2}%
^{\mathbf{\lambda}_{2}}\right)  )\\
& \leq & \mathbb{E}(\mathbb{E}\left(  \left.  e^{-qr_{2}}V(X_{r_{2}}%
^{1}-L_{r_{2}},\lambda_{r_{_{2}}}^{1})\right\vert \mathcal{F}_{r2}%
^{\mathbf{\lambda}_{2}}\right)  )\\
& \leq & \mathbb{E}\left[  e^{-qr_{2}}V(X_{r_{2}}^{1}-L_{r_{2}},\lambda
_{r_{_{2}}}^{1})\right] \\
& \leq & \mathbb{E}\left[  I_{r_{2}>T}e^{-qr_{2}}V(\sum_{m=1}^{\widehat
{N}_{r_{2}}}\widehat{U}_{m},\lambda_{r_{_{2}}}^{1})\right]  +\mathbb{E}\left[
I_{r_{2}\leq T}e^{-qr_{2}}V(X_{r_{2}}^{1}-L_{r_{2}},\lambda_{r_{_{2}}}%
^{1})\right] \\
& \leq & \varepsilon.
\end{array}
\]
From $V(y,\lambda_{r_{_{2}}}^{1})\leq V(y,\underline{\lambda})\leq
y+\frac{p}{q}$ and%
\[
\mathbb{E}(\widehat{N}_{t})=\left(  \lambda_{2}-\lambda_{1}\right)
\frac{1-e^{-dt}}{d}
\]
we get%
\[%
\begin{array}
[c]{ccl}%
\mathbb{E}\left[  I_{r_{2}>T}e^{-qr_{2}}V(\sum_{m=1}^{\widehat{N}_{r_{2}}%
}\widehat{U}_{m},\lambda_{r_{_{2}}}^{1})\right]  & \leq & \mathbb{E}\left[
I_{r_{2}>T}e^{-qr_{2}}\left(  \sum_{m=1}^{\widehat{N}_{r_{2}}}\widehat{U}%
_{m}+\frac{p}{q}\right)  \right] \\
& \leq & e^{-qT}\mathbb{E}\left[  \left(  \sum_{m=1}^{\widehat{N}_{r_{2}}%
}\widehat{U}_{m}+\frac{p}{q}\right)  \right] \\
& \leq & e^{-qT}\left(  \frac{1}{d}\left(  \lambda_{2}-\lambda_{1}\right)
\mathbb{E(}U_{1}\mathbb{)+}\frac{p}{q}\right) \\
& \leq & e^{-qT}\left(  \frac{1}{d}\mathbb{E(}U_{1}\mathbb{)+}\frac{p}%
{q}\right)  +\frac{\varepsilon}{2}\\
& \leq & \frac{\varepsilon}{2}.%
\end{array}
\]
Also, if $\widehat{N}_{T}=0$ then $X_{r_{2}}^{1}-L_{r_{2}}=X_{r_{2}}%
^{2}-L_{r_{2}}<0$ and so $V(X_{r_{2}}^{1}-L_{r_{2}},\lambda_{r_{_{2}}}%
^{1})=0.$ Hence, taking%
\[
\lambda_{2}-\lambda_{1}\leq\frac{-\log(1-\frac{\varepsilon}{2\left(  \frac
{1}{d}\mathbb{E(}U_{1}\mathbb{)+}p/q\right)  })}{T\frac{(1-e^{-dT})}{dT}},
\]
we have%
\[
1-\mathbb{P}(\widehat{N}_{T}=0)=1-e^{-\left(  \lambda_{2}-\lambda_{1}\right)
t\frac{(1-e^{-dt})}{dt}}<\frac{\varepsilon}{2\left(  \frac{1}{d}%
\mathbb{E(}U_{1}\mathbb{)+}p/q\right)  }%
\]
and so%
\[%
\begin{array}
[c]{lll}%
\mathbb{E}\left[  I_{r_{2}<T}e^{-qr_{2}}V(X_{r_{2}}^{1}-L_{r_{2}}%
,\lambda_{r_{_{2}}}^{1})\right]  & \leq & \mathbb{E}\left[  I_{r_{2}%
<T}e^{-qr_{2}}V(\sum_{m=1}^{\widehat{N}_{r_{2}}}\widehat{U}_{m},\lambda
_{r_{_{2}}}^{1})I_{\left\{  \widehat{N}_{T}>0\right\}  }\right] \\
& \leq & \mathbb{E}\left[  I_{r_{2}<T}V(\sum_{m=1}^{\widehat{N}_{T}}%
\widehat{U}_{m},\lambda_{r_{_{2}}}^{1})I_{\left\{  \widehat{N}_{T}>0\right\}
}\right]  \mathbb{P}\left[  \widehat{N}_{T}>0\right] \\
& \leq & \mathbb{E}\left[  \left(  \sum_{m=1}^{\widehat{N}_{T}}\widehat{U}%
_{m}+p/q\right)  \right]  \mathbb{P}\left[  \widehat{N}_{T}>0\right] \\
& \leq & \left(  \frac{1}{d}\mathbb{E(}U_{1}\mathbb{)+}p/q\right)
\mathbb{P}\left[  \widehat{N}_{T}>0\right] \\
& \leq & \frac{\varepsilon}{2}.~
\end{array}
\]
Hence, the result follows. 
\hfill$\blacksquare$

\bigskip

\noindent\textit{Proof of Proposition \ref{Locally Lipschitz x}.}
It is straightforward to show that $V(\cdot,\lambda)$ is non-decreasing
because
\[
V(x_{2},\lambda)\geq x_{2}-x_{1}+V(x_{1},\lambda).
\]
Let us prove the other inequality. Take $t_{0}=(x_{2}-x_{1})/p$ $<1$ and
consider $\lambda_{t_{0}}=\underline{\lambda}+e^{-dt_{0}}\left(
\lambda-\underline{\lambda}\right)  $. Given an initial surplus $x\geq0$ and
$\varepsilon>0$, consider an admissible strategy $L=(L_{t})_{t\geq0}\in
\Pi_{x_{2},\lambda_{t_{0}}}$ such that $J(L;x_{2},\lambda_{t_{0}})\geq
V(x_{2},\lambda_{t_{0}})-\varepsilon$ for any $x_{2}>x_{1}$. Take now the
strategy $\widetilde{L}\in\Pi_{x_{1},\lambda}$ that starts with surplus
$x_{1}$, pays no dividends if $X_{t}^{\widetilde{L}}<x_{2}$ and follows strategy
$L$ after the current reserve reaches $\left(  x_{2},\lambda_{t_{0}}\right)
$, that is%
\[
\widetilde{L}_{t}=\left\{
\begin{array}
[c]{lll}%
0 & \text{if} & t\leq t_{0}\text{ }\wedge\tau_{1}\wedge T_{1}\\
L_{t-t_{0}} & \text{if} & t\geq t_{0}\text{ and }\tau_{1}\wedge T_{1}>t_{0}\\
0 & \text{if} & t>\tau_{1}\wedge T_{1}\ \text{and }t_{0}\text{ }\geq\tau
_{1}\wedge T_{1}.%
\end{array}
\right.
\]
The strategy $\widetilde{L}$ is admissible, and we get
\[%
\begin{array}
[c]{lll}%
V(x_{1},\lambda) & \geq & J(\widetilde{L};x_{1},\lambda)\\
& \geq & \mathbb{P}(\tau_{1}\wedge T_{1}>t_{0})J(L;x_{2},\lambda_{t_{0}%
})e^{-qt_{0}}\\
& \geq & \left(  V(x_{2},\lambda_{t_{0}})-\varepsilon\right)  e^{-qt_{0}%
}\mathbb{P}(\tau_{1}\wedge T_{1}>t_{0}),
\end{array}
\]
and%
\[%
\begin{array}
[c]{ccc}%
V(x_{2},\lambda)-V(x_{1},\lambda) & \leq & V(x_{2},\lambda)-\left(
V(x_{2},\lambda_{t_{0}})-\varepsilon\right)  \mathbb{P}(\tau_{1}\wedge
T_{1}>t_{0})e^{-qt_{0}}.%
\end{array}
\]
So, using that $V$ is non-increasing on $\lambda$ and $t_{0}<1,$
\[%
\begin{array}
[c]{lcl}%
V(x_{2},\lambda)-V(x_{1},\lambda) & \leq & V(x_{2},\lambda)-V(x_{2}%
,\lambda_{t_{0}})\mathbb{P}(\tau_{1}\wedge T_{1}>t_{0})e^{-qt_{0}}\\
& \leq & V(x_{2},\lambda)-V(x_{2},\lambda)\mathbb{P}(\tau_{1}\wedge
T_{1}>t_{0})e^{-qt_{0}}\\
& \leq & V(x_{2},\lambda)(1-\mathbb{P}(\tau_{1}\wedge T_{1}>t_{0})e^{-qt_{0}%
})\\
& \leq & V(x_{2},\lambda)\frac{\beta\lambda_{2}+q}{p}(x_{2}-x_{1})\\
& \leq & V(x_{2},\underline{\lambda})\frac{\beta\lambda_{2}+q}{p}(x_{2}%
-x_{1}).
\end{array}
\]
\hfill$\blacksquare$\\

\noindent\textit{Proof of Proposition \ref{Condicion de Lipschitz en Lambda}.}
From Proposition \ref{Continuidad Uniforme en Lambda}, we have $0\leq
V(x,\lambda_{1})-V(x,\lambda_{2})$. Let us prove the second inequality. Take
$L\in\Pi_{x,\lambda_{1}}$ such that $V(x,\lambda_{1})\leq J(L;x,\lambda
_{1})+\varepsilon/2$, and the associated controlled process $(X_{t}%
^{L},\lambda_{t})$ starting at $(x,\lambda_{1}).$ Consider $\delta_{0}%
=\frac{1}{d}\log(\frac{\lambda_{2}-\underline{\lambda}}{\lambda_{1}%
-\underline{\lambda}})$ so $\underline{\lambda}+e^{-d\delta_{0}}\left(
\lambda_{2}-\underline{\lambda}\right)  =\lambda_{1}$, then we have that
\[
\delta_{0}\leq\frac{1}{d(\lambda_{1}-\underline{\lambda})}\left(  \lambda
_{2}-\lambda_{1}\right)  .
\]
Define $\widetilde{L}\in\Pi_{x,\lambda_{2}}$ as
\[
\widetilde{L}_{t}=\left\{
\begin{array}
[c]{lll}%
pt & \text{if} & t\leq\delta_{0}\text{ }\wedge\tau_{1}\wedge T_{1}\\
L_{t-\delta_{0}} & \text{if} & t\geq\delta_{0}\text{ and }\tau_{1}\wedge
T_{1}>\delta_{0}\\
p(\tau_{1}\wedge T_{1}) & \text{if} & t>\tau_{1}\wedge T_{1}\ \text{and
}\delta_{0}\text{ }\geq\tau_{1}\wedge T_{1}.
\end{array}
\right.
\]
We have that%
\[%
\begin{array}
[c]{lll}%
J(\widetilde{L};x,\lambda_{2}) & \geq & J(L;x,\lambda_{1})e^{-q\delta_{0}%
}\mathbb{P}(\tau_{1}\wedge T_{1}>\delta_{0})\\
& = & J(L;x,\lambda_{1})e^{-q\delta_{0}}(1-\mathbb{P}(\tau_{1}\wedge T_{1}%
\leq\delta_{0}))\\
& \geq & J(L;x,\lambda_{1})e^{-q\delta_{0}}(1-(1-e^{-\beta\delta_{0}%
)})(1-e^{-\lambda_{2}\delta_{0}}))\\
& \geq & J(L;x,\lambda_{1})\left(  1-q\delta_{0}\right)  (1-\beta\lambda
_{2}\delta_{0}^{2})\allowbreak
\end{array}
\]
because
\[
\int_{0}^{\delta_{0}}\text{ }\left(  \underline{\lambda}+e^{-ds}\left(
\lambda_{2}-\underline{\lambda}\right)  \right)  ds\leq\lambda_{2}\delta_{0},
\]
and so, taking $\lambda_{2}\ $and$~\lambda_{1}$ close enough so that $\delta_{0}%
<1$,
\[%
\begin{array}
[c]{lll}%
V(x,\lambda_{1})-V(x,\lambda_{2}) & \leq & J(L;x,\lambda_{1})-J(\widetilde
{L},x,\lambda_{2})+\varepsilon/2\\
& \leq & J(L;x,\lambda_{1})\left(  1-\left(  1-q\delta_{0}\right)
(1-\beta\lambda_{2}\delta_{0}^{2})\allowbreak\right)  +\varepsilon/2\\
& \leq & V(x,\lambda_{1})\left(  1-\left(  1-q\delta_{0}\right)
(1-\beta\lambda_{2}\delta_{0}^{2})\allowbreak\right)  +\varepsilon/2\\
& \leq & V(x,\lambda_{1})\allowbreak\frac{\left(  \beta\lambda_{2}+q\right)
}{d(\lambda_{1}-\underline{\lambda})}\left(  \lambda_{2}-\lambda_{1}\right)
+\varepsilon/2,
\end{array}
\]
and the result follows. \hfill$\blacksquare$

\bigskip \noindent\textit{Proof of Proposition \ref{Convergencia puntual con lambda a infinito}.}
Let us define%

\[
t^{\ast}\left(  \lambda\right)  :=\frac{1}{d}\ln(\frac{\lambda-\underline
{\lambda}}{\ln(\lambda)-\underline{\lambda}})
\]
so that%

\[
\underline{\lambda}+e^{-dt^{\ast}\left(  \lambda\right)  }\left(
\lambda-\underline{\lambda}\right)  =\ln(\lambda)
\]
and $t^{\ast}\left(  \lambda\right)  \rightarrow\infty$ as $\lambda
\rightarrow\infty$. Take a near optimal strategy $L=(L_{t})_{t\geq0}\in
\Pi_{x,\lambda}$ such that $V(x,\lambda)\leq J(L;x,\lambda)+\varepsilon$,
hence $\lambda_{t}\geq\ln(\lambda)$ for $t\leq t^{\ast}\left(  \lambda\right)
$. Then, using Lemma \ref{DPP}, Proposition
\ref{Continuidad Uniforme en Lambda} and Remark \ref{Crecimiento lineal V},%

\[%
\begin{array}
[c]{lll}%
V(x,\lambda)-\varepsilon & \leq & J(L;x,\lambda)\\
& = & \mathbb{E}(\int_{0^{-}}^{t^{\ast}\left(  \lambda\right)  \wedge\tau^{L}%
}e^{-qs}dL_{s})+e^{-qt^{\ast}\left(  \lambda\right)  }\mathbb{E}\left(
I_{t^{\ast}\left(  \lambda\right)  >\tau^{L}}V(X_{t^{\ast}\left(
\lambda\right)  }^{L},\lambda_{t^{\ast}\left(  \lambda\right)  })\right) \\
& \leq & v^{\ln(\lambda)}(x)+e^{-qt^{\ast}\left(  \lambda\right)
}v^{\underline{\lambda}}(x+pt^{\ast}\left(  \lambda\right)  ).
\end{array}
\]
By Remark \ref{Optima Unidimensional}, we have that $\lim_{\lambda
\rightarrow\infty}v^{\ln(\lambda)}(x)=x$ and
\[
\lim_{\lambda\rightarrow\infty}e^{-qt^{\ast}\left(  \lambda\right)
}v^{\underline{\lambda}}(x+pt^{\ast}\left(  \lambda\right)  )\leq\lim
_{\lambda\rightarrow\infty}e^{-qt^{\ast}\left(  \lambda\right)  }\left(
x+pt^{\ast}\left(  \lambda\right)  +K\right)  =0,
\]
so we have the result. \hfill$\blacksquare$

\bigskip

\subsection{Proofs of Section \ref{Section HJB}~\label{Appendix Section HJB}}

\textit{Proof of Proposition \ref{Prop V is a viscosity solution}.}
Let us prove first that $V$ is a viscosity supersolution. Given initial values
$\left(  x,\lambda\right)  \in(0,\infty)\times(\underline{\lambda},\infty)$
and any $l$ $\geq0$, let us consider the admissible strategy $L\in
\Pi_{x,\lambda}$ where the company pays dividends with constant rate $l$ and
consider $\tau^{L}$ defined as in (\ref{DefinicionTauRuina}). Let $\varphi$ be
a test function for supersolution of (\ref{HJB}) at $(x,\lambda).$ We have
$X_{t}^{L}=x+\left(  p-l\right)  t$ and $\lambda_{t}=\underline{\lambda
}+e^{-dt}\left(  \lambda-\underline{\lambda}\right)  $ for $t<\tau_{1}\wedge
T_{1}$. Applying Lemma \ref{DPP} with stopping time $\tau_{1}\wedge
T_{1}\wedge h$, we obtain%
\[%
\begin{array}
[c]{lll}%
0 & = & V(x,\lambda)-\varphi(x,\lambda)\\
& \geq & \mathbb{E}\left(  \int_{0}^{\tau_{1}\wedge T_{1}\wedge h}%
e^{-qt}ldt\right) \\[0.1cm]
&  & +\mathbb{E}\left(  I_{\tau_{1}<T_{1}\wedge h}e^{-q\tau_{1}}%
V(x+(p-l)\tau_{1}-U_{1},\underline{\lambda}+e^{-d\tau_{1}}\left(
\lambda-\underline{\lambda}\right)  )\right) \\[0.1cm]
&  & +\mathbb{E}\left(  I_{T_{1}<\tau_{1}\wedge h}e^{-qT_{1}}V(x+(p-l)T_{1}%
,\underline{\lambda}+e^{-dT_{1}}\left(  \lambda-\underline{\lambda}\right)
+Y_{1}))\right) \\[0.1cm]
&  & +\mathbb{E}\left(  I_{h<\tau_{1}\wedge T_{1}}e^{-qh}V(x+(p-l)h,\underline
{\lambda}+e^{-dh}\left(  \lambda-\underline{\lambda}\right)  )\right)
-\varphi(x,\lambda)\\[0.1cm]
& \geq & \mathbb{E}(\int_{0}^{\tau_{1}\wedge T_{1}\wedge h}e^{-qt}ldt)\\[0.1cm]
&  & +\mathbb{E}\left(  I_{\tau_{1}<T_{1}\wedge h}e^{-q\tau_{1}}%
\varphi(x+(p-l)\tau_{1}-U_{1},\underline{\lambda}+e^{-d\tau_{1}}\left(
\lambda-\underline{\lambda}\right)  )\right) \\[0.1cm]
&  & +\mathbb{E}\left(  I_{T_{1}<\tau_{1}\wedge h}e^{-qT_{1}}\varphi
(x+(p-l)T_{1},\underline{\lambda}+e^{-dT_{1}}\left(  \lambda-\underline
{\lambda}\right)  +Y_{1}))\right) \\[0.1cm]
&  & +\mathbb{E}\left(  I_{h<\tau_{1}\wedge T_{1}}e^{-qh}\varphi
(x+(p-l)h,\underline{\lambda}+e^{-dh}\left(  \lambda-\underline{\lambda
}\right)  )\right)  -\varphi(x,\lambda).
\end{array}
\]

So, dividing by $h$ and taking $h\rightarrow0^{+},$ we obtain%
\[
\mathcal{L}(\varphi)(x,\lambda)+l(1-\varphi_{x}(x,\lambda))\leq0.
\]
For $l\rightarrow\infty,$ we obtain%
\[
\max\{\mathcal{L}(\mathbb{\varphi})(x,\lambda),1-\mathbb{\varphi}%
_{x}(x,\lambda)\}\leq0.
\]
Hence $V$ is a viscosity supersolution at $(x,\lambda)$.

Let us prove now that $V$ is a viscosity subsolution. \textit{\ }Arguing by
contradiction, we assume that $V$ is not a subsolution of (\ref{HJB}) at
$(x_{0},\lambda_{0})\in(0,\infty)\times(\underline{\lambda},\infty)$. As in
the proof of Proposition 3.8 of {\cite{AM 2005}} but extended to two
variables as in Proposition 3.2 in {\cite{AAM 2017}}, we can find a
continuously differentiable function $\psi:(0,\infty)\times(\underline
{\lambda},\infty)\rightarrow\mathbb{R}$ such that $\psi$ is a test function
for subsolution of equation (\ref{HJB}) at $\left(  x_{0},\lambda_{0}\right)
$ , for $h<x_{0}\wedge\left(  \lambda_{0}-\underline{\lambda}\right)  $ small
enough so%

\begin{equation}
\psi_{x}(x,\lambda)\geq1 \label{Desig1aprima}%
\end{equation}
for $(x,\lambda)\in\lbrack0,x_{0}+h]\times(\underline{\lambda},\infty)$,
\begin{equation}
\mathcal{L}\left(  \psi\right)  (x,\lambda)\leq-\varepsilon q \label{Desig1a}%
\end{equation}
for $(x,\lambda)\in\lbrack x_{0}-h,x_{0}+h]\times\lbrack\lambda_{0}%
-h,\lambda_{0}+h]$, and%

\begin{equation}
V(x,\lambda)\leq\psi(x,\lambda)-\varepsilon\label{Desig2a}%
\end{equation}
for $\left(  x,\lambda\right)  \in\lbrack0,x_{0}+h]\times(\underline{\lambda
},\infty)-[x_{0}-h,x_{0}+h]\times\lbrack\lambda_{0}-h,\lambda_{0}+h].$

Let us take any admissible strategy $L\in\Pi_{x_{0},\lambda_{0}}$. Consider
the corresponding controlled risk process $\left(  X_{t}^{L},\lambda
_{t}\right)  $ starting at $(x_{0},\lambda_{0})$, and define the stopping time%

\[
\tau^{\ast}=\inf\{t>0:\left(  X_{t}^{L},\lambda_{t}\right)  \in(0,\infty
)\times(\underline{\lambda},\infty)-[x_{0}-h,x_{0}+h]\times\lbrack\lambda
_{0}-h,\lambda_{0}+h]\}.
\]
From (\ref{Desig2a}), we obtain that if $\tau^{\ast}<\tau^{L}$,%
\begin{equation}
V(X_{\tau^{\ast}}^{L},\lambda_{\tau^{\ast}})\leq\psi(X_{\tau^{\ast}}%
^{L},\lambda_{\tau^{\ast}})-2\varepsilon\text{.} \label{Vborde}%
\end{equation}
We can write%
\[
d\lambda_{t}=-d\left(  \lambda_{t}-\underline{\lambda}\right)  +P_{t}%
^{Y}\text{, }dX_{t}^{L}=pdt+P_{t}^{U}-dL_{t},
\]
where $P_{t}^{Y}=\sum_{T_{i}}I_{T_{i}=t}Y_{i}$ and $P_{t}^{U}=\sum_{\tau_{i}%
}I_{\tau_{i}=t}Y_{i}$. Here we assume that $P_{t}^{Y}$ and $P_{t}^{U}$ do not
jump simultaneously (because they jump at exponential times that are
independent). Using Theorem 31 of {\cite{Protter}}, we get%

\[%
\begin{array}
[c]{l}%
\psi(X_{\tau^{\ast}}^{L},\lambda_{\tau^{\ast}})e^{-q\tau^{\ast}}-\psi
(x_{0},\lambda_{0})\\%
\begin{array}
[c]{ll}%
\leq & \int\nolimits_{0}^{\tau^{\ast}}\mathcal{L}\left(  \psi\right)
(X_{s^{-}}^{L},\lambda_{s^{-}})e^{-qs}ds+M_{\tau^{\ast}}^{1}+M_{\tau^{\ast}%
}^{2}-\int\nolimits_{0}^{\tau^{\ast}}e^{-qs}dL_{s},
\end{array}
\end{array}
\]
where%
\[
M_{t}^{1}=%
{\textstyle\sum\limits_{\substack{P_{s}^{U}\neq0 \\s\leq t}}}
\left(  \psi(X_{s^{-}}^{L}-P_{s}^{U},\lambda_{s})-\psi(X_{s^{-}}^{L}%
,\lambda_{s^{-}})\right)  e^{-qs}-\int\nolimits_{0}^{t}\lambda_{s^{-}}%
{\textstyle\int\nolimits_{0}^{X_{s^{-}}^{L}}}
\psi(X_{s}^{L}-\alpha,\lambda_{s^{-}})dF_{U}(\alpha)e^{-qs}ds
\]
and%
\[
M_{t}^{2}=%
{\textstyle\sum\limits_{\substack{P_{s}^{Y}\neq0 \\s\leq t}}}
\left(  \psi(X_{s^{-}}^{L},\lambda_{s^{-}}-P_{s}^{Y})-\psi(X_{s^{-}}%
^{L},\lambda_{s^{-}})\right)  e^{-qs}-\int\nolimits_{0}^{t}\beta%
{\textstyle\int\nolimits_{0}^{X_{s^{-}}^{L}}}
\psi(X_{s^{-}}^{L},\lambda_{s^{-}}+\gamma)dF_{Y}(\gamma)e^{-qs}ds
\]
are martingales with zero expectation. Hence, we obtain%

\begin{equation}
\mathbb{E}(\psi(X_{\tau^{\ast}}^{L},\lambda_{\tau^{\ast}})e^{-q\tau^{\ast}%
}-\psi(x_{0},\lambda_{0}))\leq\mathbb{E}(\int\nolimits_{0}^{\tau^{\ast}%
}\mathcal{L}(\psi)(X_{s^{-}}^{L},\lambda_{s^{-}})e^{-qs})-\mathbb{E}%
(\int_{0^{-}}^{\tau^{\ast}}e^{-qs}dL_{s}). \label{nueva1}%
\end{equation}
Using (\ref{Desig1a}), we get 
\begin{equation}
\int\nolimits_{0}^{\tau^{\ast}}\mathcal{L}(\psi)(X_{s^{-}}^{L},\lambda_{s^{-}%
})e^{-qs}ds\leq-\varepsilon q\int\nolimits_{0}^{\tau^{\ast}}e^{-qs}ds.
\label{nueva2}%
\end{equation}
From (\ref{Desig}), Lemma \ref{DPP}, (\ref{Vborde}), (\ref{nueva1}) and
(\ref{nueva2}) it follows that

\begin{equation}%
\begin{array}
[c]{lll}%
V(x_{0},\lambda_{0}) & = & \sup\nolimits_{L}\mathbb{E}\left(  \int_{0^{-}%
}^{\tau^{\ast}}e^{-qs}dL_{s}+e^{-q\tau^{\ast}}V(X_{\tau^{\ast}}^{L}%
,\lambda_{\tau^{\ast}})\right) \\
& \leq & \sup\nolimits_{L}\mathbb{E}\left(  \int_{0^{-}}^{\tau^{\ast}}%
e^{-qs}dL_{s}+e^{-q\tau^{\ast}}\left(  \psi(X_{\tau^{\ast}}^{L},\lambda
_{\tau^{\ast}})-\varepsilon\right)  I_{\tau^{\ast}<\tau^{L}}\right) \\
& \leq & \psi(x_{0},\lambda_{0})+\sup\nolimits_{L}\mathbb{E}\left(
\int\nolimits_{0}^{\tau^{\ast}}\mathcal{L}\left(  \psi\right)  (X_{s^{-}}%
^{L},\lambda_{s^{-}})e^{-qs}ds-\varepsilon e^{-q\tau^{\ast}}I_{\tau^{\ast
}<\tau^{L}}\right) \\
& \leq & \psi(x_{0},\lambda_{0})+\sup\nolimits_{L}\mathbb{E}\left(
-\varepsilon(1-e^{-q\tau^{\ast}})-\varepsilon e^{-q\tau^{\ast}}I_{\tau^{\ast
}<\tau^{L}}\right) \\
& \leq & \psi(x_{0},\lambda_{0})-\varepsilon+\varepsilon\mathbb{E}%
(e^{-q\left(  \tau_{1}\wedge T_{1}\right)  })\\[0.1cm]
& \leq & \psi(x_{0},\lambda_{0})-\varepsilon+\varepsilon\left(  \lambda
_{0}+\beta\right)  /\left(  q+\lambda_{0}+\beta\right) \\[0.1cm]
& < & \psi(x_{0},\lambda_{0}).
\end{array}
\label{Desig}%
\end{equation}
But the latter contradicts the assumption that $V(x_{0},\lambda_{0})=\psi
(x_{0},\lambda_{0})$.\smallskip\ Hence $V$ is a viscosity subsolution at
$(x_{0},\lambda_{0})$ and this complete the proof. \hfill$\blacksquare$\\

The next lemma will be used to prove Proposition {\ref{MenorSuper}. }
\begin{lemma}
\label{A.1} Let $\overline{u}$\ be a non-negative supersolution of (\ref{HJB})
satisfying the growth condition (\ref{Growth Condition}) and non-increasing in
$\lambda$.\ We can find a sequence of positive functions $\overline{u}%
^{m}:(0,\infty)\times(\underline{\lambda},\infty)\rightarrow\mathbb{R}$\ such that:

(a) $\overline{u}^{m}$\ is continuously differentiable.

(b) $\overline{u}^{m}(x,\lambda)\leq K+x$ and $\overline{u}^{m}$ is
non-increasing in $\lambda$.

(c)$\ 1\leq\overline{u}_{x}^{m}(x,\lambda)\leq(\left(  q+\lambda+\beta\right)
/p)\overline{u}(x+1/m,\lambda)$ for $(x,\lambda)\in (0,\infty)\times(\underline{\lambda
},\infty)$.

(d) $\overline{u}^{m}$\ $\rightarrow\overline{u}$\ uniformly on compact sets
in $(0,\infty)\times(\underline{\lambda},\infty)$ and $\nabla\overline{u}^{m}%
$\ converges to $\nabla\overline{u}$\ a.e. in $(0,\infty)\times(\underline
{\lambda},\infty)$.

(e) There exists a sequence $c_{m}$ with $\lim\limits_{m\rightarrow\infty
}c_{m}=0$ such that
\[
\sup\nolimits_{\left(  x,\lambda\right)  \in A_{0}}\mathcal{L}(\overline
{u}^{m})\left(  x,\lambda\right)  \leq c_{m},\text{where }A_{0}=[0,x_{0}%
]\times\lbrack\lambda_{0},\lambda_{1}],
\]
where $x_{0}>0$ and $\lambda_{0},\lambda_{1}\in(\underline{\lambda},\infty).$
\end{lemma}

\noindent\textit{Proof of Lemma \ref{A.1}. }Let us define the set
\[
D=\left\{  (x,\lambda)\in(0,\infty)\times(\underline{\lambda},\infty)\text{
s.t. }\overline{u}\text{ differentiable in }(x,\lambda)\right\}  .
\]
Since $\overline{u}$ is a supersolution of (\ref{HJB}), we have that
\begin{equation}
p\overline{u}_{x}(x,\lambda)-d\left(  \lambda-\underline{\lambda}\right)  \overline
{u}_{\lambda}(x,\lambda)\leq(q+\lambda+\beta)\overline{u}(x,\lambda)
\label{UesPositivoDividendos}%
\end{equation}
and $\overline{u}_{x}\geq1$ for all $(x,\lambda)\in D$ and so a.e. in
$(0,\infty)\times(\underline{\lambda},\infty).$

Let $\phi$ be a nonnegative continuously differentiable function with support
included in $(0,1)$ such that $\int_{0}^{1}\phi(x)dx=1$, we define
$\overline{u}^{m}:(0,\infty)\times(\underline{\lambda},\infty)\rightarrow
\mathbb{R}\ $ as the convolution
\begin{equation}
\overline{u}^{m}(x,\lambda)=m^{2}\int\nolimits_{-\infty}^{\infty}%
\int\nolimits_{-\infty}^{\infty}\overline{u}(x+s,\lambda+t)\phi(ms)\phi
(mt)dsdt. \label{arreglado1Dividendos}%
\end{equation}
By definition, $\overline{u}^{m}(x,\lambda)$ is a weighted average of values
of $\overline{u}$ in $A_{m}=[x,x+1/m]\times\lbrack\lambda,\lambda+1/m]$; (a),
(b) and (d) follow by standard techniques, because $\overline{u}^{m}%
\geq\overline{u}$ and $\overline{u}$ is absolutely continuous in
$(0,\infty)\times(\underline{\lambda},\infty)$ and satisfies the growth
condition (\ref{Growth Condition}) (see for instance \cite{WheedenZygmund}).
From Equation (\ref{HJB}) we have that $\overline{u}_{x}\geq1$ a.e., also
since for all $(x,\lambda)\in D$, $\mathcal{L}(\overline{u})\left(
x,\lambda\right)  \leq0$ \ we have%

\[
-(q+\lambda+\beta)\overline{u}(x,\lambda)\leq p\overline{u}_{x}(x,\lambda
)-d\left(  \lambda-\underline{\lambda}\right)  \overline{u}_{\lambda
}(x,\lambda)\leq(q+\lambda+\beta)\overline{u}(x,\lambda),
\]
so that we conclude (c).

Let us define for $(x,\lambda)\in A_{0}$ the function
\begin{equation}
\xi_{m}(x,\lambda)=\sup_{(x,\lambda)\in A_{m}\cap D}\left(  p\overline{u}%
_{x}(x,\lambda)-d\left(  \lambda-\underline{\lambda}\right)  \overline
{u}_{\lambda}(x,\lambda)\right)  \text{.} \label{arreglado2Dividendos}%
\end{equation}
We have that for all $(x,\lambda)\in D$
\begin{equation}
p\overline{u}_{x}(x,\lambda)-d\left(  \lambda-\underline{\lambda}\right)
\overline{u}_{\lambda}(x,\lambda)\leq\xi_{m}(x,\lambda)\leq\left(
q+\lambda+\beta\right)  \overline{u}(x+1/m,\lambda)\text{.}
\label{arreglado3Dividendos}%
\end{equation}
From (\ref{arreglado2Dividendos}), for any $\left(  x,\lambda\right)  \in$
$A_{0}$ there exists $(x_{m},\lambda_{m})\in A_{m}\cap D$ such that
\begin{equation}
p\overline{u}_{x}(x_{m},\lambda_{m})-d\left(  \lambda-\underline{\lambda
}\right)  \overline{u}_{\lambda}(x_{m},\lambda_{m})\geq\text{ }\xi
_{m}(x,\lambda)-\frac{1}{m}. \label{arreglado4Dividendos}%
\end{equation}
So, 
\[%
\begin{array}
[c]{l}%
p\overline{u}_{x}^{m}(x,\lambda)-d\left(  \lambda-\underline{\lambda}\right)
\overline{u}_{\lambda}^{m}(x,\lambda)-(p\overline{u}_{x}(x_{m},\lambda
_{m})-d\left(  \lambda-\underline{\lambda}\right)  \overline{u}_{\lambda
}(x_{m},\lambda_{m}))\text{ }\\[0.1cm]
=m^{2}%
{\textstyle\iint\nolimits_{D}}
\left(  p\overline{u}_{x}(x+s,\lambda+t)-d\left(  \lambda+t-\underline
{\lambda}\right)  \overline{u}_{\lambda}(x+s,\lambda+t)\right)  \phi
(ms)\phi(mt)dsdt\\[0.1cm]
-(p\overline{u}_{x}(x_{m},\lambda_{m})-d\left(  \lambda_{m}-\underline
{\lambda}\right)  \overline{u}_{\lambda}(x_{m},\lambda_{m}))\\[0.1cm]
\leq\xi_{m}(x,\lambda)-(\xi_{m}(x,\lambda)-\frac{1}{m})\\[0.1cm]
\leq\frac{1}{m}.
\end{array}
\]

From (\ref{UesPositivoDividendos}), (\ref{arreglado3Dividendos}),
(\ref{arreglado4Dividendos}) and using that $\overline{u}$ and $\overline
{u}^{m}$ are uniformly continuous in compact sets and that it is
non-decreasing in $\lambda,$ we have the result. \hfill$\blacksquare$

\bigskip

\noindent\textit{Proof of Proposition \ref{MenorSuper}.}
\textit{\ }Let $\overline{u}$\ be a non-negative supersolution of (\ref{HJB})
satisfying the growth condition (\ref{Growth Condition}) and take an
admissible strategy $L\in\Pi_{x,\lambda}$, define $\left(  X_{t}^{L}%
,\lambda_{t}\right)  $ as the corresponding controlled risk process starting
at $\left(  x,\lambda\right)  $. Let us consider the function $\overline
{u}^{m}$ as defined in Lemma \ref{A.1} in $(0,\infty)\times(\underline
{\lambda},\infty)$ and let us extend this function as $\overline{u}%
^{m}(x,\lambda)=0$ otherwise, as in (\ref{nueva1}) in the proof of Proposition
\ref{Prop V is a viscosity solution}, we obtain (using $\overline{u}_{x}%
^{m}\geq1)$,%

\begin{equation}%
\begin{array}
[c]{l}%
\mathbb{E}\left(  \overline{u}^{m}(X_{t\wedge\tau^{L}}^{L},\lambda
_{t\wedge\tau^{L}})e^{-q(t\wedge\tau^{L})}\right)  -\overline{u}^{m}%
(x,\lambda)\\
\leq\mathbb{E}\left(  \int\nolimits_{0}^{t\wedge\tau^{L}}\mathcal{L}%
(\overline{u}^{m})(X_{s^{-}}^{L},\lambda_{s^{-}})e^{-qs}ds\right)
-\mathbb{E}\left(  \int\nolimits_{0^{-}}^{t\wedge\tau^{L}}e^{-qs}%
dL_{s}\right)  \text{.}%
\end{array}
\label{ItoUnMenorSuper}%
\end{equation}
Since $L_{t}$ is a non-decreasing process we get, by the monotone
convergence theorem, that
\begin{equation}
\lim\limits_{t\rightarrow\infty}\mathbb{E}\left(  \int\nolimits_{0^{-}%
}^{t\wedge\tau^{L}}e^{-qs}dL_{s}\right)  =J(L;x,\lambda).\nonumber
\end{equation}
From Lemma \ref{A.1}(c), we have%

\begin{equation}
-(q+\lambda+\beta)\overline{u}(x+\frac{1}{m},\lambda)\leq\mathcal{L}%
(\overline{u}^{m})(x,\lambda)\leq\left(  q+\lambda+\beta\right)  \overline
{u}(x+\frac{1}{m},\lambda)+\beta\overline{u}(x,\lambda). \label{acotacion L}%
\end{equation}
But using Lemma \ref{A.1}(b) and the inequalities $X_{s}^{L}\leq x+ps$,
$\lambda_{s}>\underline{\lambda}$ we obtain
\begin{equation}
\overline{u}(X_{s}^{L},\lambda_{s})\leq K+ps. \label{Acotacionu}%
\end{equation}
So, by the bounded convergence theorem, we get
\begin{equation}
\lim\limits_{t\rightarrow\infty}\mathbb{E}\left(  \int\nolimits_{0}%
^{t\wedge\tau^{L}}\mathcal{L}(\overline{u}^{m})(X_{s^{-}}^{L},\lambda_{s^{-}%
})e^{-qs}ds\right)  =\mathbb{E}\left(  \int\nolimits_{0}^{\tau^{L}}%
\mathcal{L}(\overline{u}^{m})(X_{s^{-}}^{L},\lambda_{s^{-}})e^{-qs}ds\right)
. \label{monotone2}%
\end{equation}
From (\ref{ItoUnMenorSuper}) and (\ref{monotone2}), we have
\begin{equation}%
\begin{array}
[c]{l}%
\lim\limits_{t\rightarrow\infty}\mathbb{E}\left(  \overline{u}^{m}%
(X_{t\wedge\tau^{L}}^{L},\lambda_{t\wedge\tau^{L}})e^{-q(t\wedge\tau^{L}%
)}I_{\tau^{L}<t}\right)  -\overline{u}^{m}(x,\lambda)\\
=-\overline{u}^{m}(x,\lambda)\\
\leq\mathbb{E}\left(  \int\nolimits_{0}^{\tau^{L}}\mathcal{L}(\overline{u}%
^{m})(X_{s^{-}}^{L},\lambda_{s^{-}})e^{-qs}ds\right)  -J(L;x,\lambda).
\end{array}
\label{limite0}%
\end{equation}
Let us prove now that
\begin{equation}
\limsup\limits_{m\rightarrow\infty}\mathbb{E}\left(  \int\nolimits_{0}%
^{\tau^{L}}\mathcal{L}(\overline{u}^{m})(X_{s^{-}}^{L},\lambda_{s^{-}}%
)e^{-qs}ds\right)  \leq0. \label{limite2}%
\end{equation}
Because $\overline{u}^{m}(x,\lambda)\leq K+x$, Lemma \ref{A.1}(b) and Lemma \ref{A.1}(b)(c) give%
\[%
\begin{array}
[c]{lll}%
\left\vert \mathcal{L}(\overline{u}^{m})(X_{s^{-}}^{L},\lambda_{s^{-}%
})\right\vert  & \leq & \left(  q+\lambda+\beta\right)  \overline{u}(X_{s^{-}%
}^{L}+\frac{1}{m},\lambda_{s^{-}})+\beta\overline{u}(X_{s^{-}}^{L}%
,\lambda_{s^{-}})\\[0.1cm]
& \leq & \left(  q+\lambda+\beta\right)  \overline{u}(X_{s^{-}}^{L}+\frac
{1}{m},\underline{\lambda})+\beta\overline{u}(X_{s^{-}}^{L},\underline
{\lambda})\\[0.1cm]
& \leq & \left(  q+\lambda+2\beta\right)  \left(  x+\frac{1}{m}+ps\right).
\end{array}
\]
Then, given any $\varepsilon>0,$ we can find $T$ such that
\begin{equation}
\int\nolimits_{T}^{\infty}\mathcal{L}(\overline{u}^{m})(X_{s^{-}}^{L}%
,\lambda_{s^{-}})e^{-qs}ds<\frac{\varepsilon}{4} \label{arreglado5}%
\end{equation}
for any $m\geq1.$ Note that for $s\leq T$, we have that $X_{s^{-}}^{L}$ $\leq
x_{0}:=x+pT$ , $\lambda_{s^{-}}$ $\geq\lambda_{1}:=\underline{\lambda}+\left(
\lambda-\underline{\lambda}\right)  e^{-dt}$. \ From (\ref{Lambda t corrida}),
there exists $M$ large enough such that
\[
\mathbb{P}(\sup_{s\leq T}\lambda_{s}>M)<\frac{\varepsilon}{4\left(
q+\lambda+2\beta\right)  \left(  x+\frac{1}{m}+ps\right)  }.
\]

Now, take the compact set $\left[  0,x_{0}\right]  \times\left[  \lambda
_{1},M\right]  $ and the set%

\[
A^{T,M}=\left\{  \left(  \lambda_{s}\right)  _{s\leq T}:\sup_{s\leq T}%
\lambda_{s}\leq M\right\}  .
\]
From Lemma \ref{A.1}(e), we can find $m_{0}$ large enough such that for any
$m\geq m_{0},$%
\[
I_{A^{T,M}}\int\nolimits_{0}^{T}\mathcal{L}(\overline{u}^{m})(X_{s^{-}}%
^{L},\lambda_{s^{-}})e^{-qs}ds\leq c_{m}\int\nolimits_{0}^{T}e^{-qs}%
ds\leq\frac{c_{m}}{q}\leq\frac{\varepsilon}{2},
\]
and so taking the expectation we get (\ref{limite2}). Then, from
(\ref{limite0}) and using Lemma \ref{A.1}(d), we finally obtain%
\begin{equation}
\overline{u}(x,\lambda)=\lim\limits_{m\rightarrow\infty}\overline{u}%
^{m}(x,\lambda)\geq J(L;x,\lambda). \label{supersolucionMayorqueVL}%
\end{equation}
Since $V$ is a viscosity solution of (\ref{HJB}), the result follows.\hfill 
$\blacksquare$

\subsection{Proofs of Section \ref{Section Asymptotic} \label{Appendix Asymptotic}%
}

\textit{Proof of Proposition \ref{Proposicion acotacion de retirar}.} 
Let us define for $M>0$,%
\[
V_{M}(x,\lambda)=\left\{
\begin{array}
[c]{lll}%
V(x,\lambda) & \text{if} & x\leq M,\\
V(M,\lambda)+x-M & \text{if} & x>M.
\end{array}
\right.
\]
$V_{M}(x,\lambda)$ is a limit of value functions of strategies. From
Propositions \ref{Continuidad Uniforme en Lambda}, \ref{Locally Lipschitz x}
and \ref{Condicion de Lipschitz en Lambda}, $V_{M}$ is locally Lipschitz,
increasing on $x$ and non-increasing on $\lambda$. For $x<M$, $V_{M}$ is a
viscosity supersolution of (\ref{HJB}). For $x>M$ , $\partial_{x}%
V_{M}(x,\lambda)=1$. In order to see that $V^{M}$ is a viscosity supersolution
of (\ref{HJB}) for $x>M$ and $M$ large enough, we need to show that
$\mathcal{L}(V_{M})(x,\lambda)\leq0$ for $x>M$. Indeed,%

\[%
\begin{array}
[c]{l}%
\mathcal{L}(V_{M})(x,\lambda)\\[0.1cm]
\leq p-(q+\lambda)V_{M}(x,\lambda)+\lambda%
{\textstyle\int\nolimits_{0}^{x}}
V_{M}(x-\alpha,\lambda)dF_{U}(\alpha)+\beta\left(
{\textstyle\int\nolimits_{0}^{\infty}}
V_{M}(x,\lambda)dF_{Y}(\gamma)-V_{M}(x,\lambda)\right) \\[0.1cm]
\leq p-(q+\lambda)V_{M}(x,\lambda)+\lambda%
{\textstyle\int\nolimits_{0}^{x}}
V_{M}(x-\alpha,\lambda)dF_{U}(\alpha)\\[0.1cm]
\leq p-(q+\lambda)(V(M,\lambda)+x-M)+\lambda%
{\textstyle\int\nolimits_{0}^{x}}
(V(M,\lambda)+x-\alpha-M)dF_{U}(\alpha)\\[0.1cm]
\leq p-qV(M,\lambda)\\
\leq0
\end{array}
\]
for any $M\geq p/q$ because $V(x,\lambda)\geq x$. Consider the following dense
set $B$ in $\left[  0,\infty\right)  \times\left[  \underline{\lambda}%
,\infty\right)  $ 
\[
B:=\{(x,\lambda)\in\left[  0,\infty\right)  \times\left[  \underline{\lambda
},\infty\right)  :V~\text{is differentiable}\}.
\]
For any point $(M,\lambda)\in B$ with $M\geq\frac{p}{q}$, we have that
$\partial_{x}V_{M}(M,\lambda)=1$ and so, by Corollary
\ref{VerificacionDividendos}, $V_{M}$ coincides with $V.$
\hfill$\blacksquare$\\

\noindent\textit{Proof of Proposition \ref{Convergencia Uniforme con Lambda}.} 
From Proposition \ref{Proposicion acotacion de retirar}, $V(x,\lambda
)=x+A(\lambda)$ for $x\geq p/q$, where
\[
A(\lambda):=V(p/q,\lambda)-p/q\text{, }%
\]
and since from Proposition \ref{Convergencia puntual con lambda a infinito},
$\lim_{\lambda\rightarrow\infty}V(p/q,\lambda)-p/q=0$ and $V(p/q,\lambda)$ is
non-increasing on $\lambda$, this implies $\lim_{\lambda\rightarrow\infty
}A(\lambda)\searrow0$ and so
\[
\lim_{\lambda\rightarrow\infty}\left(  \sup_{x\geq p/q}V(x,\lambda)-x\right)
=0.
\]
Also, for all $x\geq0$ such that $V_{x}(x,\lambda)$ exists, $V_{x}%
(x,\lambda)\geq1$ and so $h(x,\lambda):=V(x,\lambda)-x$ is non-decreasing on
$x$. Hence,
\[
V(x,\lambda)-x\leq V(p/q,\lambda)-p/q\leq A(\lambda)
\]
and that implies the result. \hfill$\blacksquare$

\subsection{Proofs of Section \ref{Section Approximation Disc Surplus}
\label{Appendix Disc Surplus}}

\textit{Proof of Proposition \ref{Lipschitz Vdelta}}.
We have that%
\[
V^{\delta}(x_{n+1}^{\delta},\lambda)-V^{\delta}(x_{n}^{\delta},\lambda)\leq
V^{\delta}(x_{n}^{\delta})e^{\left(  \beta+q\right)  \delta+\int_{0}^{\delta
}\lambda_{u}^{c}du}-V^{\delta}(x_{n}^{\delta}).
\]
Then, denoting $x_{n}^{\delta}=\rho^{\delta}(x_{2})$ and $x_{m}^{\delta}$
$=\rho^{\delta}(x_{1}),$%

\[%
\begin{array}
[c]{lll}%
V^{\delta}(x_{2},\lambda)-V^{\delta}(x_{1},\lambda) & \leq & V^{\delta}%
(\rho^{\delta}(x_{2}),\lambda)-V^{\delta}(\rho^{\delta}(x_{1}),\lambda
)+(x_{2}-\rho^{\delta}(x_{2}))\\[0.1cm]
& \leq & V^{\delta}(x_{n}^{\delta},\lambda)-V^{\delta}(x_{m}^{\delta}%
,\lambda)+\delta p\\
& \leq & \sum\nolimits_{j=m}^{n-1}\left(  e^{\left(  \beta+q\right)
\delta+\int_{0}^{\delta}\lambda_{u}^{c}du}-1\right)  V^{\delta}(x_{j}^{\delta
},\lambda)+\delta p\\
& \leq & \left(  \rho^{\delta}(x_{2})-\rho^{\delta}(x_{1})\right)
\frac{\left(  e^{\left(  \beta+q\right)  \delta+\int_{0}^{\delta}\lambda
_{u}^{c}du}-1\right)  }{p\delta}V^{\delta}(x_{2},\lambda)+\delta p.
\end{array}
\]

With a similar proof to the one of Proposition
\ref{Continuidad Uniforme en Lambda}, we have that $V^{\delta}(x,\cdot)$ is
non-increasing and so $0\leq V^{\delta}(x,\lambda_{1})-V^{\delta}%
(x,\lambda_{2}).$ It suffices to prove 
\begin{equation}\label{that}
V^{\delta}(x_{n}^{\delta},\lambda_{1})-V^{\delta}(x_{n}^{\delta},\lambda
_{2})\leq V^{\delta}(x_{n}^{\delta},\lambda_{1})\frac{\beta\lambda_{2}%
+q}{d(\lambda_{1}-\underline{\lambda})}\left(  \lambda_{2}-\lambda_{1}\right)
\end{equation}
for $0<\lambda_{2}-\lambda_{1}$ small enough.\\
In order to prove \eqref{that}, we modify the set of admissible
strategies $\widetilde{\Pi}_{x_{n}^{\delta},\lambda}^{\delta}$, adding new
local strategies which are not optimal. Let us define $\widehat{\Pi}%
_{x_{n}^{\delta},\lambda}^{\delta}$ as the set of all the strategies
with initial surplus $x_{n}^{\delta}\in\mathcal{G}_{\delta}$ which can be
obtained by a sequence of control actions in
\[
\widehat{\mathcal{E}}=\{\mathbf{E}_{F},\mathbf{E}_{1},\mathbf{E}_{0}%
,\widehat{\mathbf{E}}_{\eta}\},
\]
where the controls $\mathbf{E}_{F}$,$\ \mathbf{E}_{1}\ $and$~\mathbf{E}_{0}$
are defined in (\ref{Conjunto E de Estrategias.}). For any time $\eta>0, $
the new control action $\widehat{\mathbf{E}}_{\eta}$ consists of throwing away
the incoming premium $p$ up to time $\eta\wedge\tau_{1}\wedge T_{1}$ (note
that if $\eta<\tau_{1}\wedge T_{1}$ the final surplus and intensity of
this local control action is $(x_{n}^{\delta},\underline{\lambda}%
+(\lambda-\underline{\lambda})e^{-d\eta})\in\mathcal{G}_{\delta}\times
\lbrack0,\infty)$).
Since it is never optimal to throw away money,
\[
\sup\nolimits_{\pi\in\widehat{\Pi}_{x_{n}^{\delta},\lambda}^{\delta}}%
J(\pi;\left(  x_{n}^{\delta},\lambda\right)  )=\sup\nolimits_{\pi\in
\widetilde{\Pi}_{x_{n}^{\delta},\lambda}^{\delta}}J(\pi;\left(  x_{n}^{\delta
},\lambda\right)  )=V^{\delta}(x_{n}^{\delta},\lambda).
\]
Given any $\varepsilon>0$, take $\pi=(L,\tau_{F})\in\widetilde{\Pi}%
_{x_{n}^{\delta},\lambda_{1}}^{\delta}$ such that $V^{\delta}(x_{n}^{\delta
},\lambda_{1})-J((L,\tau_{F});(x_{n}^{\delta},\lambda_{1}))<\varepsilon/2$. By
Remark \ref{Remark Estrategias Extendidas} we can take $\tau_{F}=\infty$, and
the associated controlled process
\[
(X_{t}^{L},\lambda_{t})=\left(  x_{n}^{\delta}+pt-\sum\limits_{i=1}^{N_{t}%
}U_{i}-L_{t},\underline{\lambda}+e^{-dt}\left(  \lambda_{1}-\underline
{\lambda}\right)  +\sum_{0\leq T_{k}\leq\overline{t}}Y_{n}~~e^{-d(t-T_{k}%
)}\right).
\]
Consider $\varkappa$ such that 
\begin{equation}
\underline{\lambda}+e^{-d\varkappa}\left(
\lambda_{2}-\underline{\lambda}\right)  =\lambda_{1},  \label{Cota de delta0}%
\end{equation}
that is
\[%
\varkappa=\frac{1}{d}\log(\frac{\lambda_{2}-\underline{\lambda}}{\lambda
_{1}-\underline{\lambda}})=\frac{1}{d}\log(1+\frac{\lambda_{2}-\lambda_{1}}{\lambda
_{1}-\underline{\lambda}})\leq\frac{1}{d(\lambda_{1}-\underline{\lambda}%
)}\left(  \lambda_{2}-\lambda_{1}\right)  .
\]
Define now $\widehat{\pi}$ as the strategy to apply first the local control $\widehat
{\mathbf{E}}_{\varkappa}$ and then either $\pi=(L,\tau
_{F})\in\widetilde{\Pi}_{x_{n}^{\delta},\lambda_{1}}^{\delta}$ in case $\varkappa$ $<\tau_{1}\wedge T_{1}$, or $\mathbf{E}_{F}\ $otherwise. So
$\widehat{\pi}\in\widehat{\Pi}_{x_{n}^{\delta},\lambda_{2}}^{\delta}$ and
$\ $
\[
\widehat{L}_{t}=\left\{
\begin{array}
[c]{lll}%
0 & \text{if} & t\leq\varkappa\text{ }\wedge\tau_{1}\wedge T_{1}\\
L_{t-\varkappa} & \text{if} & t\geq\varkappa\text{ and }\tau_{1}\wedge
T_{1}>\varkappa
\end{array}
\right.  .%
\]
One has
\[%
\begin{array}
[c]{lll}%
J(\widehat{\pi};\left(  x_{n}^{\delta},\lambda_{2}\right)  ) & \geq &
J(\widehat{\pi};x_{n}^{\delta},\lambda_{1})e^{-q\varkappa}\mathbb{P}(\tau
_{1}\wedge T_{1}\geq\varkappa)\\[0.1cm]
& \geq & J(\widehat{\pi};x_{n}^{\delta},\lambda_{1}))e^{-q\varkappa
}(1-(1-e^{-\beta\varkappa)})(1-e^{-\int_{0}^{\varkappa}\text{ }\left(
\underline{\lambda}+e^{-ds}\left(  \lambda_{2}-\underline{\lambda}\right)
\right)  ds}))\\[0.1cm]
& \geq & J(\widehat{\pi};x_{n}^{\delta},\lambda_{1}))e^{-q\varkappa
}(1-(1-e^{-\beta\varkappa)})(1-e^{-\lambda_{2}\varkappa}))\\[0.1cm]
& \geq & J(\widehat{\pi};x_{n}^{\delta},\lambda_{1}))\left(  1-q\varkappa
\right)  (1-\beta\lambda_{2}\varkappa)\allowbreak
\end{array}
\
\]
because
\[
\int_{0}^{\varkappa}\text{ }\left(  \underline{\lambda}+e^{-ds}\left(
\lambda_{2}-\underline{\lambda}\right)  \right)  ds\leq\lambda_{2}\varkappa.
\]
Then, using that $\lambda_{2}-\lambda_{1}\leq d(\lambda_{1}-\underline
{\lambda})$, we get $\varkappa<1.$ So, from (\ref{Cota de delta0}),
\[%
\begin{array}
[c]{lll}%
V^{\delta}(x_{n}^{\delta},\lambda_{1})-V^{\delta}(x_{n}^{\delta},\lambda
_{2}) & \leq & J(\pi;x_{n}^{\delta},\lambda_{1})-J(\widehat{\pi};x_{n}%
^{\delta},\lambda_{2})+\varepsilon/2\\[0.1cm]
& \leq & J(\pi;x_{n}^{\delta},\lambda_{1})-J(\pi;x_{n}^{\delta},\lambda
_{1})\left(  1-q\varkappa\right)  (1-\beta\lambda_{2}\varkappa)\allowbreak
+\varepsilon/2\\[0.1cm]
& \leq & V(x_{n}^{\delta},\lambda_{1})\allowbreak\varkappa\left(  \beta
\lambda_{2}\varkappa+q\right)  +\varepsilon/2\\[0.1cm]
& \leq & V(x,\lambda_{1})\left(  \beta\lambda_{2}\frac{\left(  \lambda
_{2}-\lambda_{1}\right)  }{d(\lambda_{1}-\underline{\lambda})}+q\right)
\allowbreak\frac{\left(  \lambda_{2}-\lambda_{1}\right)  }{d(\lambda
_{1}-\underline{\lambda})}+\varepsilon/2,
\end{array}
\]
so that the result follows. \hfill$\blacksquare$\\

In order to prove Propositions \ref{Limite de Vl} and
{\ref{Menor Supersolucion Discreta}}, we need the following lemma.

\begin{lemma}
\label{Lemma tk a infinito}Given $\pi\in\widetilde{\Pi}_{x_{n}^{\delta
},\lambda}^{\delta},$ $\pi$ can be obtained by a sequence of control actions
$\mathbf{s}=(s_{k})_{k=1,...,\tilde{k}}$, where $\tilde{k}$ could be finite or
infinite. Let us define $t_{k}$ as the time when the control action $s_{k}$ is
applied. We have that $\lim_{k\rightarrow\infty}t_{k}=\infty$ a.s. within the
subset $\{\tilde{k}=\infty\}\subset\Omega. $
\end{lemma}

\noindent\textit{Proof of Lemma \ref{Lemma tk a infinito}.} Suppose that$\ \tilde{k}=\infty$. Calling $k_{m}=n+m$, let $i_{m}$ be the
number of control actions $\mathbf{E}_{0}$ in $\left(  s_{1},s_{2}%
,...,s_{k_{m}}\right)  $, then $i_{m}\geq m$. Given the two sequences of
stopping times $(\tau_{i})_{i\geq1}\ $and$~(T_{j})_{j\geq1}$, we define the
ordered union of the two sequences as
\[
(\tau_{i})_{i\geq1}\amalg(T_{j})_{j\geq1}:=(r_{h})_{h\geq1},%
\]
where $\left\{  r_{h}:h\geq1\right\}  =\left\{  \tau_{i}:i\geq1\right\}
\cup\left\{  T_{j}:j\geq1\right\}  $ and $r_{h}\leq r_{h+1}.$ We have that
$\lim_{h\rightarrow\infty}r_{h}=$ $\infty$ a.s. Let us consider the
non-decreasing sequence $(j_{m})$ defined as $j_{m}=\max\{h:r_{h}\leq
t_{k_{m}}\}$, then we have that $t_{k_{m}}\geq\max\{r_{j_{m}},(i_{m}-j_{m}%
-1)\delta\}$. If $\lim_{m\rightarrow\infty}i_{m}-j_{m}=\infty$, then
\[
\lim_{m\rightarrow\infty}t_{k_{m}}\geq\lim_{m\rightarrow\infty}(i_{m}%
-j_{m}-1)\delta=\infty;
\]
if not, $\lim_{m\rightarrow\infty}j_{m}=\infty$ and so%

\[
\lim_{m\rightarrow\infty}t_{k_{m}}\geq\lim_{m\rightarrow\infty}r_{j_{m}}%
\]
and since $\lim_{m\rightarrow\infty}r_{j_{n}}=$ $\lim_{i\rightarrow\infty
}r_{i}=$ $\infty$ a.s., since $t_{k+1}\geq t_{k}$, we have $\lim
_{m\rightarrow\infty}t_{k}=\infty$ a.s. \hfill$\blacksquare$\\

\noindent\textit{Proof of Proposition \ref{Limite de Vl}.} Let us define $W=\lim_{l\rightarrow\infty}V_{l}^{\delta}$ and let us show that
$W(x_{n}^{\delta},\lambda)=V^{\delta}(x_{n}^{\delta},\lambda).$
Given $(x_{n}^{\delta},\lambda)$ and $\varepsilon>0$, take $\pi=(L,\tau
_{F})\in\widetilde{\Pi}_{x_{n}^{\delta},\lambda}^{\delta}$ such that
$V^{\delta}(x_{n}^{\delta},\lambda)-J(\pi;x_{n}^{\delta},\lambda
)<\varepsilon/2$. $\pi$ can be obtained by a sequence of control actions
$\mathbf{s}=(s_{k})_{k=1,...,\tilde{k}}$, where $\tilde{k}$ could be finite or
infinite. Let us define $t_{k}$ as the time when the control action $s_{k}$ is
applied. By Lemma \ref{Lemma tk a infinito}, $\lim_{k\rightarrow\infty}%
t_{k}=\infty$ a.s. within the subset $\{\tilde{k}=\infty\}\subset\Omega$. Let
us take $l$ large enough such that $e^{-qt_{l}}\sup_{x\geq0}\left(
V(x,\lambda)-x\right)  \leq e^{-qt_{l}}\frac{p}{q}<\varepsilon/2$ and consider
$\pi^{l}\in\widetilde{\Pi}_{x_{n}^{\delta},\lambda}^{\delta,l}$, defined by
the sequence $\left(  s_{1},s_{2},s_{3},\ldots,s_{l-1},\mathbf{E}_{F}\right)
$ if $\tilde{k}\geq l $ and by $\mathbf{s}$ otherwise. We have that
\[%
\begin{array}
[c]{lll}%
J(\pi;x_{n}^{\delta},\lambda)-J(\pi^{l};x_{n}^{\delta},\lambda) & = &
\mathbb{E}\left(  \mathbb{E}(\left.  I_{\tau^{L}\wedge\tau_{F}>t_{l}}%
\int_{t_{l}}^{\tau^{L}}e^{-qs}dL_{s}+I_{\tau^{L}\wedge\tau_{F}>t_{l}%
}I_{\left\{  \tau^{F}<\tau^{L}\right\}  }e^{-q\tau^{F}}X_{\tau^{F}}%
^{L})-e^{-qt_{l}}X_{t_{l}}^{L})\right\vert \mathcal{F}_{t_{l}})\right) \\[0.1cm]
& \leq & e^{-qt_{l}}\mathbb{E}\left(  V(X_{t_{l}}^{L},\underline{\lambda
})-X_{t_{l}}^{L}\right) \\[0.1cm]
& < & \varepsilon/2,
\end{array}
\]
and so $V^{\delta}(x_{n}^{\delta},\lambda)$ $-V_{l}^{\delta}(x_{n}^{\delta
},\lambda)\leq\varepsilon.~$

Finally, since $V_{l}^{\delta}(x_{n}^{\delta},\lambda)\nearrow V^{\delta
}(x_{n}^{\delta},\lambda)$, from Proposition \ref{Properties of Vl}, we get
that $\mathcal{T}(V^{\delta})(x_{n}^{\delta},\lambda)=$ $V^{\delta}%
(x_{n}^{\delta},\lambda)$. \hfill$\blacksquare$

\bigskip

\noindent\textit{Proof of Proposition \ref{Menor Supersolucion Discreta}}. Assume that
$\pi=(L,\tau_{F})\in\Pi_{x_{m}^{\delta},\lambda}^{\delta}$. For any
$\omega=(\tau_{i},U_{i})_{i\geq1},(T_{j},Y_{j})_{j\geq1}$, consider the
sequence $\mathbf{s}=(s_{k})_{k=1,...,\tilde{k}}$ with $s_{k}\in\mathcal{E}$
corresponding to $\pi.$ Let $x_{m^{k}}^{\delta}\in\mathcal{G}_{\delta}\ $and
$\lambda^{k}\geq\underline{\lambda}$ be the surplus and the intensity in which
the control action $s_{k}$ is applied, $t_{k}$ be the time at which the
control action $s_{k}$ is chosen, and let $y^{k}$ be the end surplus resulting
from the control action $s_{k}$. Denote by $\left(  \kappa_{l}\right)
_{l\geq1}$ the indices of the sequence $\mathbf{s}=(s_{k})_{k=1,...,\tilde{k}%
}$, where $s_{k}$ is either $\mathbf{E}_{F}$ or $\mathbf{E}_{0}$\textbf{. }If
the sequence stops at $\tilde{k}=\kappa_{l_{0}}<\infty$, we define
\[
\kappa_{l}=\kappa_{l_{0}}\text{ for }l\geq l_{0},\text{ }t_{\kappa_{l_{0}+j}%
}=t_{\kappa_{l_{0}}}+\Delta_{\kappa_{l_{0}}}\text{ for }j\geq1;
\]
if $\tilde{k}=\infty$ we put $l_{0}=\infty$. We define, for $l\geq1$,
\[
H(l)=W(x_{m^{1+\kappa_{l}}}^{\delta},\lambda^{1+\kappa_{l}})I_{\{s_{\kappa
_{l}}=\mathbf{E}_{0}\}}I_{\{y^{k}\geq0\}}+x_{m^{\kappa_{l}}}^{\delta
}I_{\{s_{\kappa_{l}}=\mathbf{E}_{F}\}}\text{.}%
\]
If we put $H(0)=W(x_{m}^{\delta},\lambda)$, $\kappa_{0}=0$ and $t_{0}=0$, we
have, using $T_{1}(W)-W\leq0$,

\begin{equation}%
\begin{array}
[c]{lll}%
e^{-qt_{\kappa_{l+1}}}H(l)-W(x_{m}^{\delta},\lambda) & = & \sum_{j=1}%
^{l}(e^{-qt_{\kappa_{j+1}}}H(j)-e^{-qt_{\kappa_{j}}}H(j-1))\\[0.1cm]
& = & \sum_{j=1}^{l}I_{\{\kappa_{j+1}\neq\kappa_{j}\}}(e^{-qt_{\kappa_{j+1}}%
}H(j)-e^{-qt_{\kappa_{j}}}H(j-1))\\[0.1cm]
& = & \sum_{j=1}^{l}I_{\{\kappa_{j+1}\neq\kappa_{j}\}}(e^{-qt_{1+\kappa_{j-1}%
}}(\sum_{k=1+\kappa_{j-1}}^{\kappa_{j}-1}\left(  W(x_{m^{k}}^{\delta}%
-p\delta)-W(x_{m^{k}}^{\delta})\right)  ))\\[0.1cm]
&  & +\sum_{j=1}^{l}I_{\{\kappa_{j+1}\neq\kappa_{j}\}}(e^{-qt_{\kappa_{j+1}}%
}H(j)-e^{-qt_{\kappa_{j}}}W(x_{m^{k_{j}}}^{\delta}))\\[0.1cm]
& \leq & \sum_{j=1}^{l}I_{\{\kappa_{j+1}\neq\kappa_{j}\}}(\sum_{k=1+\kappa
_{j-1}}^{\kappa_{j}-1}e^{-qt_{1+\kappa_{j-1}}}(-p\delta I_{\{s_{k}%
=\mathbf{E}_{i}\}}))\\[0.1cm]
&  & +\sum_{j=1}^{l}I_{\{\kappa_{j+1}\neq\kappa_{j}\}}(e^{-qt_{\kappa_{j+1}}%
}H(j)-e^{-qt_{\kappa_{j}}}W(x_{m^{k_{j}}}^{\delta})).
\end{array}
\label{Desigualdad 1}%
\end{equation}
Since $T_{0}(W)-W\leq0$ and $T_{F}(W)-W\leq0,$ if $\kappa_{j+1}\neq
\kappa_{j}$,%
\begin{equation}%
\begin{array}
[c]{l}%
\mathbb{E}\left(  \left.  e^{-qt_{\kappa_{j+1}}}H(j)-e^{-qt_{\kappa_{j}}%
}W(x_{m^{k_{j}}}^{\delta},\lambda^{k_{j}})\right\vert \mathcal{F}%
_{t_{\kappa_{j}}}\right) \\%
\begin{array}
[c]{ll}%
= & \mathbb{E}\left(  \left.  (e^{-qt_{\kappa_{j+1}}}H(j)-e^{-qt_{\kappa_{j}}%
}W(x_{m^{k_{j}}}^{\delta},\lambda^{k_{j}}))I_{\{s_{\kappa_{j}}=\mathbf{E}%
_{0}\}}\right\vert \mathcal{F}_{t_{\kappa_{j}}}\right)  +I_{\{s_{\kappa_{j}%
}=\mathbf{E}_{F}\}}e^{-qt_{\kappa_{j}}}\left(  x_{m^{k_{j}}}^{\delta
}-W(x_{m^{k_{j}}}^{\delta},\lambda^{k_{j}})\right) \\
\leq & \mathbb{E}\left(  \left.  e^{-qt_{\kappa_{j+1}}}I_{\{s_{\kappa_{j}%
}=\mathbf{E}_{0}\}}(W(x_{m^{k_{j}+1}}^{\delta},\lambda^{k_{j}+1}%
)I_{\{\mathbf{y}^{\kappa_{j}}\geq0\}})\right\vert \mathcal{F}_{t_{\kappa_{j}}%
}\right)  -e^{-qt_{\kappa_{j}}}W(x_{m^{k_{j}}}^{\delta},\lambda^{k_{j}%
})I_{\{s_{\kappa_{j}}=\mathbf{E}_{0}\}}\\
= & e^{-qt_{\kappa_{j}}}I_{\{s_{\kappa_{j}}=\mathbf{E}_{0}\}}\left(
T_{0}(W)\left(  x_{m^{k_{j}}}^{\delta},\lambda^{k_{j}}\right)  -W(x_{m^{k_{j}%
}}^{\delta},\lambda^{k_{j}})\right) \\
\leq & 0.
\end{array}
\end{array}
\label{Desigualdad 2 Nueva}%
\end{equation}
From (\ref{Desigualdad 1}) and (\ref{Desigualdad 2 Nueva}), we have
\[
\lim\sup_{l\rightarrow\infty}\mathbb{E}\left(  e^{-qt_{\kappa_{l+1}}%
}H(l)-W(x_{m}^{\delta},\lambda)\right)  \leq-\mathbb{E}\left(  \int
_{0-}^{\left(  \tau_{L}\wedge\tau_{F}\right)  ^{-}}e^{-qs}dL_{s}\right)  .
\]
Consequently,%
\[
W(x_{m}^{\delta},\lambda)\geq J(\pi;x_{m}^{\delta},\lambda)+\lim
\sup_{l\rightarrow\infty}\mathbb{E}\left(  e^{-qt_{\kappa_{l+1}}}\left(
W(x_{m^{1+\kappa_{l}}}^{\delta},\lambda^{1+\kappa_{l}})I_{\{s_{\kappa_{l}%
}=\mathbf{E}_{0}\}}I_{\{y^{k}\geq0\}}\right)  \right)  .
\]
Since $W$ satisfies the growth condition (\ref{Growth Condition}), by Lemma
\ref{Lemma tk a infinito},
\[
\lim\sup_{l\rightarrow\infty}\mathbb{E}\left(  e^{-qt_{\kappa_{l+1}}}\left(
W(x_{m^{1+\kappa_{l}}}^{\delta},\lambda^{1+\kappa_{l}})I_{\{s_{\kappa_{l}%
}=\mathbf{E}_{0}\}}I_{\{y^{k}\geq0\}}\right)  \right)  =0,
\]
and so we have the result. \hfill$\blacksquare$\\

In order to prove Theorem \ref{Main Theorem}, we need the next definition.

\begin{definition}
We define the auxiliary function $\overline{V}:$ $[0,\infty)\times
\lbrack\underline{\lambda},\infty)\rightarrow\mathbf{R}$ as%
\[
\overline{V}(x,\lambda):=\lim\nolimits_{k\rightarrow\infty}V^{\delta_{k}%
}(x,\lambda).
\]
\end{definition}

We will prove that $\overline{V}$ is the optimal value function. In order to
do that, we will show that $\overline{V}$ is a viscosity supersolution of
(\ref{HJB}). It is straightforward to see that $\overline{V}$ is a limit of
value functions of admissible strategies in $\Pi_{x,\lambda}$ for all
$(x,\lambda)\in\lbrack0,\infty)\times\lbrack\underline{\lambda},\infty)$, so
the result will follow from Corollary \ref{VerificacionDividendos}. Since
there is no uniqueness of the solution of the HJB equation, it is essential to
show that this function is a limit of value functions of admissible
strategies. In the next lemma, we find a bound on the variation of $V^{\delta
}$ and as a consequence we obtain that $\overline{V}$ is locally Lipschitz in
$[0,\infty)\times\lbrack\underline{\lambda},\infty)$ and so it is absolutely continuous.

\begin{lemma}
\label{Vbarra Lipschitz} We have that $\overline{V}$ is locally Lipschitz in
$\mathcal{[}0,\infty)\times(\underline{\lambda},\infty)$. That is, for any
$x_{2},x_{1}\geq0\ $and for any $\lambda_{2},\lambda_{1}>\underline{\lambda}$
with $\lambda_{2}-\lambda_{1}\leq d(\lambda_{1}-\underline{\lambda}),$
\[%
\begin{array}
[c]{l}%
\left\vert \overline{V}(x_{2},\lambda_{2})-\overline{V}(x_{1},\lambda
_{1})\right\vert \\[0.1cm]
\leq\left\vert \overline{V}(x_{2},\lambda_{2})-\overline{V}(x_{1},\lambda
_{2})\right\vert +\left\vert \overline{V}(x_{1},\lambda_{2})-\overline
{V}(x_{1},\lambda_{1})\right\vert \\[0.1cm]
\leq\overline{V}(x_{2}\vee x_{1},\underline{\lambda})\frac{p+q+\left(
\lambda_{1}\vee\lambda_{2}\right)  }{p}\left\vert x_{2}-x_{1}\right\vert
+\overline{V}(x_{2}\vee x_{1},\underline{\lambda})\allowbreak\frac{\left(
\beta\left(  \lambda_{1}\vee\lambda_{2}\right)  +q\right)  }{d(\left(
\lambda_{1}\wedge\lambda_{2}\right)  -\underline{\lambda})}\left\vert
\lambda_{2}-\lambda_{1}\right\vert \\[0.1cm]
\leq\overline{V}(x_{2}\vee x_{1},\underline{\lambda})(\frac{p+q+\left(
\lambda_{1}\vee\lambda_{2}\right)  }{p}+\allowbreak\frac{\beta\left(
\lambda_{1}\vee\lambda_{2}\right)  +q}{d(\left(  \lambda_{1}\wedge\lambda
_{2}\right)  -\underline{\lambda}})(\left\vert x_{2}-x_{1}\right\vert
+\left\vert \lambda_{2}-\lambda_{1}\right\vert ).
\end{array}
\]

\end{lemma}

\noindent\textit{Proof of Lemma \ref{Vbarra Lipschitz}.} We can write, from Proposition \ref{Lipschitz Vdelta},%

\[%
\begin{array}
[c]{l}%
\overline{V}(x_{2},\lambda)-\overline{V}(x_{1},\lambda)\\[0.1cm]
=\overline{V}(x_{2},\lambda)-V^{\delta_{k}}(x_{2},\lambda)+V^{\delta_{k}%
}(x_{2},\lambda)-V^{\delta_{k}}(x_{1},\lambda)+V^{\delta_{k}}(x_{1}%
,\lambda)-\overline{V}(x_{1},\lambda)\\
\leq\overline{V}(x_{2},\lambda)-V^{\delta_{k}}(x_{2},\lambda)+\left(
\rho^{\delta_{k}}(x_{2})-\rho^{\delta_{k}}(x_{1})\right)  \frac{e^{\left(
\beta+q\right)  \delta_{k}+\int_{0}^{\delta_{k}}\lambda_{u}^{c}du}-1}%
{p\delta_{k}}V^{\delta_{k}}(x_{2},\lambda)+\delta_{k}p\\[0.1cm]
+V^{\delta_{k}}(x_{1})-\overline{V}(x_{1}).
\end{array}
\]
Also, 
\[%
\begin{array}
[c]{l}%
\overline{V}(x,\lambda_{1})-\overline{V}(x,\lambda_{2})\\[0.1cm]
=\overline{V}(x,\lambda_{1})-V^{\delta_{k}}(x,\lambda_{1})+V^{\delta_{k}%
}(x,\lambda_{1})-V^{\delta_{k}}(x,\lambda_{2})+V^{\delta_{k}}(x,\lambda
_{2})-\overline{V}(x_{1},\lambda_{2})\\[0.1cm]
\leq\overline{V}(x,\lambda_{1})-V^{\delta_{k}}(x,\lambda_{1})+V^{\delta
}(x,\lambda_{1})\allowbreak\frac{\left(  \beta\lambda_{2}+q\right)
}{d(\lambda_{1}-\underline{\lambda})}\left(  \lambda_{2}-\lambda_{1}\right)
+V^{\delta_{k}}(x,\lambda_{2})-\overline{V}(x_{1},\lambda_{2}).
\end{array}
\]
Taking the limit as $k$ goes to infinity, we obtain%
\[
\left\vert \overline{V}(x_{2},\lambda)-\overline{V}(x_{1},\lambda)\right\vert
\leq\overline{V}(x_{2}\vee x_{1},\underline{\lambda})\frac{p+q+\lambda}%
{p}\left\vert x_{2}-x_{1}\right\vert
\]
and%
\[
\left\vert \overline{V}(x,\lambda_{1})-\overline{V}(x,\lambda_{2})\right\vert
\leq\overline{V}(x,\underline{\lambda})\allowbreak\frac{\left(  \beta
\lambda_{2}+q\right)  }{d(\lambda_{1}-\underline{\lambda})}\left\vert
\lambda_{2}-\lambda_{1}\right\vert .
\]
\hfill$\blacksquare$\\

In the next lemma, we show that the convergence of $V^{\delta_{k}}$ to
$\overline{V}\ $ is locally uniformly.

\begin{lemma}
\label{Limite u barra uniforme} $V^{\delta_{k}}\nearrow\overline{V}\ $
locally uniformly as $k$ goes to infinity.
\end{lemma}

\noindent\textit{Proof of Lemma \ref{Limite u barra uniforme}.} Consider a compact set $K$ in $[0,\infty)\times(\underline{\lambda},\infty)$,
$\mathbf{(}x_{1},\lambda_{1}\mathbf{)}\in K$ and $\varepsilon>0 $. Let us
take\ $M\in\lbrack0,\infty)$ such that $M\geq x$, for all $(x,\lambda)\in K$.
We show first that there exists $k_{0}$ large enough and $\eta>0$ small enough
such that if $\left\vert x-x_{1}\right\vert +\left\vert \lambda-\lambda
_{1}\right\vert <\eta,$ and $k\geq$ $k_{0}$, then%
\begin{equation}
\overline{V}(x,\lambda)-V^{\delta_{k}}(x,\lambda)<\varepsilon.
\label{diferencia en Bolas}%
\end{equation}
Indeed, by pointwise convergence at $\mathbf{(}x_{1},\lambda_{1}\mathbf{)}$,
there exists $k_{1}$ such that
\begin{equation}
\overline{V}(x_{1},\lambda_{1})-V^{\delta_{k}}(x_{1},\lambda_{1}%
)<\varepsilon/3~\text{for }k\geq k_{1}. \label{desig 1}%
\end{equation}
By Lemma \ref{Vbarra Lipschitz}, there exists $\eta_{1}$ such that if
$\left\vert x-x_{1}\right\vert +\left\vert \lambda-\lambda_{1}\right\vert
<\eta_{1}$, then
\begin{equation}
\left\vert \overline{V}(x,\lambda)-\overline{V}(x_{1},\lambda_{1})\right\vert
<\varepsilon/3. \label{desig 2}%
\end{equation}
Also, from Proposition \ref{Lipschitz Vdelta}, there exists $\eta_{2}$ and
$k_{2}$ such that if $\left\vert x-x_{1}\right\vert +\left\vert \lambda
-\lambda_{1}\right\vert <\eta_{1}$, then%

\begin{equation}%
\begin{array}
[c]{lll}%
\left\vert V^{\delta_{k}}(x,\lambda)-V^{\delta_{k}}(x_{1},\lambda
_{1})\right\vert  & \leq & V^{\delta_{k}}(M,\underline{\lambda})(\frac
{e^{\left(  \beta+q\right)  \delta+\int_{0}^{\delta}\lambda_{u}^{c}du}%
-1}{p\delta}\\
&  & +\allowbreak\frac{\beta\left(  \lambda_{1}\vee\lambda_{2}\right)
+q}{d(\left(  \lambda_{1}\wedge\lambda_{2}\right)  -\underline{\lambda}%
)})(\rho^{\delta}(x_{2})-\rho^{\delta}(x_{1})+\left\vert \lambda_{2}%
-\lambda_{1}\right\vert )+\delta p\\
& < & \varepsilon/3
\end{array}
\label{desig 3}%
\end{equation}
for $k\geq k_{2}$. Therefore, taking $\eta:=\eta_{1}\wedge\eta_{2}$, for
$k\geq k_{0}:=k_{1}\vee k_{2}$, we obtain (\ref{diferencia en Bolas}) from
(\ref{desig 1}), (\ref{desig 2}) and (\ref{desig 3}).

Finally, we conclude the result by taking a finite covering of the compact set
$K$. \hfill$\blacksquare$\\

The next lemma is the key argument in the proof of Theorem \ref{Main Theorem}.

\begin{lemma}
\label{vbarra es supersolucion} $\overline{V}$ is a viscosity supersolution of
(\ref{HJB}) in $(0,\infty)\times(\underline{\lambda},\infty)$, and so
$\overline{V}=\lim\nolimits_{k\rightarrow\infty}V^{\delta_{k}}=V.$
\end{lemma}

\noindent\textit{Proof of Lemma \ref{vbarra es supersolucion}.} Take $(x^{0},\lambda^{0})\in\mathbf{(}0\mathbf{,\infty)\times(}\underline
{\lambda}\mathbf{,\infty)}$ and a differentiable test function $\varphi
:\mathbf{[}0\mathbf{,\infty)\times\lbrack}\underline{\lambda}\mathbf{,\infty
)}\rightarrow\mathbb{R}$ for a viscosity supersolution of (\ref{HJB}) at
$(x^{0},\lambda^{0})$, that is
\begin{equation}
\overline{V}(x\mathbf{,}\lambda)\geq\varphi(x,\lambda)\text{ and }\overline
{V}(x^{0},\lambda^{0})=\varphi(x^{0},\lambda^{0}). \label{Comparacion}%
\end{equation}
Consider the sets $K_{1}=[x^{0},x^{0}+\delta_{1}]\subset(0,\infty)$,
$K_{2}=[\lambda^{0},\lambda^{0}+1]\subset(\underline{\lambda
},\infty)$ and $K^{\delta_{k}}=(\mathcal{G}%
_{\delta_{k}}\cap K_{1})\times K_{2}\subset(0,\infty)\times
(\underline{\lambda},\infty)$. 
In order to prove that $\mathcal{L}(\varphi)(x^{0},\lambda^{0})\leq0,$
consider now, for $\eta>0$ small enough,
\[
\varphi_{\eta}(x\mathbf{,}\lambda)=\varphi(x\mathbf{,}\lambda)-\eta\left(
\left(  x-x^{0}\right)  ^{2}\mathbf{+}(\lambda-\lambda^{0})^{2}\right)  .
\]
Given $k\geq0$, the set $K^{\delta_{k}}$ is non-empty and compact, so we can
define
\begin{equation}
a_{k}^{\eta}:=\min\nolimits_{K^{\delta_{k}}}\{V^{\delta_{k}}(x\mathbf{,}%
\lambda)-\varphi_{\eta}(x\mathbf{,}\lambda)\} \label{minimo V_delta-Fi}%
\end{equation}
and
\begin{equation}
\left(  x_{k}^{\eta},\lambda_{k}^{\eta}\right)  :=\arg\min\nolimits_{K^{\delta
_{k}}}\{V^{\delta_{k}}(x\mathbf{,}\lambda)-\varphi_{\eta}(x\mathbf{,}%
\lambda)\}\in K^{\delta_{k}} \label{Definicion  argmin}.
\end{equation}
Since $V^{\delta_{k}}\leq\overline{V}$, we have from (\ref{Comparacion}), that
$a_{k}^{\eta}\leq0$. Taking%
\[
0\leq b_{k}^{\eta}:=\max\nolimits_{K^{\delta_{k}}}\{\overline{V}%
(x\mathbf{,}\lambda)-V^{\delta_{k}}(x\mathbf{,}\lambda)\},
\]
by Lemma \ref{Limite u barra uniforme}, $a_{k}^{\eta}\rightarrow0$ and
$b_{k}^{\eta}\rightarrow0$ as $k\rightarrow\infty$. Moreover, for all
$(x\mathbf{,}\lambda)\in K^{\delta_{k}}$, we get from (\ref{Comparacion}),
(\ref{minimo V_delta-Fi}) and (\ref{Definicion argmin}) that
\[%
\begin{array}
[c]{l}%
a_{k}^{\eta}=V^{\delta_{k}}(x_{k}^{\eta},\lambda_{k}^{\eta})-\varphi_{\eta
}(x_{k}^{\eta},\lambda_{k}^{\eta})\\[0.1cm]
=V^{\delta_{k}}(x_{k}^{\eta},\lambda_{k}^{\eta})-\overline{V}(x_{k}^{\eta
},\lambda_{k}^{\eta})+\overline{V}(x_{k}^{\eta},\lambda_{k}^{\eta}%
)-\varphi(x_{k}^{\eta},\lambda_{k}^{\eta})+\eta\left(  \left(  x_{k}^{\eta
}-x^{0}\right)  ^{2}\mathbf{+}(\lambda_{k}^{\eta}-\lambda^{0})^{2}\right) \\[0.1cm]
\geq-b_{k}+\eta\left(  \left(  x_{k}^{\eta}-x^{0}\right)  ^{2}\mathbf{+}%
(\lambda_{k}^{\eta}-\lambda^{0})^{2}\right).
\end{array}
\]
Then, the minimum argument in (\ref{minimo V_delta-Fi}) is attained at
$\left(  x_{k}^{\eta},\lambda_{k}^{\eta}\right)  \in K^{\delta_{k}}$ such
that
\[
\left(  x_{k}^{\eta}-x^{0}\right)  ^{2}\mathbf{+}(\lambda_{k}^{\eta}%
-\lambda^{0})^{2}\leq(b_{k}+a_{k})/\eta.
\]
Hence, $\left(  x_{k}^{\eta},\lambda_{k}^{\eta}\right)  \rightarrow(\left(
x^{0}\right)  ^{+},\left(  \lambda^{0}\right)  ^{+})$ as $k$ goes to infinity.
So
\begin{equation}
V^{\delta_{k}}(x\mathbf{,}\lambda)\geq\varphi_{\eta}(x\mathbf{,}\lambda
)+a_{k}^{\eta}\text{ for }(x\mathbf{,}\lambda)\in K^{\delta_{k}}\text{ and
}V^{\delta_{k}}\left(  x_{k}^{\eta},\lambda_{k}^{\eta}\right)  =\varphi_{\eta
}\left(  x_{k}^{\eta},\lambda_{k}^{\eta}\right)  +a_{k}^{\eta}.
\label{Desigualdad Vdelta Fieta}%
\end{equation}
Since
\[
\mathcal{T}_{0}(V^{\delta_{k}})\left(  x_{k}^{\eta},\lambda_{k}^{\eta}\right)
-V^{\delta_{k}}\left(  x_{k}^{\eta},\lambda_{k}^{\eta}\right)  \leq0,
\]
we obtain%
\[%
\begin{array}
[c]{lll}%
0 & \geq & \lim_{k\rightarrow\infty}\frac{\mathcal{T}_{0}^{\delta_{k}}%
(\varphi_{\eta})(x_{k}^{\eta},\lambda_{k}^{\eta})-\varphi_{\eta}(x_{k}^{\eta
},\lambda_{k}^{\eta})-a_{k}^{\eta}\left(  1-e^{-q\delta_{k}}\right)  }%
{\delta_{k}}\\
& = & \lim_{k\rightarrow\infty}\frac{\mathcal{T}_{0}^{\delta_{k}}%
(\varphi_{\eta})(x_{k}^{\eta},\lambda_{k}^{\eta})-\varphi_{\eta}(x_{k}^{\eta
},\lambda_{k}^{\eta})}{\delta_{k}}\\[0.1cm]
& = & \mathcal{L}(\varphi_{\eta})(x^{0},\lambda^{0}).
\end{array}
\]
Since $\partial_{x}(\varphi_{\eta})(x^{0},\lambda^{0})=\partial_{x}%
(\varphi)(x^{0},\lambda^{0})$ and $\partial_{\lambda}(\varphi_{\eta}%
)(x^{0},\lambda^{0})=\partial_{\lambda}(\varphi)(x^{0},\lambda^{0})$ and
$\varphi_{\eta}\nearrow\varphi$ as $\eta\searrow0$, we obtain that
$\mathcal{L}(\varphi)(x^{0},\lambda^{0})\leq0$ and the result follows.
\hfill$\blacksquare$\\

\noindent\textit{Proof of Theorem \ref{Main Theorem}.}
From Lemmas \ref{Limite u barra uniforme} and \ref{vbarra es supersolucion} we
have that for any $\delta>0$, the functions $V^{\delta_{k}}\nearrow
\overline{V}=V$ locally uniformly as $k$ goes to infinity. From Proposition
\ref{Convergencia Uniforme con Lambda}, it is enough to show that
$V^{\delta_{k}}\nearrow V$ uniformly in $[0,\infty)\times\lbrack
\underline{\lambda},\lambda_{1}]$ for any $\lambda_{1}>\underline{\lambda}.$
From Proposition \ref{Proposicion acotacion de retirar}, take $x>x^{\ast
}=p/q+\delta_{k}$, then%
\[
V^{\delta_{k}}(x^{\ast},\lambda)+(x-x^{\ast})\leq V^{\delta_{k}}%
(x,\lambda)\leq V(x,\lambda)=V(x^{\ast},\lambda)+(x-x^{\ast}),
\]
so
\[%
\begin{array}
[c]{lll}%
V(x,\lambda)-V^{\delta_{k}}(x,\lambda) & = & V(x^{\ast},\lambda)+(x-x^{\ast
})-V^{\delta_{k}}(x,\lambda)\\[0.1cm]
& \leq & V(x^{\ast},\lambda)+(x-x^{\ast})-(V^{\delta_{k}}(x^{\ast}%
,\lambda)+(x-x^{\ast}))\\[0.1cm]
& = & V(x^{\ast},\lambda)-V^{\delta_{k}}(x^{\ast},\lambda),
\end{array}
\]
and the result follows. \hfill$\blacksquare$

\subsection{Proofs of Section \ref{Section Disc Intensity}%
\label{Appendix Disc Intensity}}

\textit{Proof of Proposition \ref{cercania funciones especiales}. }
Since $\widehat{\lambda}_{t}\geq\lambda_{t}$ for all $t\geq0$, with a proof
analogous to the one of Proposition \ref{Continuidad Uniforme en Lambda}-(1),
one can prove that $\widehat{V}^{\delta,\Delta}\leq V^{\delta}$. Let us prove
now that $\lim_{\Delta\rightarrow0}\left(  \sup_{x\geq0,\lambda\geq
\underline{\lambda}}V^{\delta}(x,\lambda)-\widehat{V}^{\delta,\Delta
}(x,\lambda)\right)  =0$. It is enough to do the proof for $\left(
x,\lambda\right)  \in\mathcal{G}_{\delta}\times\mathcal{H}_{\Delta}$. Take the
optimal $\mathcal{G}_{\delta}$-strategy $\pi=\left(  L,\infty\right)
\in\widetilde{\Pi}_{x,\lambda}$ \ such that
\[
V^{\delta}(x,\lambda)=J(\pi;x,\lambda)
\]
and let us call the corresponding ruin time of the controlled process
$X_{t}^{L}=X_{t}-L_{t}$ as $\tau^{L}.$

Given $\delta$ and $\Delta$, since $0\leq\widehat{\lambda}_{t}-\lambda_{t}%
\leq\Delta$, we can write the Poisson process $\widehat{N}_{t}$ as
$\widehat{N}_{t}=N_{t}+\overline{N}_{t}$ where $\overline{N}_{t}$ is a Poisson
process independent of $N_{t}$ and intensity $\widehat{\lambda}_{t}%
-\lambda_{t}$. Therefore,
\begin{equation}
\widehat{X}_{t}=X_{t}-\sum_{m=1}^{\overline{N}_{t}}\overline{U}_{m}.
\label{Relacion Xt y Xt con intensidad discreta}%
\end{equation}
Define
\[
\widehat{\tau}_{1}=\sup\{t:\widehat{X}_{t}-L_{t}\geq0\},
\]%
\[
\widehat{L}_{t}=L_{t}\ I_{\{t<\widehat{\tau}_{1}\}}+L_{\widehat{\tau}_{1}%
}\ I_{\{t\geq\widehat{\tau}_{1}\}}%
\]
and
\[
\widehat{\tau}^{F}=\widehat{\tau}_{1}I_{\{X_{\widehat{\tau}_{1}}%
-L_{\widehat{\tau}_{1}^{-}}\geq0\}}+\infty I_{\{X_{\widehat{\tau}_{1}%
}-L_{\widehat{\tau}_{1}^{-}}<0\}}.
\]
Then, taking $\widehat{\pi}=\left(  \widehat{L},\widehat{\tau}^{F}\right)$, it holds that $\widehat{\pi}\in\widehat{\Pi}_{x,\lambda}^{\delta,\Delta}$. That is, it is admissible for the discrete intensity process $\widehat
{\lambda}_{t}$. Denote the ruin time of as $\widehat{\tau}^{L}$ and
$\widehat{\tau}_{2}=\widehat{\tau}^{L}\wedge$ $\widehat{\tau}^{F}$. So we have
from (\ref{Relacion Xt y Xt con intensidad discreta}) $\tau^{L}\geq
\widehat{\tau}_{2}$ and%
\[
X_{\widehat{\tau}_{2}}^{L}=X_{\widehat{\tau}_{2}}-L_{\widehat{\tau}_{2}%
}=\widehat{X}_{\widehat{\tau}_{2}}+\sum_{m=1}^{\overline{N}_{\widehat{\tau
}_{2}}}\overline{U}_{m}-L_{\widehat{\tau}_{2}}\leq\sum_{m=1}^{\overline
{N}_{\widehat{\tau}_{2}}}\overline{U}_{m}\text{.}%
\]
We therefore can write
\begin{equation}%
\begin{array}
[c]{lll}%
J(\pi;x,\lambda)-J_{\widehat{\lambda}}(\widehat{\pi};x,\lambda) & = &
\mathbb{E}(\int_{\widehat{\tau}_{2}{}^{-}}^{\tau^{L}}e^{-qs}dL_{s}%
)-\mathbb{E}(I_{\left\{  \widehat{\tau}^{F}<\widehat{\tau}^{L}\right\}
}e^{-q\tau^{F}}(\widehat{X}_{\widehat{\tau}^{F}}-L_{\left(  \widehat{\tau}%
^{F}\right)  ^{-}}))\\
& \leq & \mathbb{E}(\int_{\widehat{\tau}_{2}{}^{-}}^{\tau^{L}}e^{-qs}dL_{s})\\
& \leq & \mathbb{E}\left(  e^{-q\widehat{\tau}_{2}}I_{\left\{  \widehat{\tau
}_{2}<\tau^{L}\right\}  }V(\sum_{m=1}^{\overline{N}_{\widehat{\tau}_{2}}%
}\overline{U}_{m},\lambda_{\widehat{\tau}_{2}})))\right) \\
& \leq & \mathbb{E}\left(  e^{-q\widehat{\tau}_{2}}I_{\left\{  \widehat{\tau
}_{2}<\tau^{L}\right\}  }(\sum_{m=1}^{\overline{N}_{\widehat{\tau}_{2}}%
}\overline{U}_{m}+\frac{p}{q})\right)  .
\end{array}
\label{Diferencia valor estrategia con intensidad discreta}%
\end{equation}

From $\widehat{\lambda}_{t}-\lambda_{t}\leq\Delta,$ we get that $\mathbb{E}%
\left(  \overline{N}_{t}\right)  \leq t\Delta$ for any $t\geq0$. So, given
$\varepsilon>0$ and taking $\Delta\leq1$ and $T$ large enough such that
$e^{-qT}\left(  T\Delta\mathbb{E(}U_{1}\mathbb{)+}\frac{p}{q}\right)
\leq\frac{\varepsilon}{2}$, we get%

\[%
\begin{array}
[c]{ccl}%
\mathbb{E}\left(  I_{\left\{  \widehat{\tau}_{2}<\tau^{L}\right\}
}I_{\left\{  \widehat{\tau}_{2}>T\right\}  }e^{-q\widehat{\tau}_{2}}\sum
_{m=1}^{\overline{N}_{\widehat{\tau}_{2}}}\overline{U}_{m}+\frac{p}{q})\right)
& \leq & e^{-qT}\mathbb{E}\left(  \sum_{m=1}^{\overline{N}_{\widehat{\tau}%
_{2}}}\overline{U}_{m}+\frac{p}{q}\right) \\
& \leq & e^{-qT}\left(  T\Delta~\mathbb{E(}U_{1}\mathbb{)+}\frac{p}{q}\right)
\\
& \leq & \frac{\varepsilon}{2}.
\end{array}
\]

Moreover, $\widehat{\tau}_{2}<\tau^{L}$ implies $\overline{N}_{\widehat{\tau
}_{2}}\geq1,$ so choosing $\Delta\leq\varepsilon/(4\left(  \mathbb{E(}%
U_{1}\mathbb{)}T+\frac{p}{q}\right)  )$ and since $\mathbb{P}(\overline{N}%
_{t}\geq1)\leq1-e^{-\Delta t}$, we obtain%

\[%
\begin{array}
[c]{l}%
\mathbb{E}\left(  I_{\left\{  \widehat{\tau}_{2}<\tau^{L}\right\}
}I_{\left\{  \widehat{\tau}_{2}\leq T\right\}  }e^{-q\widehat{\tau}_{2}}%
(\sum_{m=1}^{\overline{N}_{\widehat{\tau}_{2}}}\overline{U}_{m}+\frac{p}%
{q})\right) \\
\leq\mathbb{E}\left(  I_{\overline{N}_{T}\geq1}(\sum_{m=1}^{\overline{N}_{T}%
}\overline{U}_{m}+\frac{p}{q})\right) \\
\leq\mathbb{P}\left[  \overline{N}_{T}\geq1\right]  ~\mathbb{E}\left(
\sum_{m=1}^{\overline{N}_{T}}\overline{U}_{m}+\frac{p}{q}\right) \\
\leq(1-e^{-\Delta t})\left(  T\Delta\mathbb{E(}U_{1}\mathbb{)}+\frac{p}%
{q}\right) \\
\leq\frac{\varepsilon}{2}.
\end{array}
\]

So, from (\ref{Diferencia valor estrategia con intensidad discreta}),
\[
J(\pi;x,\lambda)-J_{\widehat{\lambda}}(\widehat{\pi};x,\lambda)\leq
\varepsilon\text{ for }\Delta\leq\frac{\varepsilon}{4\left(  T\mathbb{E(}%
U_{1}\mathbb{)}+\frac{p}{q}\right)  },
\]
and we get the result.\hfill$\blacksquare$\\

\noindent\textit{Proof of Theorem} \ref{Teorema Aproximacion Uniforme}.
On the one hand, since $\widehat{\lambda}_{t}\geq\lambda_{t}$ for all $t\geq
0$, one can prove that $\widehat{V}^{\delta,\Delta}\leq W_{\mathcal{P}%
^{\delta,\Delta}}$ in $[0,\infty)\times\lbrack\underline{\lambda},\infty
)\ $with a proof analogous to the one of Proposition
\ref{Continuidad Uniforme en Lambda}-(1). On the other hand, given any
$\varepsilon>0$, from Theorem \ref{Main Theorem} and from Proposition
\ref{cercania funciones especiales} there exists $\delta$ and $\Delta$ small
enough so that
\[
0\leq V-\widehat{V}^{\delta,\Delta}(x,\lambda)\leq\varepsilon
\]
in $[0,\infty)\times\lbrack\underline{\lambda},\infty), $ which establishes the
result. \hfill$\blacksquare$

\bigskip

\bigskip

\bigskip

\end{document}